\documentclass[fleqn,usenatbib]{mnras}

\usepackage{newtxtext,newtxmath}
\usepackage[T1]{fontenc}

\DeclareRobustCommand{\VAN}[3]{#2}
\let\VANthebibliography\thebibliography
\def\thebibliography{\DeclareRobustCommand{\VAN}[3]{##3}\VANthebibliography}

\usepackage{graphicx} % Including figure files
\usepackage{amsmath} % Advanced maths commands
\usepackage{hyperref} % Including as Arxiv no longer does
\usepackage{xspace} % Adaptive spaces for macros
\usepackage{orcidlink} % ORCiD links for author list

\newcommand{\coion}{$^{13}$CO\xspace} % shortcut for 13CO
\newcommand{\htwo}{H$_2$\xspace} % shortcut for H2
\newcommand{\htwoco}{H$_2$CO\xspace} % shortcut for H2CO
\newcommand{\rc}{R\mbox{-}C\xspace} % shortcut for R-C with an unbreakable hyphen

\title[Detecting HISA using Neural Networks]{Detecting \ion{H}{i} Self-Absorption using Neural Networks}

\author[E. G. M. Muller et al.]{Eric G. M. Muller,$^{1}$\thanks{Corresponding author: eric.muller@anu.edu.au}\orcidlink{0000-0001-5621-1577}
Naomi M. McClure-Griffiths,$^{1,2}$\orcidlink{0000-0003-2730-957X}
Hiep Nguyen,$^{1}$\orcidlink{0000-0002-2712-4156}
Matthew J. Alger,$^{3}$\orcidlink{0000-0001-5110-8845}
\newauthor
Frances Buckland-Willis,$^{4}$\orcidlink{0000-0003-4213-8094}
J. R. Dawson,$^{5,6}$\orcidlink{0000-0003-0235-3347}
Min-Young Lee,$^{7}$\orcidlink{0000-0002-9888-0784}
Antoine Marchal$^{4}$\orcidlink{0000-0002-5501-232X}
\\
$^{1}$Research School of Astronomy and Astrophysics, The Australian National University, Canberra, ACT 2611, Australia\\
$^{2}$SKA Observatory, Jodrell Bank, Lower Withington, Macclesfield, SK11 9FT, UK\\
$^{3}$Google Australia, Pyrmont, Australia\\
$^{4}$Laboratoire de Physique de l’Ecole Normale Supérieure, ENS, Université PSL, CNRS, Sorbonne Université, Université de Paris, 24 rue Lhomond,\\\hspace{0.1cm} 75005 Paris Cedex 05, France\\
$^{5}$School of Mathematical and Physical Sciences and Astrophysics and Space Technologies Research Centre, Macquarie University, Sydney 2109, Australia\\
$^{6}$Australia Telescope National Facility, CSIRO Space \& Astronomy, PO Box 76, Epping, NSW 1710, Australia\\
$^{7}$Korea Astronomy and Space Science Institute, 776, Daedeokdae-ro, Yuseong-gu, Daejeon 34055, Republic of Korea}

\date{Accepted XXX. Received YYY; in original form ZZZ}

\pubyear{\the\year{}}

\begin{document}
\label{firstpage}
\pagerange{\pageref{firstpage}--\pageref{lastpage}}
\maketitle

\begin{abstract}
Cold atomic hydrogen plays a crucial role in the life cycle of interstellar gas. It serves as the intermediary phase in the condensation and cooling processes that bridge the warm diffuse gas in and around galaxies, and the cold molecular gas that drives star formation. \ion{H}{i} self-absorption in the 21-cm emission line is the most direct observational tracer of cold \ion{H}{i} gas that does not require a continuum background source, yet its systematic extraction remains a long-standing challenge. Existing methods of self-absorption identification rely on subjective by-eye inspection or modelling of the underlying emission, making repeatable, thorough, and large-scale applications difficult. We present a lightweight convolutional neural network designed to detect self-absorption features and infer their velocities via a post-hoc process, without assumptions about the underlying emission or the use of ancillary data. Trained on synthetic emission spectra, the neural network achieves 96.5 per cent accuracy, 96.0 per cent precision, and 97.0 per cent recall on held-out synthetic data. Applied to observed 21-cm emission data from the Riegel--Crutcher cloud and giant molecular filament regions towards the Galactic Plane, the network recovers the spatial distributions and velocities of known self-absorption structures when compared to previous analyses and ancillary \coion emission data. Crucially, the neural network is computationally efficient, processing detections for $\sim15,000$ spectra per second on a single consumer laptop GPU, enabling real-time cold \ion{H}{i} detection at the data rates anticipated by next-generation facilities such as the Square Kilometre Array.
\end{abstract}


%%%%%%%%%%%%%%%%%%%%%%%%%%%%%%%%%%%%%%%%%%%%%%%%%%

%%%%%%%%%%%%%%%%% BODY OF PAPER %%%%%%%%%%%%%%%%%%

\section{Introduction}\label{sec:introduction}

Neutral hydrogen (\ion{H}{i}) is the dominant baryonic component of the interstellar medium (ISM). The ISM is a thermally inhomogeneous medium in which the reservoir of \ion{H}{i} gas exists in a multiphase structure \citep{wolfire1995, wolfire2003}. The gas occupies two thermally stable phases and a third unstable phase \citep{field1969, mckee1977, mcclure-griffiths2023}. The stable warm neutral medium (WNM, $T \sim 5,000$--10,000~K, $n \sim 0.1$--$1~{\rm cm}^{-3}$) transitions through the unstable neutral medium (UNM, $T \sim 250$--5,000~K) and into the stable cold neutral medium (CNM, $T \lesssim 250$~K, $n \sim 10$--$100~{\rm cm}^{-3}$) in a thermodynamic cascade that begins with diffuse atomic gas and provides the cold material needed to form dense molecular gas \citep[\htwo;][]{wolfire2003, audit2005a, inutsuka2016}.

\begin{figure}
    \centering
    \includegraphics[width=\linewidth]{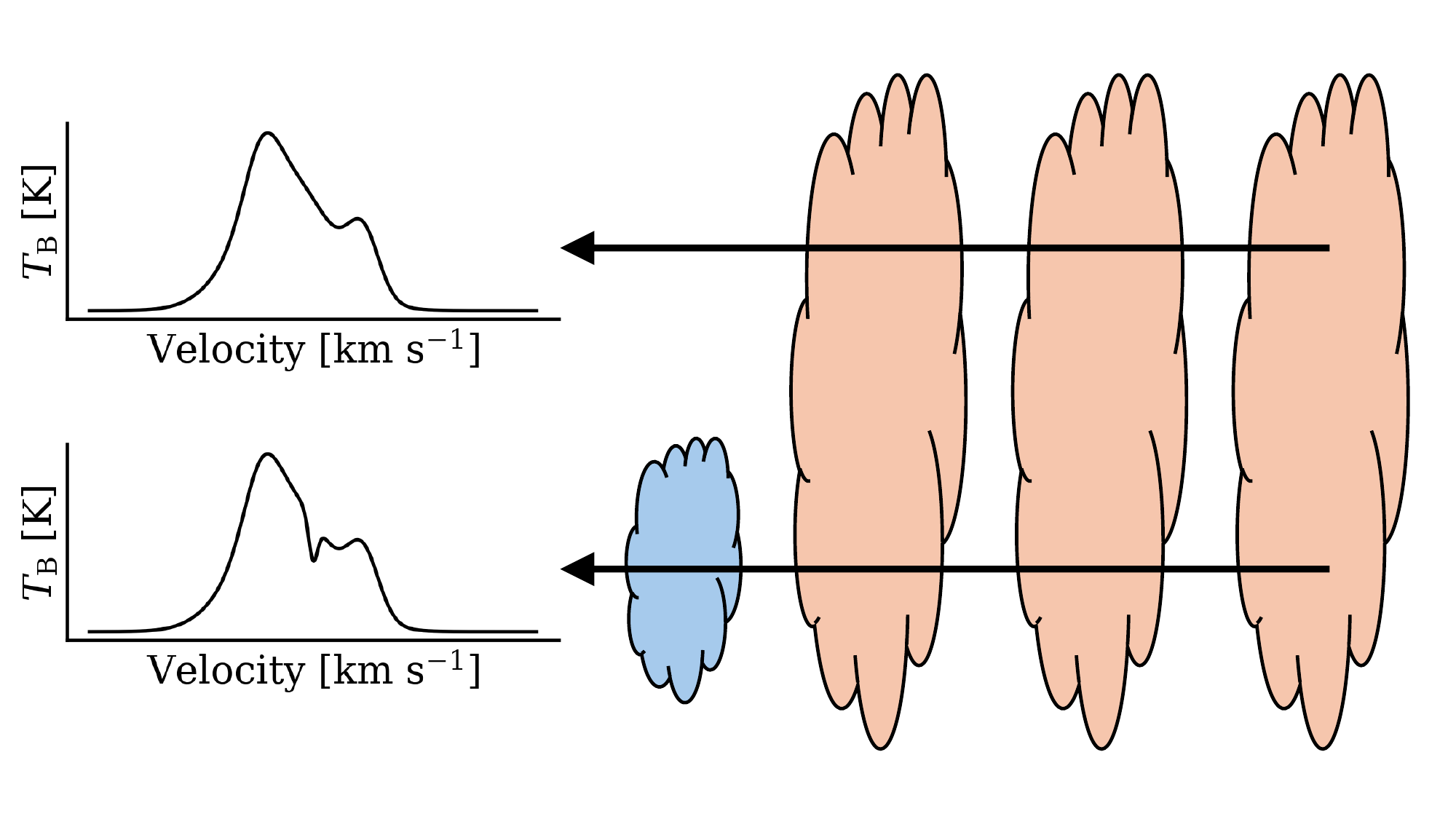}
    \caption{An example of \ion{H}{i} self-absorption arising from the presence of a cold cloud of CNM in front of a warm background. Clouds of normal \ion{H}{i} are shown in orange, and the HISA cloud is shown in blue. The resulting spectra on the left demonstrate the effect of introducing the HISA cloud in front of the other clouds in the column, inducing self-absorption against the emission background. Alt text: Graphical representation showing arrows through three orange clouds and one blue cloud, and showing two simple spectra that highlight an absorption feature caused by the blue cloud.}
    \label{fig:mock cloud diagram}
\end{figure}

\ion{H}{i} self-absorption \citep[HISA;][]{heeschen1955, gibson2000} offers a direct approach to identifying CNM gas that does not require a background continuum source like traditional absorption measurements. It also does not require an explicit model of the CNM structure at inference time. HISA arises via the same mechanism as traditional 21-cm absorption spectra, but instead warm \ion{H}{i} emission itself is the background. Consequently, HISA can be detected in emission along any line-of-sight where the column contains a cold cloud in front of a warm background cloud at the same Doppler velocity \citep{dickey1990}. This simple requirement translates directly into a large number of usable sightlines, especially in the case of regions with strong emission, and can lead to complete spatial coverage of cold gas. HISA manifests with the same structure as typical 21-cm absorption, often consisting of a narrow inverted line profile feature. Unlike absorption measurements against continuum background sources, the non-uniform structure of the \ion{H}{i} emission background can complicate the structure of the HISA feature, and foreground \ion{H}{i} emission at the same velocity can `fill-in' regions of the absorption feature. Figure~\ref{fig:mock cloud diagram} illustrates the narrow absorption feature produced by a cold \ion{H}{i} cloud in front of a warm column. The prominence of the feature is due to the simple morphology of the background and the lack of any foreground emission, and as such represents a best-case scenario.

Atomic gas traced by HISA is known to be genuinely cold: typical spin temperatures inferred from HISA features are below 100~K, and frequently below even 50~K, well within the CNM regime \citep{gibson2000, mcclure-griffiths2006, moss2012, syed2023}. A method capable of detecting and localising self-absorption in emission spectra could address limitations of existing approaches for detecting cold \ion{H}{i} gas, which typically rely on specific assumptions about the underlying emission or the properties of the CNM. Despite these advantages, reliably identifying HISA in \ion{H}{i} emission spectra has remained challenging. No existing approach has proven adequate for large-scale automated applications. Early identifications were made by-eye in cases of prominent, isolated features, and were often confirmed using molecular observations of species such as \htwoco and OH \citep[e.g.][]{knapp1974}. However, this procedure is neither efficient nor reproducible due to detections not being consistent between analyses. \citet{gibson2000} codified a set of criteria that could be used to identify HISA features in an attempt to make identifications of self-absorption more consistent, and while a significant advance, these criteria introduced selection effects that biased detections towards narrow, compact features \citep[HINSA;][]{li2003} with angular extents smaller than typical cold \ion{H}{i} clouds seen in emission.

An automated implementation using a modified CLEAN algorithm to iteratively remove large-scale emission structures and flag small-scale negative residuals as self-absorption features extended this approach to process large data volumes but inherited the same selection biases \citep{gibson2005}. An alternative class of methods employs derivative spectroscopy to enhance the contrast of absorption features against the background spectrum \citep{krco2008, denes2018}, which removes the need for explicit feature morphology criteria. Other methods attempt instead to model a theoretical `unabsorbed' background emission profile, after which taking the difference between the modelled and observed spectrum would reveal HISA features: \citet{mcclure-griffiths2006} used this method to derive the properties of the self-absorption in the case of a very prominent and visible example of HISA, while \citet{syed2023} more recently focused on applying this method more broadly as a way to detect less prominent features in large datasets, using machine learning to estimate the background emission more accurately.

Existing methods share limitations that become increasingly significant in the era of large-scale \ion{H}{i} surveys. Criteria-based methods depend on assumptions about the characteristics of the self-absorption features, and their sensitivity degrades when the background spectrum varies steeply or when the self-absorption is weak. While derivative-based methods offer a more unbiased detection algorithm, higher-order derivatives greatly amplify the noise in observed spectra, meaning that confident detections typically require confirmation from external tracers like molecular data. Methods that model the unabsorbed emission, ranging from techniques as simple as linear interpolation to as complex as Gaussian processes, make assumptions about the structure of both the emission and any HISA features. In all cases, current catalogues of identified HISA features are expected to be incomplete, and computational cost is a practical constraint at modern survey data volumes: methods that require iterative fitting on a spectrum-by-spectrum basis are challenging to scale to tens or hundreds of millions of spectra.

The issues with previous approaches to CNM mapping motivate an approach that is purely data-driven, requiring no explicit physical model of the background emission and no ancillary tracers, but that is also applicable to large data volumes. We present a supervised one-dimensional convolutional neural network (CNN) designed to detect HISA features in \ion{H}{i} 21-cm emission spectra as a way to observationally trace the CNM. CNNs have demonstrated strong performance in extracting and learning features from one-dimensional spectral data, as they learn localised spectral patterns in a translation-invariant manner (i.e., recognise HISA features across different positions in the spectrum). They are especially well-suited to astrophysical contexts \citep[e.g.][]{murray2020, rozanski2022, nguyen2025, kessler2025}, and more specifically for the detection of cold gas in the ISM: \citet{murray2020} and \citet{nguyen2025} showed that convolutional architectures can reliably extract cold gas signatures from \ion{H}{i} emission spectra, motivating their application here to the related but distinct problem of direct HISA detection. 

The translation-invariant architecture of the CNN allows it to learn the characteristic absorption morphology of HISA across a range of emission backgrounds, without fitting an explicit parametric model to the spectra at inference time or relying on ancillary tracers of the CNM. This is the principal advantage of the approach over the methods discussed above: HISA detection can be performed directly from the \ion{H}{i} emission data and can be efficiently scaled to large data sets.

This approach is not, however, free from physical assumptions. In particular, the choices made when generating synthetic training spectra from models of \ion{H}{i} clouds determine the range of physical and observational conditions represented in the training set and hence influence the network's performance. The limitation becomes important when applying the trained network to real observations whose statistical properties, noise characteristics, or HISA morphology differ from those spanned in the training data.

The remainder of this paper is organized as follows. Section~\ref{sec:ml framework} details the machine learning framework and the synthetic dataset, as well as the training process and the velocity localisation method. Section~\ref{sec:observed data} describes the observational datasets on which we apply our trained neural network. Section~\ref{sec:results} presents results on the synthetic spectra, as well as two observed datasets towards the Galactic Plane (the Riegel--Crutcher cloud and several giant molecular filaments). Section~\ref{sec:discussion} discusses the advantages and limitations of the approach, as well as future applications and work. Section~\ref{sec:conclusion} summarises our findings.

\section{Supervised Machine Learning for HISA Detection}\label{sec:ml framework}

We design a binary classifier for the detection of HISA in 21-cm emission spectra as a way to trace CNM gas in \ion{H}{i} observations. The classifier is designed as a CNN binary classifier that assigns each input spectrum a probability of containing a HISA feature and the corresponding probability of not containing a HISA feature ($\mathbb{P}({\rm HISA})$ and $\mathbb{P}(\neg{\rm HISA})$, respectively), which then determines a final classification. In addition to this detection step, we develop a post-hoc method based on saliency mapping \citep{simonyan2014} to infer the location of detected features in velocity space. This architecture provides a detection with a standard forward pass of the neural network, and a velocity estimate of the detection from a backward pass of the neural network, without any additional implementations in the neural network architecture, meaning the velocity localisation can be used for any neural network architecture that detects similar features. Below, we detail the synthetic dataset used to train and evaluate our neural network, the architectural design process, and the performance of the final network on synthetic data.

\subsection{Synthetic dataset}\label{sec:synthetic data}

Neural networks designed for classification tasks typically use supervised training to learn the predictive features of the inputs. This requires comparing the output of the classifier with the label of the input from the training dataset. To detect HISA in spectra, this requires a dataset where every emission spectrum is definitively labelled as containing a real HISA feature or not. While small datasets of real HISA detections exist, they all suffer from observational effects and biases depending on the method of identification used. Furthermore, the rarity of \ion{H}{i} self-absorption makes generating a sufficiently large and balanced training set directly from ISM simulations impractical without substantial oversampling or a much larger simulation volume.

Previous \ion{H}{i} analyses using neural networks such as \citet{murray2020} used ISM simulations to construct training data because their target quantities could be determined for every line of sight, without requiring any knowledge of the gas structure, physical configuration, or phase.  HISA, however, presents a different challenge: it requires sufficiently cold gas to lie in front of bright background HI emission at approximately the same velocity. Consequently, HISA occurs along only a small fraction of sightlines in typical ISM simulations. To quantify this limitation, we performed a simple HISA detection analysis on a simulated ISM cube from the TIGRESS suite \citep{kim2017, kado-fong2020} which has previously been used to generate training data for \ion{H}{i} CNNs. We found that, on average, in the absence of spectral noise, only 0.91 per cent of spectra contained a HISA feature of any magnitude.

This strong class imbalance would cause a neural network trained directly on the simulated spectra to be dominated by the non-HISA class and would therefore require substantial oversampling or other forms of data augmentation to obtain a balanced training set. We therefore construct a new synthetic dataset of spectra, independent of existing ISM simulations, for training, validation, and testing of the neural network. This approach allows HISA to be introduced into exactly half of the spectra, providing a naturally balanced dataset. We additionally impose a minimum HISA depth threshold, which reduces ambiguity in the labels by excluding very weak or marginal features and facilitates robust network training.

We generate a large dataset of synthetic \ion{H}{i} emission spectra using a single-sightline, cloud-based model, calculating radiative transfer through a column of simulated \ion{H}{i} clouds. Each line of sight consists of several `normal' \ion{H}{i} clouds in the background and either one or no HISA cloud in front of the rest of the column of normal clouds. Here, we adopt a single foreground HISA cloud per sightline for tractability, meaning there is at most a single self-absorption feature per spectrum. Each cloud is assumed to be homogeneous and isothermal. Each sightline is constructed by uniformly drawing between 50 and 150 normal \ion{H}{i} clouds, and either zero or one HISA cloud. This high density of clouds effectively simulates the overlapping emission components common in dense interstellar environments, replicating the spectral complexity found in the inner Galactic Plane. Figure~\ref{fig:mock cloud diagram} again shows a mock version of the cloud arrangement used to generate the synthetic spectra and the effect of introducing a HISA cloud into the line of sight.

The characteristics of \ion{H}{i} 21-cm emission vary greatly across the sky, reflecting the wide range of kinematic, thermal, and density conditions of atomic gas in the Milky Way. To capture this diversity and ensure the network encounters a broad range of HISA morphologies during training, the filling factor, volume density, line-of-sight depth, spin temperature, Mach number, and central velocity of each cloud are drawn from wide uniform distributions spanning the known physical conditions of atomic gas in the Galaxy \citep{heiner2008, seifried2022a, mcclure-griffiths2023}. The properties of the normal clouds in the synthetic dataset are sampled from the uniform distributions described in the first row of Table~\ref{tab:slab parameters}. The velocity range of the \ion{H}{i} clouds is chosen to match the velocities of gas in the first quadrant of the Milky Way \citep{dame2001, sparke2007}, mimicking spectra towards the Galactic Plane between Galactic longitudes of $0\degr < l < 90\degr$. When a HISA cloud is introduced in front of a column, its properties are drawn from the uniform distributions described in the second row of Table~\ref{tab:slab parameters}. The velocity of the HISA cloud is drawn from the velocities of the other clouds in the column, with a small offset randomly sampled from $-5$ to $5~{\rm km~s}^{-1}$. This offset ensures that the resulting absorption feature has sufficient background emission to absorb against, while preventing the self-absorption feature from always coinciding with an emission peak. For each spectrum in the dataset, a new column of clouds is generated independently.

\begin{table*}
    \centering
    \caption{Parameters used for synthetic \ion{H}{i} spectra generation. All reported distributions are uniform, except for the HISA cloud velocity, which is chosen randomly from the velocities of the non-HISA clouds in the column, with a randomly sampled offset between -5 and $5~{\rm km~s}^{-1}$. Therefore, it is drawn from the same range, but the distribution is no longer uniform. Velocities are chosen to mimic velocities of \ion{H}{i} clouds in the first quadrant ($0\degr < l < 90\degr$) of the Milky Way.}
    \begin{tabular}{rcccccc}
        \hline
        Cloud Type & Velocity & Filling Factor $f$ & $n_{\rm HI}$ & Depth & $T_{\rm spin}$ & $\mathcal{M}$\\
        & km~s$^{-1}$ &  & cm$^{-3}$ & pc & K\\
        \hline
        Normal \ion{H}{i} cloud & $[-30, 120]$ & $[0.5, 1.0]$ & $[0.2, 10]$ & $[15, 50]$ & $[50, 8000]$ & $[0.7, 1.3]$\\
        HISA cloud & $[-30,120]$ & $[0.8, 1.0]$ & $[20, 30]$ & $[5, 15]$ & $[20, 75]$ & $[3, 10]$\\
        \hline
    \end{tabular}
    \label{tab:slab parameters}
\end{table*}

The synthetic \ion{H}{i} brightness temperature spectrum $T_{\rm B}(v)$ for each column of clouds is generated by solving the radiative transfer equation through the column, following the radiative transfer prescription of \citet{marchal2019a} using the generalised implementation of \citet{wenger2024}. Clouds are indexed along the line of sight from nearest to furthest from the observer, such that $n=1$ corresponds to the cloud closest to the observer.

The FWHM velocity width of each cloud is calculated assuming thermal and turbulent broadening of the \ion{H}{i} emission from the sampled $T_{\rm spin}$ and sampled Mach number $\mathcal{M}$, using Equation~\ref{eqn:line broadening}.

\begin{equation}
    {\rm FWHM} = \sqrt{8\ln2\left(1+\mathcal{M}^2\right)\frac{k_B T_{\rm spin}}{m_H}}
    \label{eqn:line broadening}
\end{equation}

The FWHM is used to define the line profile of each cloud, which we assume to be a pure Gaussian. The optical depth profile $\tau(v)$ is calculated by multiplying the peak-normalised Gaussian line profile and the peak optical depth $\tau_{\rm peak}$ calculated using Equation~\ref{eqn:optical depth} \citep{mcclure-griffiths2023}.

\begin{equation}
     \tau_{\rm peak} \;=\; \frac{N_{\rm HI}~({\rm cm}^{-2})}{1.94\times10^{18}~({\rm cm}^{-2}~{\rm K}^{-1}~{\rm km}^{-1}~{\rm s})}\frac{1}{T_{\rm spin}~({\rm K})\times{\rm FWHM}~({\rm km~s}^{-1})}
     \label{eqn:optical depth}
\end{equation} 

The column density $N_{\rm HI}$ is calculated from the sampled volume density $n_{\rm HI}$ and sampled line-of-sight depth $d$, converting from parsecs to centimetres, and the FWHM is the combined thermal and turbulent line width calculated using Equation \ref{eqn:line broadening}. The emission spectrum is then calculated using Equation~\ref{eqn:radiative transfer}, which is a general form developed by \citet{wenger2024} that allows clouds to fill less than the full beam (via the filling factor $f$) and allows a background temperature to be specified, which we choose to be constant at 3.77~K following \citet{wenger2024}. Noise is then added to the synthetic emission profiles using a Gaussian model.

\begin{equation}
    T_{\rm B}(v) = \sum_{n=1}^{N} f_n \left(T_{{\rm spin},n} - T_{\rm bg}\right)\left[1 - e^{-\tau(v)_n}\right] \prod_{k=1}^{n-1}\!\left[(1-f_k) + f_k\, e^{-\tau(v)_k}\right]
    \label{eqn:radiative transfer}
\end{equation}

\begin{figure*}
    \centering
    \includegraphics[width=\linewidth]{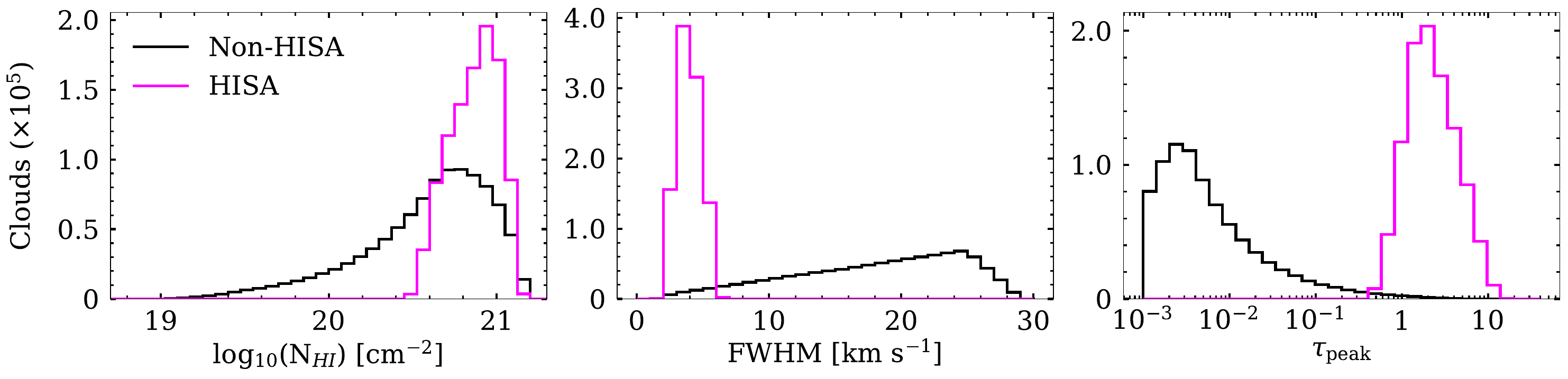}
    \caption{Histograms of the distributions of the calculated properties of two million individual \ion{H}{i} clouds generated with properties drawn from the distributions defined in Table~\ref{tab:slab parameters}. Half of the clouds are drawn from the first row of Table~\ref{tab:slab parameters} (black line) and half are drawn from the second row (magenta line). The number of non-HISA and HISA clouds has been made equal for visualisation purposes to illustrate the representative distributions of properties of the clouds, while in reality a sightline has only one or no HISA cloud and between 50 and 150 normal clouds. Alt text: graphs showing histograms of the column density, full-width half-maximum velocity width, and peak optical depth for two types of clouds.}
    \label{fig:slab property hists}
\end{figure*}

Figure~\ref{fig:slab property hists} shows the resulting distributions of the column density $N_{\rm HI}$, the FWHM velocity width, and the peak optical depth $\tau_{\rm peak}$ for one million normal \ion{H}{i} and one million HISA clouds when calculated from the uniform distributions of the parameters listed in Table~\ref{tab:slab parameters}. Although these samples do not represent the number of clouds in the synthetic dataset, they illustrate the relative distributions of the physical properties present in the two cloud populations.

The synthetic dataset is designed deliberately to favour precision over completeness. The network is intended to provide high-confidence detections of prominent HISA features that can be used to identify sightlines for subsequent spectral decomposition, rather than to produce a complete census of HISA across the sky. This choice is motivated in part by the practical difficulty of assigning reliable labels to faint or ambiguous absorption features. Such features may be obscured by noise or incorrectly labelled, and thus can degrade classifier performance \citep{frenay2014, song2023}.

We implement this precision-orientated design through two principal choices. First, we quantify the depth of each absorption feature as the maximum difference between the emission profile with and without the HISA cloud. Features with depths below 8~K are discarded, and the cloud parameters are regenerated until the resulting feature exceeds this threshold. The 8~K threshold corresponds to an S/N ratio of approximately 1.3 for the noisiest training spectra, which have an RMS noise level of 6~K. Because the threshold is applied before noise is added, the resulting training features span a range of S/N ratios rather than being selected to have a fixed S/N. The 8~K threshold therefore defines the nominal lower depth of the positive examples in the training data, while the completeness of the trained network below 8~K has not been measured and may be substantially reduced.

Second, each line of sight contains a single HISA cloud, with its parameters independently redrawn for each spectrum. This design avoids ambiguity in associating an absorption feature with a particular cloud in a multi-component spectrum and ensures that each training example has an unambiguous label. It also means, however, that the network is primarily trained to recognise a single dominant absorption feature against a given thermal background and may not generalise straightforwardly to sightlines containing multiple or overlapping HISA components.

These choices introduce a non-negligible completeness trade-off. A precision-optimised detector will preferentially identify HISA features with strong observable contrast, typically produced by relatively cold, dense gas in front of bright thermal backgrounds. It may therefore systematically under-detect weaker or more spatially extended absorption, including features in low-S/N regions and in environments where the CNM has a lower column density or contains multiple overlapping components. The implications of these selection effects for the scientific interpretation of the detected HISA sample are discussed in Section~\ref{sec:discussion}.

In total, we generate 1,440,000 spectra, half with a HISA cloud and half without. The spectra are generated over a velocity grid of 185 channels with a velocity resolution of $1.5~{\rm km~s}^{-1}$, between $\sim-113~{\rm km~s}^{-1}$ and $\sim163~{\rm km~s}^{-1}$. These properties are chosen to mimic the velocity axis of real observations of the Galactic Plane, specifically the THOR+VGPS data discussed later (Section~\ref{sec:thor data}). Additionally, such a low spectral resolution when compared to more current Galactic surveys means the neural network is performing at the lower bound of information: any increase in spectral resolution would require retraining the network but would only improve the accuracy of the network, and such a large velocity resolution allows other datasets to be down sampled on to a lower resolution grid instead of up sampled. The dataset of spectra is divided into a training set of 1,200,000 spectra, a validation set of 120,000 spectra, and a test set of 120,000 spectra. The spectra in all three datasets have Gaussian noise of 0, 2, 4, and 6~K added, with an equal number of spectra at each noise level. Although we add spectral noise to the synthetic data, we do not reproduce any additional systematic noise stemming from instrumental effects like baseline ripples or bandpass artefacts. As such, performance of the neural network is not evaluated when significant systematic baseline structure is present, but we do not expect large implications from this noise in real applications, as these features are either removed in the preprocessing of data or are relatively insignificant when compared to the kind of emission structure the neural network is trained on.

While the distributions of cloud parameters in Table~\ref{tab:slab parameters} are chosen to mimic realistic Galactic \ion{H}{i} properties, they are also chosen to avoid inducing a self-absorption feature in columns without a HISA cloud. However, clouds with a spin temperature of 50~K and a number density of $10 {\rm cm}^{-3}$ could result in high enough optical depth to produce self-absorption, although it is likely that the cloud would need to be at the front of the column for the absorption to be significant. We do not actively monitor for cases of this self-absorption, and in testing the resulting self-absorption that occurs when replacing a HISA cloud with a normal cloud with properties designed to induce self-absorption ($f=1$, $n_{\rm HI}=10~{\rm cm}^{-3}$, depth$=50~{\rm pc}$, $T_{\rm spin}=50~{\rm K}$, $\mathcal{M}=0.7$), only one in 100,000 spectra produces a HISA feature prominent enough to appear in the final spectrum. This means that out of the final 1,440,000 spectra produced, only 14 would be expected to potentially contain a HISA feature from a normal \ion{H}{i} cloud, and given that half of these would be expected to contain a HISA cloud as well, only 7 spectra would be incorrectly labelled, which will not influence the neural network's detections.

\subsection{HISA detector design and training}\label{sec:machine learning}

The architecture of the classifier is chosen by evaluating the performance of different versions of the CNN, covering a range of architecture hyperparameters. The probed hyperparameters are, in order of variation, the number of convolutional layers, kernel size, number of kernels, learning rate, and batch size. To find their optimal values, we vary one hyperparameter at a time while keeping all others fixed, as the computational resources required for a full grid search exceeded what was available. For each tested value, ten networks are trained to account for stochastic variations in training. This sequential approach is more efficient than a full grid search over all possible combinations of hyperparameters, and targets the most important hyperparameters first. The performance of each network was evaluated on the test dataset using the metrics of accuracy (correct classifications / total classifications), precision (correct positives / all predicted positives), and recall (correct positives / all true positives). Summary plots of these metrics for each hyperparameter are presented in Figure~\ref{fig:grid search}. The best-performing architecture is chosen as the network that maximises the accuracy, precision, and recall. The best-performing architecture consists of two pairs of convolutional and max-pooling layers with four 21-channel kernels with a standard stride of one and padding of ten, which reduce the 185 channels of the spectrum into four feature maps with 47 channels each. These feature maps are then flattened and passed through two fully connected layers. The final layer returns a pair of raw logits, which are converted into probabilities using a softmax operation. The probabilities of the two detection classes, non-HISA and HISA, represented as $\mathbb{P}(\neg{\rm HISA})$ and $\mathbb{P}({\rm HISA})$ respectively, are used to classify the input, where the class with the largest probability is taken as the classification. This is standard practice for classifiers and corresponds to a probability threshold of 0.5 for a binary classifier. Crucially, the final architecture has $\sim17,000$ parameters, fewer than half the parameters of typical early CNNs \citep[e.g. LeNet-5, $\sim60$ thousand parameters,][]{lecun1998a}, and several orders of magnitude fewer than modern CNNs \citep[e.g. ResNet-50, $\sim25$ million parameters,][]{he2015}. This allows the neural network to be run locally on low-power hardware, with training of the network done on HPC hardware and all further use and analysis of the network for this paper being performed on a 2023 MacBook Pro. The final CNN architecture is shown in Figure~\ref{fig:cnn architecture}.

We train the neural network on the 1.2 million spectra of the synthetic training dataset described in Section~\ref{sec:synthetic data}. The neural network's weights and biases are optimised using the Adam optimiser \citep{kingma2017a} and a cross-entropy loss function. The training process uses an exponentially decaying learning rate scheduler to reduce the learning rate by a factor of 0.1 when the training loss plateaus for 15 consecutive steps, to prevent overshooting and improve the network's performance. During training, the validation dataset is used to monitor the CNN's performance on unseen data, particularly to highlight any overfitting. The training is done for 100 epochs, with no early stopping implemented to allow the learning rate scheduling to take effect. As is typical for CNNs, we preprocess the spectra using min-max scaling. The min-max scaling divides all spectra by a fixed constant of 250~K, the maximum brightness temperature found in the synthetic training set, while the majority of the synthetic spectra have a peak $T_{\rm B}$ of $\sim110$~K. This scaling is also applied to the validation and testing sets and ensures the distribution of spectra in each dataset is the same, which aids neural networks when training and predicting.

\subsection{Evaluation on synthetic data}\label{sec:synthetic evaluation}

With the best architecture determined and trained and using the training and validation datasets, we evaluate the performance of the neural network on the test dataset. The test dataset is not seen by the network during training, which ensures the neural network cannot overfit to the labels of the test spectra, and good performance on the test dataset indicates that the network has learned to generalise to new data. The neural network achieves an accuracy of 96.5 per cent, a precision of 96.0 per cent, a recall of 97.0 per cent, and an area under the receiver operating characteristic curve (AUC-ROC) of 99.3 per cent on the test dataset. In Table~\ref{tab:confusion matrix}, we provide the confusion matrix of the neural network on the testing subset of 120,000 emission spectra. The confusion matrix shows a high number of correct detections and non-detections, and a false negative rate (3.0 per cent) lower than the false positive rate (4.1 per cent). These metrics all demonstrate that the neural network performs very well on unseen data, and has learned to generalise from the training data to the testing data. While both the percentages of incorrect detections are low, the false positive rate is higher than the false negative rate, indicating that the network may be overconfidently detecting HISA when no real HISA features are present.

\begin{table}
    \centering
    \caption{The confusion matrix of the trained neural network, applied to the 120,000 spectra of the test dataset described in Section~\ref{sec:synthetic data}.}
    \begin{tabular}{rcc}
        \hline
          & No HISA Predicted & HISA Predicted\\\hline
         No HISA Present & 57,477  (47.9\%) & 2,438  (2.0\%) \\
         HISA Present & 1,777  (1.5\%) & 58,308  (48.6\%) \\\hline
    \end{tabular}
    \label{tab:confusion matrix}
\end{table}

A low false positive rate is critical for application to large datasets where manual vetting of detections is impractical. Where a 4.1 per cent false detection rate is unacceptable, incidents of false positives from the neural network can be reduced by modifying the classification threshold for the neural network. This technique is referred to as threshold-moving, and is typically used in applications with imbalanced training labels \citep{zhou2006}. While our training labels are balanced by design (50 per cent of spectra containing a HISA feature, 50 per cent without), threshold-moving can also be used to selectively remove lower-quality detections from the final sample, without retraining the neural network. The classification threshold can be raised above 0.5 by the user to trade recall for higher purity, depending on the science application, and if we instead require a 0.9 threshold for a detection, the neural network achieves an accuracy of 96.5 per cent, precision of 99.8 per cent, and recall of 93.3 per cent. The false negative rate increases to 6.8 per cent while the false positive rate drops to 0.2 per cent. This trade-off of fewer false positives to more false negatives can be desirable specifically for our application of HISA detection, as missing HISA detections in a dataset may be preferable to contaminating a detected sample with false positives that could bias subsequent analyses of cold-gas properties \citep[see][for a discussion]{gibson2005}. As such, we design the neural network to output the raw class probabilities $\mathbb{P}({\rm HISA})$ and $\mathbb{P}(\neg{\rm HISA})$ which allows a non-standard threshold to be used after the neural network is run.

To assess whether these predicted probabilities provide well-calibrated measures of classification confidence, we compute the Expected Calibration Error \citep[ECE;][]{guo2017} and Adaptive Calibration Error \citep[ACE;][]{nixon2020} on the 120,000-spectrum held-out test set. We obtain an ECE of 0.0022 and an ACE of 0.0030, indicating that the predicted probabilities closely track the empirical frequency of correct classifications over the range of confidence values represented in the test set. However, the test set is constructed using the same 8~K depth threshold as the training set (Section~\ref{sec:synthetic data}) , and therefore contains no weak or ambiguous HISA features near the classification boundary. The calibration results consequently demonstrate that the network is well calibrated within the distribution of unambiguous strong-HISA and no-HISA spectra represented in the training and test sets, but they do not establish calibration in the sub-threshold regime of more ambiguous self-absorption features.

\subsection{Velocity localisation using activation mapping}\label{sec:velocity localisation}

The metrics reported in Section~\ref{sec:synthetic evaluation} demonstrate that the trained neural network successfully identifies \ion{H}{i} self-absorption features in the synthetic dataset. CNNs are known to encode information about which parts of the input matter most to their predictions, a property typically exploited for neural network interpretability \citep{simonyan2014, selvaraju2016}. Here, we instead leverage this property directly to extract real information from the neural network, in order to localise detected HISA features along the spectral axis. This adds functionality that is not typically possible with classification neural networks, but that is commonplace in other HISA identification approaches. Velocity localisation is often explicitly encoded into approaches for CNM detection: in Gaussian decomposition methods, the velocity of each cloud is directly optimised, while baseline emission spectrum reconstruction methods model the absorption and thus its central velocity directly. Our approach extracts this information post-prediction via activation mapping, rather than encoding it directly.

\begin{figure}
    \centering
    \includegraphics[width=\linewidth]{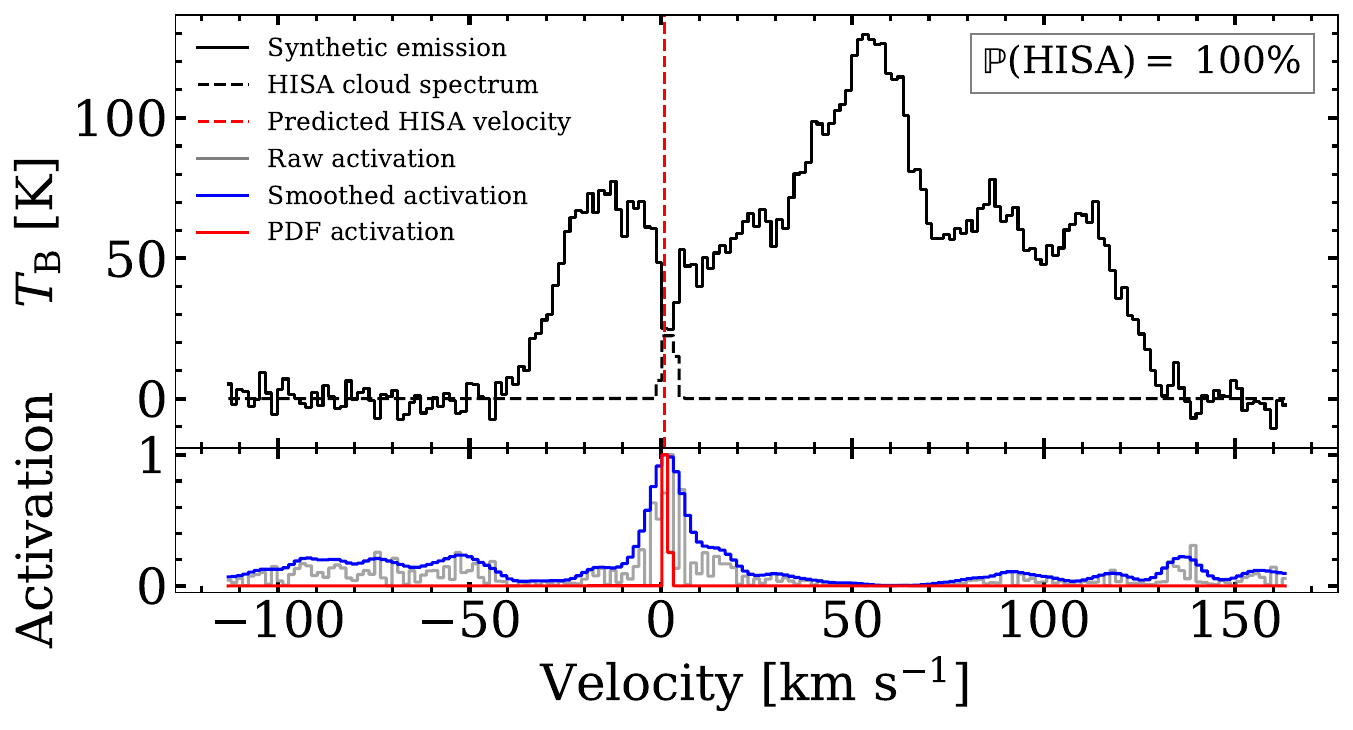}
    \caption{Example synthetic emission spectrum containing a self-absorbing cloud, and its corresponding activation spectra. The top panel shows the total emission spectrum in black; the emission spectrum of the HISA cloud is shown by the dashed black line. The lower panel shows the raw (grey), smoothed (blue), and probability density function (red) activation spectra. The PDF activation spectrum shows a strong, concentrated feature at the same location as the self-absorption, with the predicted velocity channel indicated by the red dashed line in the top panel. The difference between the true HISA cloud velocity and the predicted velocity for this example is $0.24~{\rm km~s}^{-1}$, within the channel width $(1.5~{\rm km~s}^{-1})$. Alt text: two-panel graph showing two emission spectra and three activation spectra. Text in the upper right of the graph reports the HISA probability as 100 per cent. The activation spectra all show a strong peak at the same velocity as an absorption feature in the emission spectra.}
    \label{fig:activation example}
\end{figure}

The process of activation mapping and subsequent velocity localisation is performed as a post-processing step after the network has made its predictions and is calculated as follows. First, the gradient of every data point in each input spectrum is calculated with respect to the positive class using back-propagation. This returns an activation spectrum that measures the local sensitivity of the network's HISA output to the input spectrum at each velocity. Figure~\ref{fig:activation example} shows an example emission spectrum with a HISA feature, and the corresponding activation mapping products. First, the absolute value of the gradient spectrum is taken (shown in grey in the lower panel). Then, to reduce the effect of the single-channel noise in these activation spectra, we smooth the gradient spectrum using a Gaussian kernel with a $\sigma$ value of 2 channels, corresponding to $\sim3~{\rm km~s}^{-1}$ in the velocity grid of the synthetic spectra (shown in blue in the lower panel). Finally, a softmax transformation of the smoothed activation spectra allows interpretation as a probability density function (shown in red in the lower panel). The maximum of the probability density distribution is taken as the most likely location of the feature in velocity, represented by the dashed red line in Figure~\ref{fig:activation example}, which corresponds to the real location of the HISA feature (shown by the dashed black line). For the example shown in Figure~\ref{fig:activation example}, the measured offset between the true HISA cloud velocity and the activation-derived velocity is $0.24~{\rm km~s}^{-1}$, well within a single velocity channel ($1.5~{\rm km~s}^{-1}$). However, given our implementation of this method, only a single velocity can be returned per spectrum, and will correspond to the strongest activation. This is adequate for the synthetic data that contains at most a single HISA cloud; however, in real spectra where there may be many instances of cold gas along the line of sight and therefore might contain multiple HISA features, the current method can only return the velocity of a single feature.

\begin{figure}
    \centering
    \includegraphics[width=\linewidth]{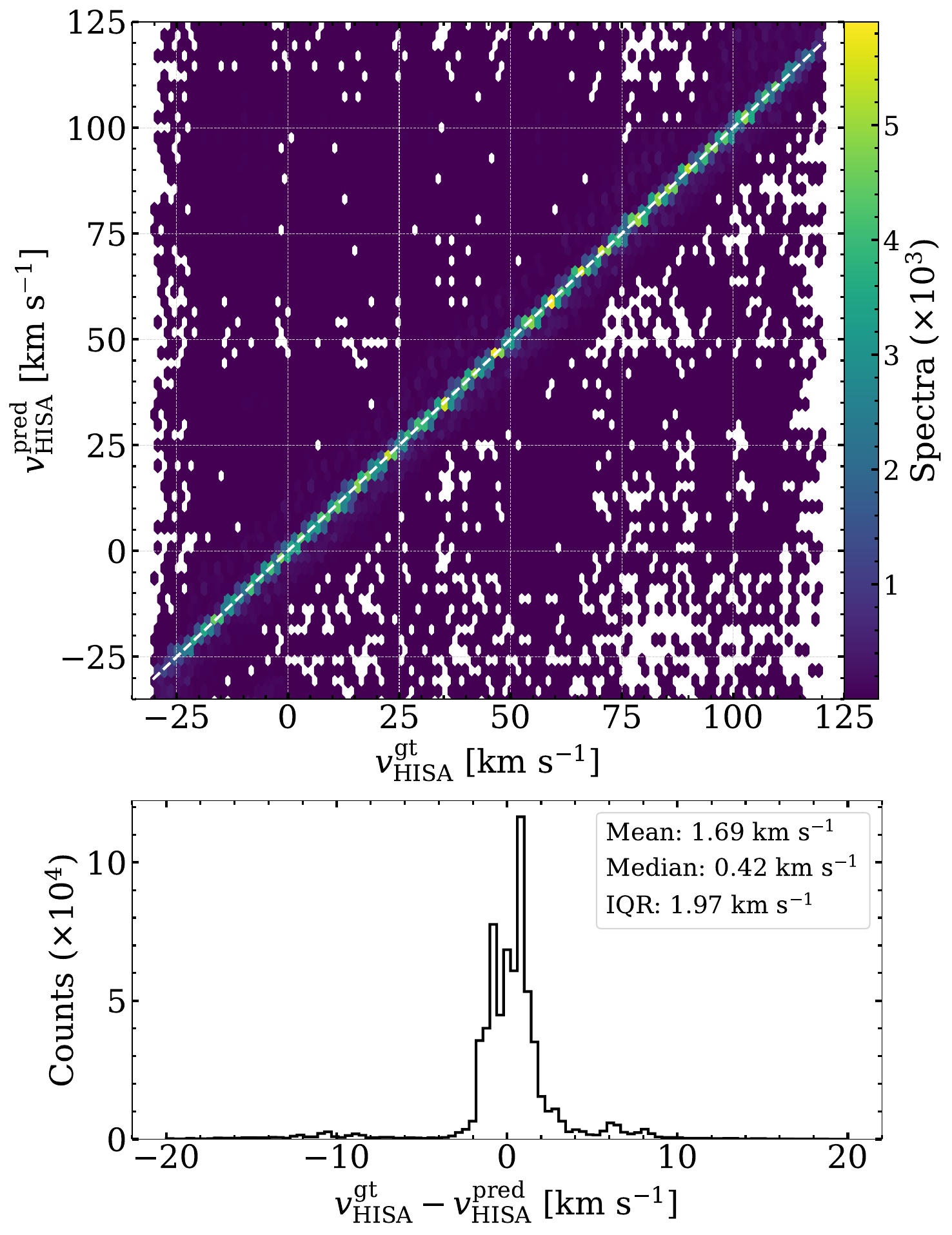}
    \caption{Top: scatter of the true ($v_{\rm HISA}^{\rm gt}$) and predicted ($v_{\rm HISA}^{\rm pred}$) HISA feature velocities of the synthetic dataset. The dashed line shows the one-to-one line. Bottom: histogram of offsets between the true and predicted HISA feature velocities. Summary statistics of the offset distribution are reported in the upper-right corner. A tight correlation between the velocities and the narrow distribution of offsets implies that the activation spectrum method for localising the HISA feature is effective at finding the real HISA features in the spectra. Alt text: a two-panel graph showing a two-dimensional histogram in the upper panel, and a one-dimensional histogram in the lower panel. The upper panel shows a strong one-to-one correlation, with bright pixels indicating high counts aligning with a dashed one-to-one line. Dark pixels indicating low counts fill almost the entire upper panel, with some empty patches throughout the plot. The lower panel shows a broad positively-skewed distribution, and text in the upper right reports the mean, median, and interquartile range of the distribution.}
    \label{fig:synthetic velocities}
\end{figure}

We now evaluate the accuracy of this activation spectrum method by comparing predicted HISA velocities ($v_{\rm HISA}^{\rm pred}$) with the true velocities ($v_{\rm HISA}^{\rm gt}$) of the HISA clouds in the synthetic dataset. Because the localisation process is a deterministic method and does not use any learned parameters, we evaluate the performance utilising the entire synthetic dataset described in Section~\ref{sec:synthetic data} (training, validation, and testing datasets), rather than just the testing dataset. However, we restrict the analysis to spectra that contain a HISA feature and that have a positive detection from the neural network ($\sim702,000$ spectra). Figure~\ref{fig:synthetic velocities} illustrates the correlation between the true and predicted HISA feature velocities and the resulting distribution of offsets. The mean (median) of the velocity offsets is $1.69~{\rm km~s}^{-1}$ ($0.42~{\rm km~s}^{-1}$), with an interquartile range (IQR) of $1.97~{\rm km~s}^{-1}$. Of the $\sim702,000$ velocity offsets, 66.3 per cent fall within a single channel of the true HISA velocity. Furthermore, the percentages of offsets within 1, 2, and $3\delta v$ of the true HISA velocity are 68.5 per cent, 82.3 per cent, and 85.1 per cent respectively.

The lower panel of Figure~\ref{fig:synthetic velocities} shows a systematically positive skew of the velocity offsets. This offset is primarily a result of the spectra in the synthetic dataset with 0~K of noise, which has the largest interquartile range of the four noise levels in the dataset ($5.7~{\rm km~s}^{-1}$ as opposed to 1.7--$2.0~{\rm km~s}^{-1}$ for the other three noise levels). This is likely the result of the method used for velocity localisation: \citet{smilkov2017} found that noiseless inputs produced worse localisation using vanilla gradient maps than when the inputs contained some noise, and the noisy spectra in our synthetic dataset work to pull the gradient spectra away from the saturated plateau of very confident detections, conditioning the gradients in the activation spectra and producing smaller velocity offsets. Despite the larger spread in the noiseless spectra, median offsets remain sub-channel across all noise levels, ranging between 0.24 and $0.55~{\rm km~s}^{-1}$.

\section{Observed Datasets}\label{sec:observed data}

Having established the network's strong performance in HISA detection and velocity localisation on synthetic data, we now apply it to two distinct ISM environments towards the Galactic Plane. These regions present examples of real spectral complexity and observational noise that test the ability of the neural network to generalise to new datasets. The trained neural network is applied to observed \ion{H}{i} emission data from two distinct Galactic environments: the well-studied Riegel--Crutcher cloud (Section~\ref{sec:rc data}) and five Giant Molecular Filament (GMF) regions in the inner Galactic Plane (Section~\ref{sec:thor data}). We select the Riegel--Crutcher cloud as it is the most clearly identified HISA structure in the Milky Way, and the GMF regions as they represent challenging regions against the Galactic Plane where HISA might be expected and has been previously found. To evaluate the accuracy of the CNN's velocity estimates, we employ two independent reference datasets. For the Riegel--Crutcher cloud data, we apply a simple linear interpolation method to identify the velocities of potential HISA features following \citet{mcclure-griffiths2006}. For the more spectrally complex GMF regions, where simple interpolation methods fail due to the increased occurrence of features mimicking self-absorption, we use \coion molecular emission from the Galactic Ring Survey (Section~\ref{sec:thor data}), as well as the results of the \textsc{astroSaber} HISA detection algorithm presented in \citet{syed2023}. We stress that neither the \coion data nor the interpolated velocities are used as inputs to the neural network at any stage; they serve exclusively as independent benchmarks against which the network's outputs are evaluated. This is what separates the neural network from previous approaches that bias the search for HISA to the presence of other tracers.

\subsection{Riegel--Crutcher cloud data}\label{sec:rc data}

The Riegel--Crutcher (hereafter \rc) cloud, first discovered by \citet{heeschen1955} and later fully mapped by \citet{riegel1969}, is notable as one of the closest examples of a singular coherent self-absorbing \ion{H}{i} cloud, at a distance of only $125\pm25$ pc \citep{crutcher1984}. It covers $\sim400~{\rm deg}^2$ of the sky, and the cloud's location towards the Galactic Centre provides a strong \ion{H}{i} emission background against which the cold, dense cloud produces self-absorption \citep{crutcher1973, crutcher1984}. The cloud has been shown to consist of a highly filamentary structure \citep{mcclure-griffiths2006, clark2014}. \citet{denes2018} used absorption measurements from the Australian Telescope Compact Array (ATCA) to show that the spin temperature of the \ion{H}{i} gas in the \rc cloud ranged from 20 to 80~K, indicating that the cloud exists mostly as CNM gas. The existence of cold atomic gas and the sub-parsec resolution of the \rc cloud due to its proximity make it optimal as a benchmark for cold-gas detection methods.

The \rc cloud observations used in this work are combined data from ATCA and the Parkes Murriyang single-dish radio telescope as part of the Southern Galactic Plane Survey \citep[SGPS;][]{mcclure-griffiths2005}. The SGPS Galactic Centre Survey \citep[SGPS GC;][]{mcclure-griffiths2012} covered $100~{\rm deg}^2$ of sky centred on the Galactic Centre, observing the lower-right quarter of the \rc cloud. The final combination of ATCA and Parkes observations produces a datacube with a pixel size of 35 arcsec and a spectral resolution of $0.824~{\rm km~s^{-1}}$. The data contain a noise level of $\sim2~{\rm K}$ in the off-line channels \citep{mcclure-griffiths2006}. 

\begin{figure}
    \centering
    \includegraphics[width=\linewidth]{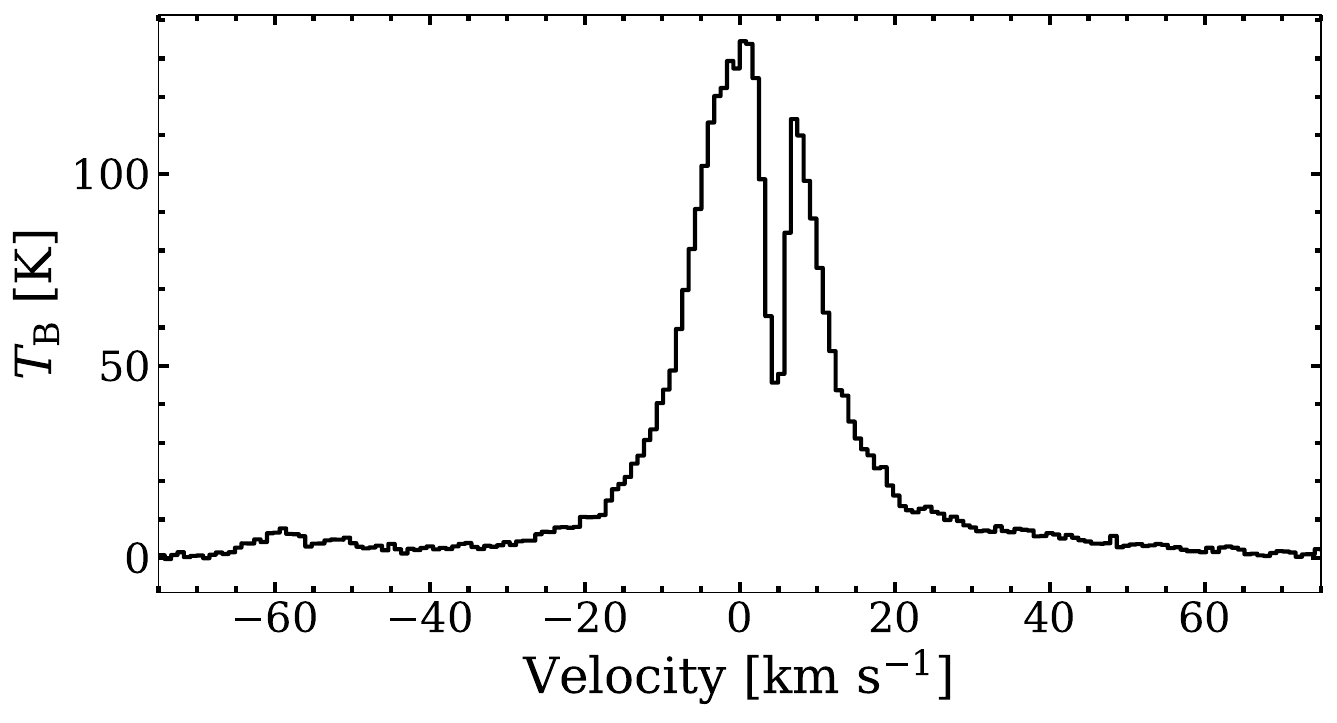}
    \caption{\ion{H}{i} emission spectrum of the Riegel--Crutcher cloud, towards $(l,\,b) = (0\fdg79, 1\fdg99)$. The spectrum exhibits the strong HISA feature produced by the \rc cloud at $\sim5~{\rm km~s}^{-1}$. Alt text: graph of an emission spectrum, showing a broad, strong emission feature, and a narrow, strong absorption feature in the centre of the emission feature.}
    \label{fig:rc spectrum}
\end{figure}

In Figure~\ref{fig:rc spectrum} we show an example emission spectrum from the \rc cloud, showing the bright background from the Galactic Centre and the prominent self-absorption feature centred at $\sim5~{\rm km~s^{-1}}$. The cloud serves as an ideal test case for our neural network, as the HISA features are strong and well-characterised and can be easily confirmed manually as well as easily modelled using interpolation methods. The spectra from the lower edge of the SGPS cube ($b\lesssim-5\degr$) are heavily affected by systematic noise from $\sim250~{\rm km~s}^{-1}$ onwards which looks like real signal to the neural network. To avoid these spectra contaminating our analysis, we remove the bottom 10 rows from the cube. We do not apply any further S/N ratio cuts to the cube as the remaining spectra all contain detectable signal and the neural network is designed to be able to handle a range of S/N ratios via the use of mixed-noise training data.

As a comparison sample for the velocity localisation of HISA features in the Riegel--Crutcher cloud, we replicate the simple linear interpolation approach of \citet{mcclure-griffiths2006} to automatically identify the most prominent absorption features. In this approach, local emission peaks are identified using a peak-finding algorithm, and narrow, deep features between neighbouring emission peaks are interpolated across. Only interpolated features with a visible depth (5~K or deeper) are retained; all other features are discarded. The deepest channel in an interpolated absorption feature is taken as the location of the HISA feature for that spectrum. While multiple interpolated features may exist across a spectrum, the velocity of the strongest interpolated feature is taken as the HISA velocity for each spectrum for consistency with the method described in Section~\ref{sec:velocity localisation}. 

The full results of the linear interpolation method are presented in Figures~\ref{fig:rc linear interpolation slice}--\ref{fig:rc linear interpolation velocity map}. We provide examples of reconstructed unabsorbed emission slices and $\Delta T_{\rm B}$ velocity slices for comparison with \citet{mcclure-griffiths2006}. We verify the interpolation localisations qualitatively by comparing the interpolated velocity map against the published figures of \citet{mcclure-griffiths2006}. The spatial distribution of detected features and the recovered velocity gradient ($\sim3$--$8~{\rm km~s^{-1}}$ from north to south across the cloud) are consistent with their fig. 3, showing that our implementation serves as a valid independent reference for the CNN velocity localisation. We note, however, that differences between our implementation and the implementation from \citet{mcclure-griffiths2006} due to data differences (namely continuum subtraction around the Galactic Centre) and lack of information about the original method mean this is not a pixel-level reproduction of the original method. Any differences between our implementation and the original implementation will affect the detection/non-detection of self-absorption features, and the predicted velocity of the detected features. The former has a small effect on the overall comparison; however, we expect that the majority of potential self-absorption features will be detected by both implementations, and only the detection of very subtle features outside the body of the \rc cloud will be affected. While the latter more directly affects our analysis, we do not expect it to significantly alter the interpolated velocities. Because we define the velocity of an interpolated feature as that of the maximum $\Delta T$ channel, this peak will always be included in the interpolation provided both implementations cover the same central channels, regardless of any differences in the peak-finding algorithms used to define the features. Therefore, while the HISA velocities may vary slightly per spectrum, the overall distribution of any differences in the HISA velocities from the two implementations is expected to be narrow and centred about $0~{\rm km~s^{-1}}$.

\subsection{Giant molecular filament data}\label{sec:thor data}

Giant molecular filaments are large ($\gtrsim100$ pc), elongated structures of dense molecular gas located at the arm and inter-arm regions of the Milky Way \citep{beuther2011, ragan2014, abreu-vicente2016, zucker2018, ge2023}. These regions of molecular gas mark the final stage of the cascade of atomic material and thus are expected to be promising regions in which to search for the cold atomic hydrogen that condenses into molecular gas. \citet{ragan2014} used mid-infrared extinction measurements from the GLIMPSE survey and \coion data from the Galactic Ring Survey to identify a sample of seven GMFs along the Galactic Plane that exhibit strong velocity coherence across their full length. Full details of the GMF regions, including their distances, angular extents, and systemic velocities, are provided in table 2 of \citet{ragan2014}. Six of these seven GMFs have been observed in \ion{H}{i} emission by The HI/OH/Recombination line survey of the Milky Way \citep[THOR;][]{beuther2016}, which used the Karl G.\ Jansky Very Large Array (VLA) to observe the Northern Galactic Plane. The THOR survey data (VLA C-configuration) were combined with observations from the Very Large Array Galactic Plane Survey \citep[VGPS, VLA D-configuration;][]{stil2006} to improve sensitivity to extended \ion{H}{i} emission on large angular scales. The resulting THOR+VGPS datacubes cover $\sim132~{\rm deg}^2$ of the inner Galactic Plane with a pixel size of $\sim10$ arcsec, and a spectral resolution of $1.5~{\rm km~s^{-1}}$. The data contain a noise level of $\sim4~{\rm K}$ in the off-line channels.

The six GMFs covered by the THOR survey have been previously studied for the presence of HISA using the \textsc{astroSaber} algorithm in \citet{syed2023}. All of the GMF fields studied in that work are included here: GMF20, GMF26, GMF38 (38a and 38b filaments), GMF41, and GMF54. The GMF38 field contains a pair of \ion{H}{i} structures separated in velocity, GMF38a and GMF38b. For simplicity, we treat GMF38a and GMF38b as a single region (GMF38) and process all \ion{H}{i} spectra in the field without splitting by velocity range, except when using \coion data to investigate HISA detections around the individual filaments.

The Giant Molecular Filament regions were mapped in \coion $J=1\rightarrow0$ emission as part of the Galactic Ring Survey \citep[GRS;][]{jackson2006}. The GRS used the Five College Radio Astronomy Observatory (FCRAO) to observe the first Galactic quadrant, covering $l = 18^\circ$--$55.7^\circ$ and $|b| \lid 1^\circ$, with a spectral resolution of $0.21~{\rm km~s}^{-1}$ and an angular resolution of 46 arcsec with 22 arcsec sampling, producing a fully sampled map. Gaussian decomposition of the GRS \coion spectra across the GMF regions was performed by \citet{riener2020} using the \textsc{GaussPy+} algorithm \citep{lindner2015, riener2019}, providing centroid velocities for discrete molecular emission components along each line of sight.

Cold atomic hydrogen co-spatial with molecular material is expected to share a similar line-of-sight velocity, as the atomic gas is believed to condense to molecular \htwo, which allows other molecular species like \coion to form in the shielded region \citep{wolfire2010, sternberg2014}. However, it should be noted that the \coion traces molecular gas while the CNN detects cold atomic hydrogen in its envelope; a modest velocity offset between the two tracers is therefore physically expected in at least some sightlines, independent of any localisation error in the CNN. The \coion velocities from \citet{riener2020} are used as a reference against which to assess the HISA velocities predicted by the neural network. However, the GRS \coion data do not cover an identical sky area or velocity range as the THOR survey; comparisons between \coion and \ion{H}{i} velocities are therefore restricted to sightlines where both surveys provide coverage. We emphasize again that the \coion data and the derived molecular velocities are used solely for this post-hoc comparison: they are not provided to the neural network, do not constrain the velocity window searched for self-absorption, and play no role in either the detection or the localisation output of the CNN. This contrasts with previous HISA detection methods in which molecular emission data are used to constrain the velocity window within which found self-absorption features are deemed to be real. This not only allows the method to be used where ancillary molecular data are unavailable, unreliable, or incomplete (e.g., outside the GRS coverage, in the outer Galaxy, or in the Magellanic Clouds), but it also removes any inherent biases from detections only located coincident to molecular data.

\section{Results}\label{sec:results}

With the network trained and validated on synthetic data (Section~\ref{sec:ml framework}) and the observed datasets described (Section~\ref{sec:observed data}), we now present results of applying the CNN to the two observed regions. For each dataset, we first describe the preprocessing applied to prepare the spectra for the network, then present the HISA detections and the predicted velocity maps, as well as a quantitative comparison of the predicted velocities with an independent reference method.

\subsection{Riegel--Crutcher cloud}\label{sec:rc results}

We apply the convolutional neural network to the SGPS GC datacube to search for HISA across the \rc cloud. The \rc emission spectra in the SGPS data cube have a larger channel width and a higher number of channels than the velocity grid of the synthetic data the neural network was trained on. Prior to processing, the spectra are therefore prepared in two steps. First, the spectral axis of the SGPS GC datacube is regridded on to the velocity range of the synthetic training data ($-113$ to $163~{\rm km~s}^{-1}$) by cropping channels outside this range. Second, the spectra are resampled from the native SGPS spectral resolution of $0.824~{\rm km~s^{-1}}$ to the $1.5~{\rm km~s^{-1}}$ resolution of the synthetic grid using linear interpolation, resulting in a fixed 185-channel input consistent with the network's design. The need for down sampling and cropping is a limitation of the structure of convolutional neural networks, which must be designed with a fixed input size. Additionally, the HISA features that the neural network learns to detect are learned with some fixed channel width, not velocity width. Thus, the spectra must be down sampled so that features of a given velocity width correspond to the correct number of channels.

The resampled spectra are then normalised with the same min-max scaling applied during training, dividing by the maximum brightness temperature of the training dataset (250~K), before being passed through the neural network. Only 0.07 per cent of spectra in the SGPS GC datacube have a peak brightness temperature exceeding 250~K; these are all located towards the Galactic Centre itself, and they contribute to the noise in that region and should be treated with caution. The standard HISA detection threshold of 0.5 is used.

\subsubsection{HISA detection in the Riegel--Crutcher cloud}\label{sec:rc detection}

\begin{figure}
    \centering
    \includegraphics[width=\linewidth]{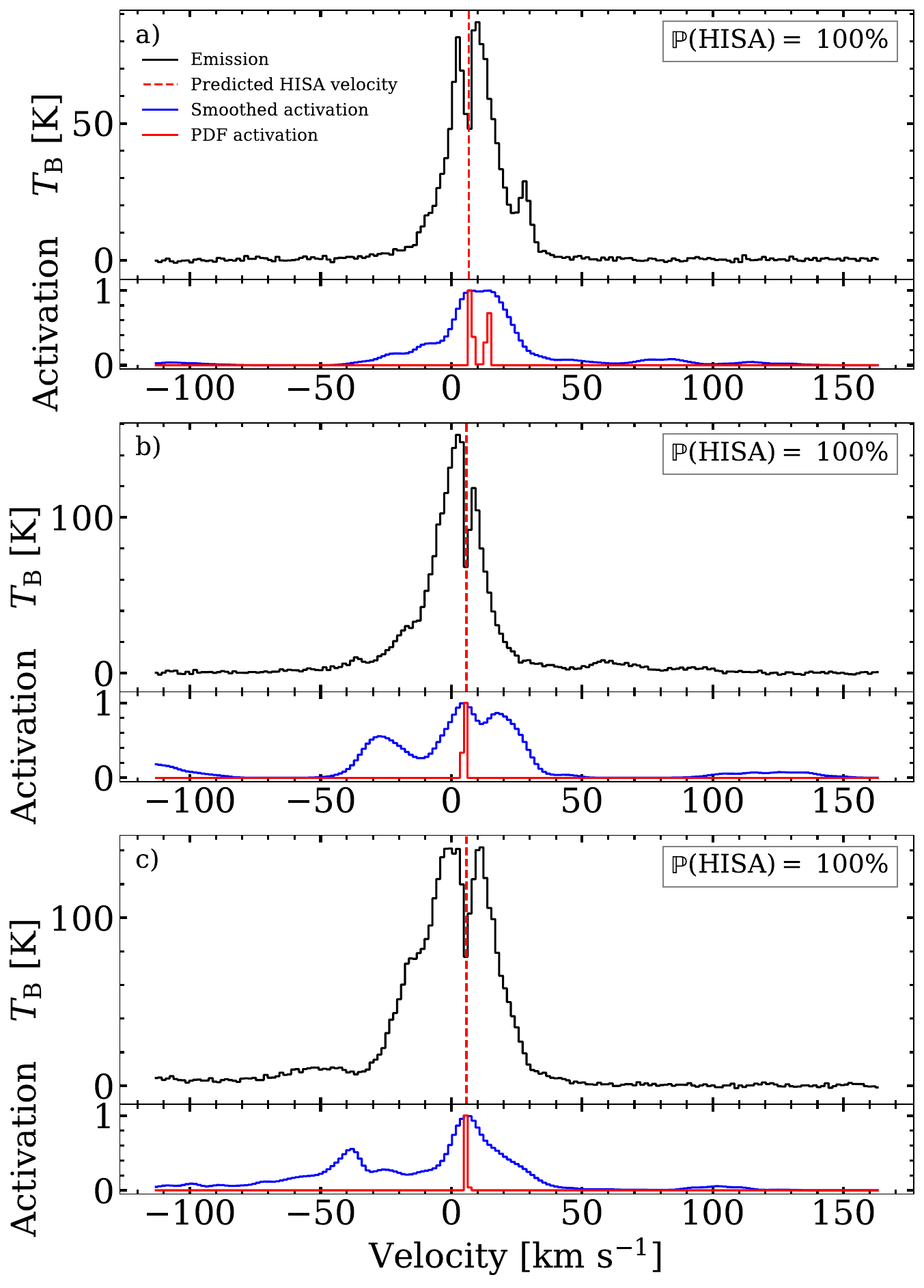}
    \caption{Three spectra with strong HISA detections from the \rc cloud. The panel labels correspond to position indicators in Figure~\ref{fig:rc detection map}. All three spectra show strong activation and velocity localisation at the known self-absorption feature produced by the \rc cloud, between 0 and $10~{\rm km~s}^{-1}$, despite the differences in the background emission morphology across the field. Alt text: three graphs showing emission spectra and activation spectra, labelled A, B, and C. All three emission spectra show broad strong emission in the middle, and narrow strong absorption features at the peak of the emission. The activation of all three panels peak at the same location as the absorption feature. Text in the upper right of each panel reports a 100 per cent HISA probability.}
    \label{fig:rc representative spectra}
\end{figure}

Figure~\ref{fig:rc representative spectra} shows three example emission spectra and their corresponding activation spectra from different positions in the \rc cloud. All three panels show 100 per cent HISA probabilities corresponding to the strong self-absorption feature known to exist from the \rc cloud, between 0 and $10~{\rm km~s}^{-1}$. While the activation spectra all show strong responses away from the HISA feature, notably even in the absence of any visible self-absorption, the final activation probability spectrum discards these peaks and activates only on the real HISA feature.

\begin{figure*}
    \centering
    \includegraphics[width=\linewidth]{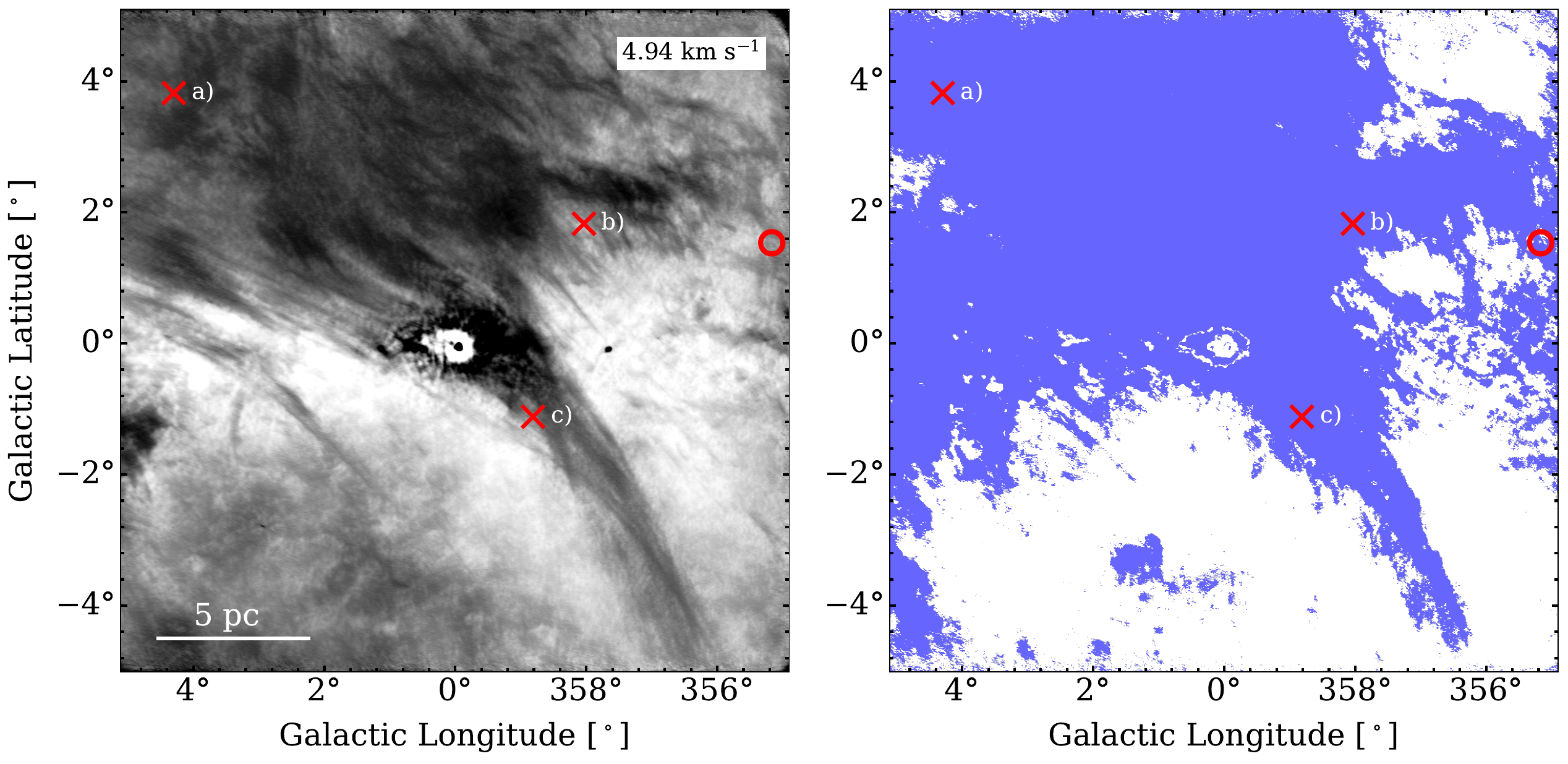}
    \caption{Left: Velocity slice of the SGPS GC \ion{H}{i} emission data at $v=4.94~{\rm km~s}^{-1}$, showing the main body of the \rc cloud in the field. The grey scale is linear between 30 (white) and 150~K (black). Right: HISA detection map of the Riegel--Crutcher cloud from the neural network, where coloured pixels indicate a positive HISA classification using a threshold of 0.5. Red crosses indicate the positions of three representative strong detections in the \rc cloud, shown in Figure~\ref{fig:rc representative spectra}, while the red circle indicates the location of a pair of non-detection and detection spectra at the edge of the cloud, shown in Figure~\ref{fig:rc spectra comparison}. The network recovers the main body of the \rc cloud, including the southern clump at $l\sim1\fdg5$, $b\sim-3\fdg5$ (not visible in the left panel), and examples of the filamentary structure identified in previous studies. Non-detections near the Galactic Centre reflect the complexity of low-latitude sightlines that differ from the synthetic training spectra. Structures traced by the HISA detections from the neural network in the right panel are generally also real structures from the main body of the \rc cloud, but at velocities not shown in the left panel, as the \rc cloud covers velocities of $\sim0$--$10~{\rm km~s}^{-1}$. Alt text: two panels, the left showing a grey-scale image of a cloud, and the right showing the same structure in a single colour, with red crosses labelled A, B, and C, and an unlabelled red circle plotted at different points on the cloud.}
    \label{fig:rc detection map}
\end{figure*}

Figure~\ref{fig:rc detection map} shows a velocity slice of the SGPS data at $\sim5~{\rm km~s}^{-1}$, alongside the full HISA detection map from the neural network. The velocity slice shows one of the most prominent velocities of the absorption feature caused by the \rc cloud. The detection map is created from the neural network output HISA probability, using a detection threshold of 0.5. The spatial distribution of the CNN detections broadly traces the known extent of the \rc cloud, recovering both the main body of the cloud, as well as more distinct regions such as the southern clump at ($l\sim1\fdg5$, $b\sim-3\fdg5$) visible in the integrated emission maps of \citet{mcclure-griffiths2006}. The detection map also reveals the prominent filamentary structure of the \rc cloud, consistent with the magnetically aligned filaments studied in \citet{mcclure-griffiths2006} and \citet{clark2014}. The red crosses in the figure and their labels correspond to the panels of Figure~\ref{fig:rc representative spectra}, while the red circle indicates the position of the spectra shown in Figure~\ref{fig:rc spectra comparison}.

Despite the performance of the neural network over the main body of the \rc cloud, there are regions where the network performs poorly. Most visibly, some filamentary structures of the \rc cloud are not fully recovered. The region near $l=1\fdg8$, $b=-1\fdg8$ is an example where the HISA feature depth almost completely disappears midway along the filamentary structure because the background emission at those sightlines is weaker than typical, falling by $\sim40~{\rm K}$. Additionally, several compact radio continuum sources lie projected against the \rc cloud, most prominently the Galactic Centre itself. At these positions, the continuum absorption that is not well-deconvolved causes profiles that differ fundamentally from the emission spectra on which the network was trained, and the network output at these positions should not be interpreted. Users applying the network to regions containing known bright continuum sources can choose to mask those positions prior to processing or use the positions to flag detections for quality checking after the fact; a continuum intensity map provides a straightforward basis for identifying such sightlines.

\subsubsection{HISA velocity localisation}\label{sec:rc velocities}

After the \rc cloud spectra are processed through the neural network, we use the activation mapping method described in Section~\ref{sec:velocity localisation} to infer the HISA feature velocities. Unlike the synthetic data used to train the neural network, real \ion{H}{i} emission data contain structures of cold gas that arise naturally in both position and velocity; the structure of the \rc cloud, particularly the spatial filaments and the velocity gradient across the bulk of the cloud, therefore provides an important test of whether the network responds to genuine physical features rather than to noise or instrumental artefacts. The prominence of the HISA features in the \rc cloud makes it an ideal dataset for evaluating the velocity localisation, as the true feature velocity has been independently well-constrained by both the interpolation method of \citet{mcclure-griffiths2006} and previous targeted absorption studies \citep{denes2018}, with both works finding HISA velocities spanning $\sim0$--$10~{\rm km~s}^{-1}$ across the cloud.

The velocity localisation produces a map of predicted HISA velocities across the \rc cloud. The distribution of these predicted velocities spans between $\sim0$ and $10~{\rm km~s}^{-1}$ with the bulk of the cloud concentrated near $5~{\rm km~s}^{-1}$, consistent with previous measurements of the \ion{H}{i} velocity of the \rc cloud \citep{riegel1969, montgomery1995, mcclure-griffiths2006}. However, a minority of spectra yield predicted velocities that are outliers with respect to this distribution, concentrated in regions of complex or low-contrast emission at the periphery of the cloud and along the Galactic Plane.

\begin{figure}
    \centering
    \includegraphics[width=\linewidth]{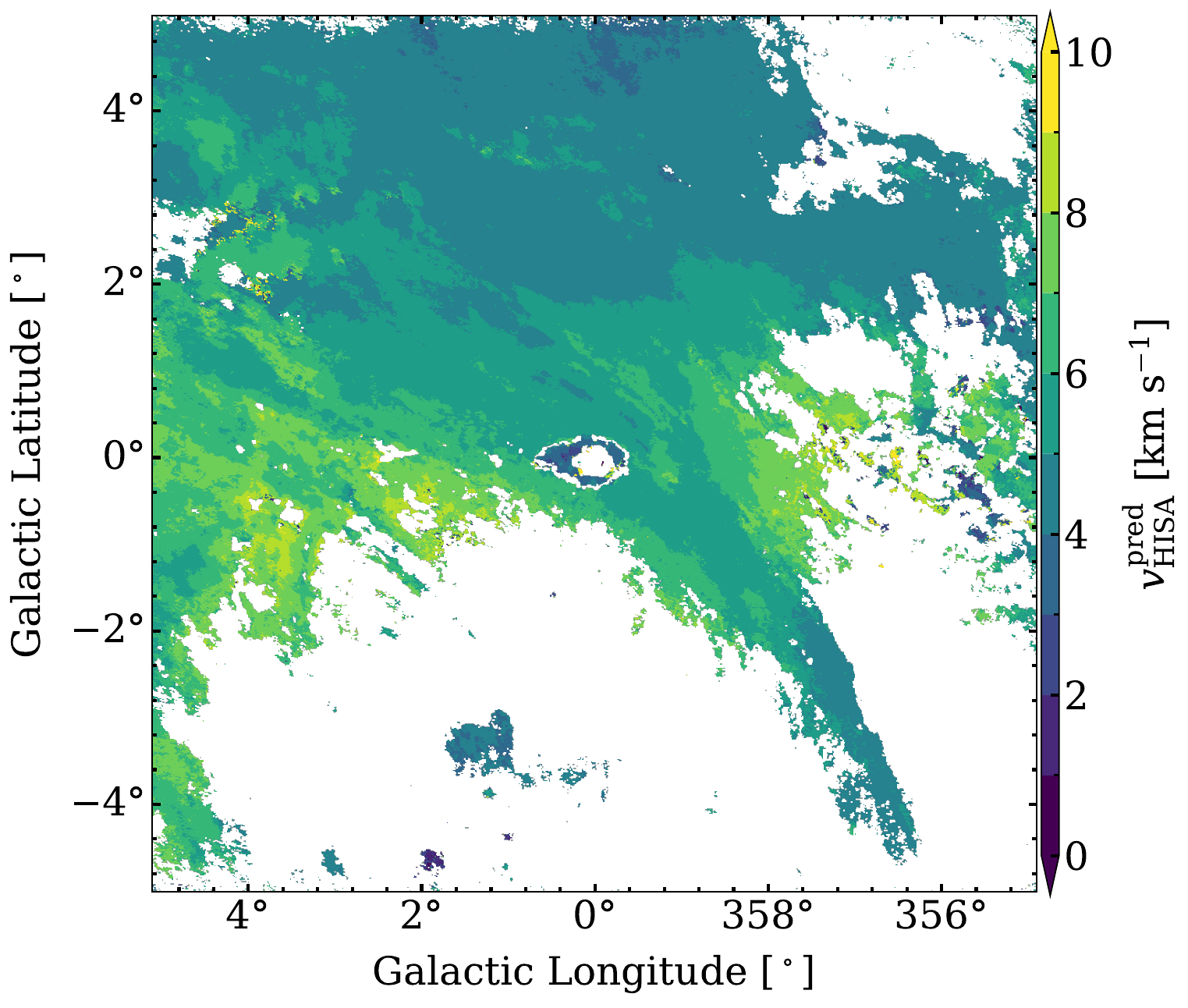}
    \caption{Predicted HISA velocities in the \rc cloud. Only pixels with a positive HISA detection are shown. Coherent velocity structures are present across the cloud, but there are regions of non-detection and noisy HISA velocities present. The clump of absorbing gas in the lower region of the image ($l\sim1\fdg5$, $b\sim-3\fdg5$) observed in \citet{mcclure-griffiths2006} has been recovered in both detections and velocity. The Galactic Centre ($l=0\degr$, $b=0\degr$) shows a region of non-detections, surrounded by a ring of detections, then a second ring of non-detections. These detections should not be interpreted as they stem from the complex spectra of the radio emission of the Galactic Centre but are shown here for completeness. Alt text: graph showing a cloud, with a colour gradient that changes across the cloud from top to bottom, representing the velocity of the cloud.}
    \label{fig:rc velocity map}
\end{figure}

Figure~\ref{fig:rc velocity map} shows the predicted HISA velocities for the detected features in the \rc cloud. Again, a restriction on the current velocity localisation method described in Section~\ref{sec:velocity localisation} is that it will only localise the strongest feature, meaning a single velocity is returned for each spectrum, and we discard velocities for non-detection spectra ($\mathbb{P}({\rm HISA})<0.5$). Strikingly, the predicted velocity structure is spatially coherent across large portions of the cloud, with velocities varying smoothly from $\sim3$--$4~{\rm km~s}^{-1}$ in the northern and eastern portions of the cloud to $\sim6$--$8~{\rm km~s}^{-1}$ towards the south, consistent with the known velocity gradient of the region of the \rc cloud towards the Galactic Centre, identified by \citet{mcclure-griffiths2006}. This coherent velocity structure emerges despite the fact that the neural network processes each spectrum entirely independently, with no positional or spectral context from neighbouring pixels. The southern clump at ($l\sim1\fdg5$, $b\sim-3\fdg5$), visible in the integrated emission maps of \citet{mcclure-griffiths2006}, is recovered with predicted velocities in good agreement with the interpolation-based velocities for that sub-region of $\sim3$--$5~{\rm km~s}^{-1}$.

We do not implement any spatial regularisation in our algorithm; the coherent velocity structure seen in Figure~\ref{fig:rc velocity map} arises purely from the consistency of the spectral features across neighbouring sightlines through the same physical cloud. This is in contrast to emission decomposition methods such as ROHSA \citep{marchal2019a}, which explicitly enforce spatial smoothness as a regularisation prior (although they aren't designed for detecting self-absorption). The recovery of spatially coherent velocity structure without such regularisation lends confidence that the CNN is genuinely responding to the HISA features, rather than noise or artefacts in the data.

While there is an overall coherent structure in the predicted velocities, there are distinct regions along the Galactic Plane and towards the upper-left of the \rc cloud that display incoherent velocity structure. On the left side of Figure~\ref{fig:rc velocity map}, the complex Galactic Plane emission ($l>1\degr$, $b=0\degr$) is strongly absorbed by the \rc cloud, and a clear HISA feature is created over the emission. However, on the right side of the same figure, there is no significant absorption from a cloud in front of the complex lines of sight, causing the neural network to assign self-absorption detections and velocities to features that may not originate from the \ion{H}{i} gas in the \rc cloud, leading to predicted velocities that appear random ($l<358\degr$, $|b|\lid1\degr$). There is also a region of incoherent velocity detections towards the upper-left of Figure~\ref{fig:rc velocity map} ($l=4\degr$, $b=1\fdg6$--$3\degr$), but this region of non-detections and incoherent velocities is also present in \citet{mcclure-griffiths2006}, and it is likely the result of a real void in the \ion{H}{i} gas of the \rc cloud causing background emission to bleed through.

\subsubsection{Spectral comparison at the cloud boundary}\label{sec:rc spectra comparison}

\begin{figure}
    \centering
    \includegraphics[width=\linewidth]{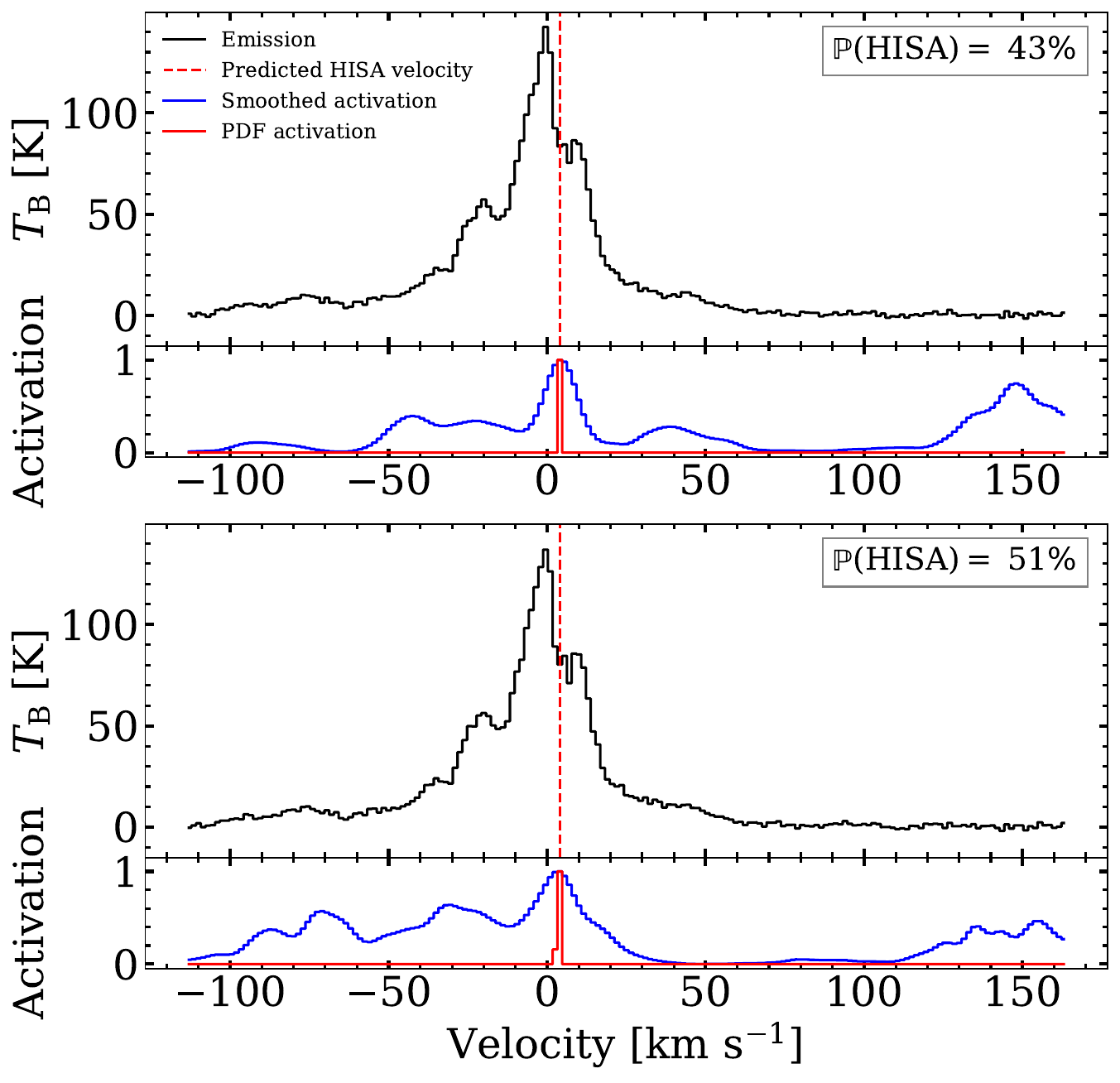}
    \caption{Neighbouring spectra from the Riegel--Crutcher cloud towards $l=355\fdg2$, $b=1\fdg5$ that have different classifications. Both spectra (black lines) and their HISA probabilities, as well as the predicted HISA velocity (dashed red lines) from the smoothed activation spectra (blue lines) and PDF activation spectra (red lines), are shown. Top: the non-detection (HISA probability of 43 per cent) spectrum. Bottom: the detection (HISA probability of 51 per cent) spectrum. The spectra are very similar, both showing potential self-absorption features between $\sim0$--$10~{\rm km~s}^{-1}$. The neural network localises an absorption feature at $\sim4~{\rm km~s}^{-1}$ in both spectra, despite the difference in classification. Alt text: two graphs showing emission spectra and activation spectra. The two emission spectra appear almost exactly the same, showing broad strong emission in the middle, and an off-centre narrow strong absorption feature. The activation spectra of both graphs peak at the same location as the absorption feature. Text in the upper right of each panel reports a 43 per cent HISA probability for the top graph and a 51 per cent HISA probability for the bottom graph.}
    \label{fig:rc spectra comparison}
\end{figure}

The spatial boundary of the \rc cloud, where the self-absorption transitions from clearly present to absent, provides an informative test of the classifier's behaviour near the detection threshold. Figure~\ref{fig:rc spectra comparison} shows two neighbouring spectra from the edge of the \rc cloud (shown by the red circle in Figure~\ref{fig:rc detection map}) that have different classifications: one spectrum is assigned a HISA probability of 43 per cent and the other 51 per cent (classified as a non-detection and detection, respectively, with a 0.5 threshold). Both spectra display a prominent absorption feature at $\sim4~{\rm km~s}^{-1}$. The near-identical appearance of the two spectra highlights that the CNN assigns a continuous-valued output probability, and the 0.5 binary classification threshold falls between the two values in this case. 

This pair of spectra also demonstrates the behaviour of the neural network at the edges of the physical structure of the cloud, where the HISA signal transitions continuously from detectable to undetectable, and no discrete threshold can perfectly separate the two cases. Users who require a higher-purity detected sample can raise the classification threshold at the cost of reduced completeness. Notably, despite the different classifications, the predicted velocity is consistent between both spectra at $\sim2~{\rm km~s}^{-1}$, indicating that the network identifies the same spectral feature in both cases; only the probability assigned to that feature differs.

\subsubsection{Comparison with the linear interpolation method}\label{sec:rc interpolation comparison}

\begin{figure}
    \centering
    \includegraphics[width=\linewidth]{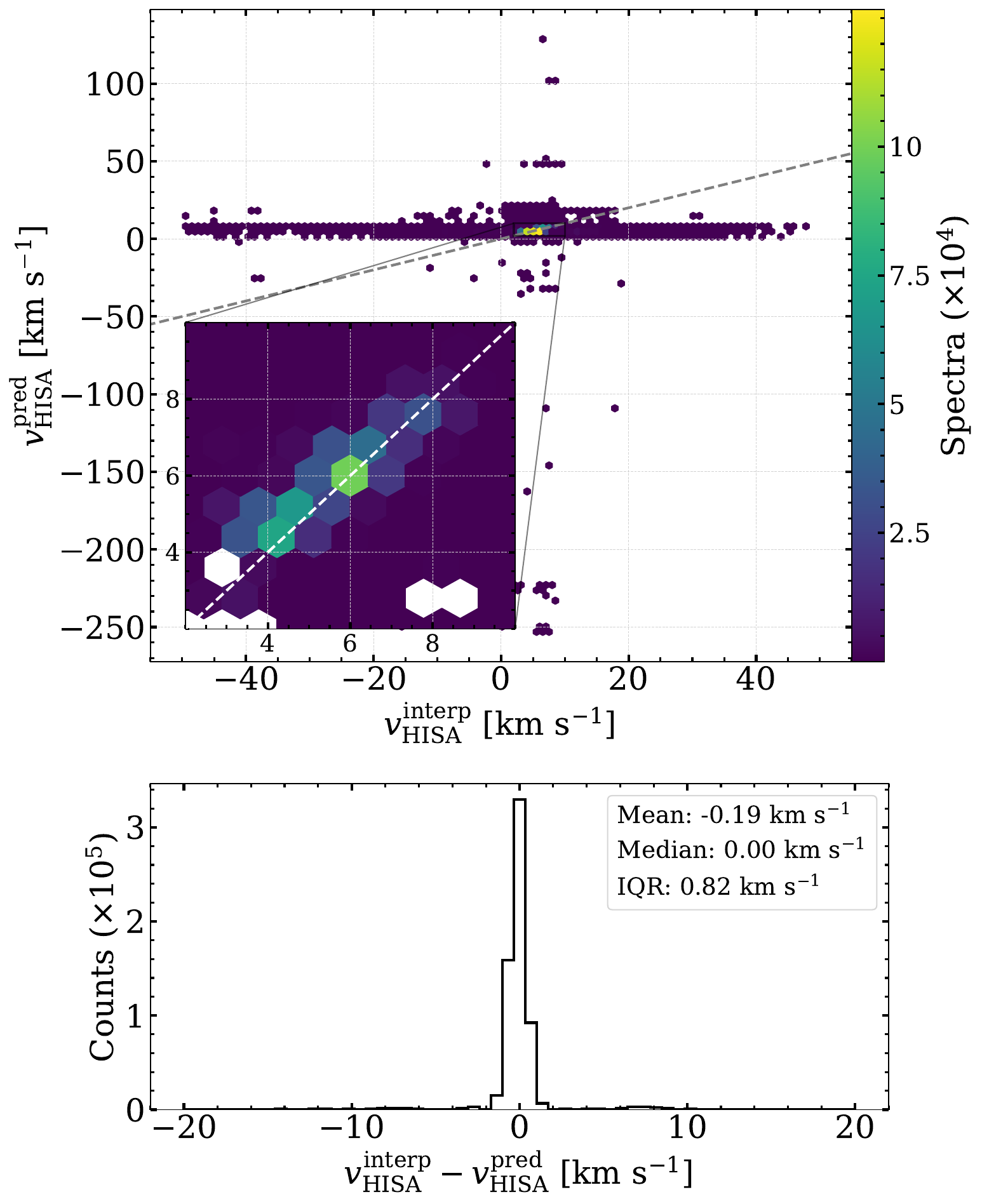}
    \caption{Top: scatter of the interpolated ($v_{\rm HISA}^{\rm interp}$) using the linear background emission modelling method (Section~\ref{sec:rc data}) and predicted ($v_{\rm HISA}^{\rm pred}$) HISA feature velocities from the neural network activation mapping method of the \rc cloud. The inset covers the known velocities of the \rc cloud, where the majority of the agreeing velocities between the two methods exist. The dashed lines show the one-to-one lines. Bottom: histogram of offsets between the interpolated and predicted HISA feature velocities of the \rc cloud. Summary statistics of the offset distribution are reported in the upper-right corner. The velocities show a strong correlation with each other and a narrow distribution of offsets. Alt text: a two-panel graph showing a two-dimensional histogram in the upper panel, and a one-dimensional histogram in the lower panel. The upper panel shows a large spread of dark pixels, indicating low counts, with a narrow horizontal band of low counts across the graph. The upper panel also shows a broad spread of a few dark pixels vertically. The upper panel has an inset figure, showing a small central region that has a strong one-to-one correlation of bright pixels, aligning with a dashed one-to-one line. The lower panel shows a narrow distribution, and text in the upper right reports the mean, median, and interquartile range of the distribution.}
    \label{fig:rc velocities}
\end{figure}

To provide a quantitative comparison for the CNN velocity estimates in the \rc cloud, we apply the linear interpolation method described in Section~\ref{sec:rc data} to the SGPS GC data. Figure~\ref{fig:rc velocities} shows the scatter between the velocities of the HISA features detected using linear interpolation ($v_{\rm HISA}^{\rm interp}$) and the predicted HISA feature velocities from the CNN ($v_{\rm HISA}^{\rm pred}$), and the distribution of the offsets between them. The mean (median) of the velocity offsets is $-0.19~{\rm km~s}^{-1}$ ($0~{\rm km~s}^{-1}$), with an interquartile range of $0.82~{\rm km~s}^{-1}$. Table~\ref{tab:rc cross detections} summarises the cross-detection statistics between the CNN and the linear interpolation method across all spectra in the SGPS GC datacube. Of the 1,092,000 spectra, 58.8 per cent are classified as containing HISA by both methods, 0.4 per cent by the CNN alone, 31.2 per cent by the interpolation method alone, and 9.5 per cent by neither. The overall agreement rate between the two methods is 68.3 per cent. The velocity comparison in Figure~\ref{fig:rc velocities} is computed exclusively from the 642,306 sightlines where both methods detect HISA, and the 88.9 per cent within-one-channel agreement should therefore be understood as characterising the regime where both methods are most confident, rather than the full performance of either method across all sightlines.

\begin{table}
    \centering
    \caption{Cross-detection statistics between the CNN with a threshold of 0.5 and the linear interpolation method for all spectra in the SGPS GC datacube. Fractions are given relative to the total number of spectra in the cube.}
    \begin{tabular}{rcc}
        \hline
          & Interpolation: No HISA & Interpolation: HISA \\\hline
         CNN: No HISA & 103,937 (9.5\%) & 341,164 (31.2\%) \\
         CNN: HISA & 4,593 (0.4\%) & 642,306 (58.8\%) \\\hline
    \end{tabular}
    \label{tab:rc cross detections}
\end{table}

The $4,593$ CNN-only detections are located sporadically around the field, with concentrations around the periphery of the southern clump and other smaller clumps towards the lower edge of the field, as well as a region in the upper-right of the field. Figure~\ref{fig:cnn only detection map} illustrates the locations of these CNN-only detections. The spatial distribution of the CNN-only detections is suggestive of a physical origin, and the sightlines around the southern clumps and throughout the remainder of the cloud may represent genuine HISA candidates missed by the interpolation criteria. However, visual inspection of the spectra reveals that the detected features manifest either as very steep, single-channel gradients that the network infers are the result of self-absorption, or as very narrow and deep features not located at the peak of an emission component. Both of these cases will be missed by the interpolation method by design if no suitable peak is found on either side of the feature, either because there is only a single strong emission feature with a steep gradient on one side, or because the narrow feature lies between two peaks that are too subtle for the peak-finding algorithm to detect.

The $\mathbb{P}({\rm HISA})$ distribution for these detections is tightly concentrated near unity: the median is $1.00$, the interquartile range is $0.031$, the 10th percentile is $0.77$, and the 90th percentile is $1.00$. Thus, the network assigns high confidence to nearly all CNN-only detections, with relatively few intermediate-confidence predictions. One plausible explanation is that the features triggering the CNN classification are shallower than the 5~K depth threshold used by the interpolation-based detection method. The spatial distribution/clustering of these CNN-only detections is also suggestive of a physical origin: their concentration around the southern clumps and their presence throughout the remainder of the cloud indicate that some may represent genuine HISA candidates that are missed by the interpolation criteria. 

The concentration of $\mathbb{P}({\rm HISA})$ near unity is also consistent with a statistical interpretation that is not mutually exclusive with the physical one described above. Because the training set contains only strong HISA features (depth $\geq8$~K, Section~\ref{sec:synthetic data}) and spectra without an inserted absorber, the network is not exposed during training to the boundary between shallow self-absorption and noise. When presented with features in this intermediate regime, the network must therefore extrapolate its learned decision function beyond the distribution represented in the training data. This extrapolation can result in extreme predicted probabilities rather than intermediate values that reflect genuine uncertainty.

This behaviour is not inconsistent with the low aggregate calibration errors reported in Section~\ref{sec:synthetic evaluation}. Those metrics were evaluated on a test set drawn from the same distribution as the training data and containing no weak or ambiguous HISA features. They therefore assess calibration only within the regime represented by the training distribution and cannot reveal miscalibration in an unrepresented regime. The CNN-only detections examined here fall precisely within such an unexplored regime. Their near-unity $\mathbb{P}({\rm HISA})$ values are therefore consistent with good calibration on the training distribution while providing no evidence that the probabilities are well calibrated for sub-threshold or ambiguous HISA features.

The $341,164$ interpolation-only detections represent features the CNN missed, but that were detected in some way by the linear interpolation. These sightlines lie around the edges of the cloud and in much of the field not covered by the CNN detections, and consist on average of features that would not be considered self-absorption if inspected by-eye. Figure~\ref{fig:linear interpolation only map} shows the locations of these detections. These sightlines have a median CNN probability of $\mathbb{P}({\rm HISA}) = 0$, consistent with the network extrapolating its decision function into the sub-threshold regime in the opposite direction: where the CNN-only detections receive $\mathbb{P}({\rm HISA}) \approx 1$ by overconfident extrapolation, the interpolation-only detections receive $\mathbb{P}({\rm HISA}) \approx 0$ by the same mechanism. The network is not producing the intermediate confidence values that would be expected at the edge of its training distribution, but instead is saturating at both extremes. This is further evidence that $\mathbb{P}({\rm HISA})$ is not a calibrated probability outside the high-confidence regime.

The median integrated $\Delta T_{\rm B}$ of these spectra from the linear interpolation is $35.75~{\rm K~km~s}^{-1}$, compared to $184.99~{\rm K~km~s}^{-1}$ for jointly detected features, indicating that many of these sightlines contain absorption below the network's effective 8~K HISA completeness floor. Beyond depth, however, the dominant cause appears to be morphological, with by-eye inspections revealing that the spectra do not contain obvious self-absorption, but rather contain emission features with steep gradients lying between two peaks at very different $T_{\rm B}$ values. These produce interpolated $\Delta T_{\rm B}$ profiles that pass the broadness check defined in Section~\ref{sec:rc data}, but have a large $\Delta T_{\rm B}$ between the two peaks, resulting in a spurious detection by the interpolation method rather than a missed detection by the CNN.

\subsubsection{Performance on the Riegel--Crutcher cloud}\label{sec:rc performance}

When presented with a simple dataset such as the \rc cloud, the neural network demonstrates that it can transfer its ability for HISA detection without needing to be trained on data that mimics the properties of the \rc cloud explicitly. The spatial distribution of detections closely matches the known extent of the cloud, including its filamentary substructure and the separate southern clump at ($l\sim1\fdg5$, $b\sim-3\fdg5$). The velocity localisation shows excellent agreement with the linear interpolation method from \citet{mcclure-griffiths2006}, with 88.9 per cent of detections lying within a single channel of the reference velocities derived from the linear interpolation method (Section~\ref{sec:rc data}) and a median offset of $0~{\rm km~s}^{-1}$. The spatially coherent velocity gradient recovered without any positional context to the network is a promising validation: the physical structure of the cloud is reflected in the spatial coherence of the HISA features, and the network responds consistently across neighbouring sightlines.

\subsection{Giant molecular filament regions}\label{sec:gmf results}

We now apply the trained CNN to the five GMF regions from the THOR+VGPS survey described in Section~\ref{sec:thor data}. The GMF regions represent a considerably more demanding test of the neural network than the \rc cloud: inner Galactic Plane sightlines contain multiple overlapping \ion{H}{i} emission components at a wide range of velocities, and genuine HISA features can be superimposed on complex, structured backgrounds with no single dominant emission component. However, we try to mimic this exact kind of structure in our synthetic dataset to increase the ability of the CNN to perform in these challenging regimes. Unlike the SGPS GC data, the THOR data already match the $1.5~{\rm km~s^{-1}}$ spectral resolution and the 185-channel length of the synthetic training spectra by design; therefore, no spectral resampling is required. We apply the same min-max brightness temperature scaling used for the synthetic and \rc datasets and pass the emission spectra through the network. We initially use the same HISA detection threshold of 0.5 for the GMF regions; however, with this threshold, the CNN classifies 88.4 per cent of sightlines across the five regions as containing a HISA feature. This fraction is likely inflated relative to the true HISA prevalence in these fields, primarily as a consequence of the complex sightlines of the inner Galactic Plane that contain many dips, some HISA but mostly the result of two close Gaussian features, which are not easily understood by the network. At a classification threshold of 0.9, the detection fraction falls to 77.6 per cent, and at a classification threshold of 1.0, the detection fraction falls further to 48.6 per cent. For comparison, the \textsc{astroSaber} algorithm of \citet{syed2023}, applied to the same five fields, recovered a HISA detection fraction of 35.9 per cent when restricting their detections to the velocity range of each GMF reported in \citet{ragan2014}, which likely led to missed HISA detections in the fields. The HISA detection fraction is therefore strongly influenced by the chosen classification threshold, and users applying the network to the inner Galactic Plane should adopt a higher detection threshold than used for the \rc cloud. Given these detection fractions, we limit detections in the GMF fields to sightlines with a HISA detection threshold of 1.0, which we adopt for all GMF analyses in the remainder of this paper. Because the real GMF spectra differ substantially from the synthetic training distribution, this threshold should be interpreted as a stringent numerical-confidence cut rather than a calibrated purity guarantee.

\subsubsection{HISA detection in giant molecular filaments}\label{sec:gmf detection}

\begin{figure}
    \centering
    \includegraphics[width=\linewidth]{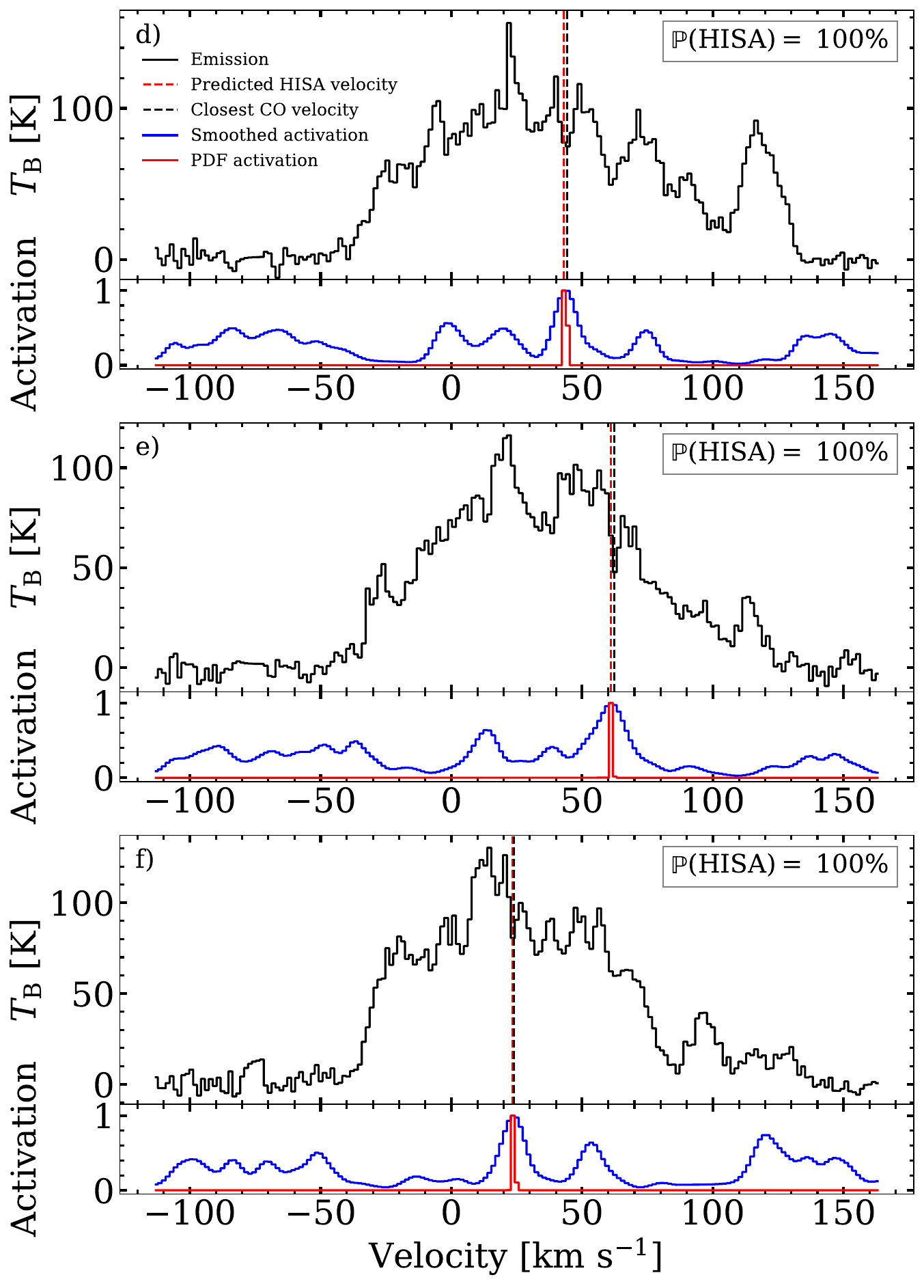}
    \caption{Three spectra with strong HISA detections from the GMF20 field. The panel labels correspond to position indicators in Figure~\ref{fig:gmf20 detection map}. Red dashed lines indicate the predicted HISA velocity, while dashed black lines indicate the closest \coion velocity to the predicted HISA velocity. All three spectra show strong activation and velocity localisation at absorption features that correspond with known \coion signal. However, there are false activation spikes that appear outside of the signal as a result of the increased S/N ratio, although these are ultimately suppressed in the PDF activation spectra. Alt text: three graphs showing emission spectra and activation spectra, labelled D, E, and F. All three emission spectra show a very broad range of emission, spanning almost the full range of the spectra, and multiple narrow emission and absorption features. The activation of the three graphs varies across the spectra, but peaks at a detected absorption feature. Vertical dashed red lines appear very close to vertical dashed black lines, at the location of the peak of the activation spectra. Text in the upper right of each panel reports a 100 per cent HISA probability.}
    \label{fig:gmf20 strong detections}
\end{figure}

Figure~\ref{fig:gmf20 strong detections} shows three example spectra and their corresponding activation spectra from different positions in the GMF20 field. All three spectra have 100 per cent HISA detection probabilities, corresponding to absorption features at three different velocities shown by the red dashed lines. The panels also show the closest \coion velocity to the predicted HISA velocity from the Gaussian decomposition of the GRS data discussed in Section~\ref{sec:thor data} in black. While the \rc cloud spectra shown in Figure~\ref{fig:rc representative spectra} demonstrate activation peaks in other regions of the spectrum away from the absorption feature, the spectra from the GMF data show this much more commonly, as a result of the spectral complexity and the increased S/N ratio. While the majority of activation peaks in Figure~\ref{fig:gmf20 strong detections} correspond to potential HISA features in the main signal ($-30$--$120~{\rm km~s}^{-1}$), the activation peaks outside of this velocity range correspond only to features produced by noise.

\begin{figure*}
    \centering
    \includegraphics[width=0.9\linewidth]{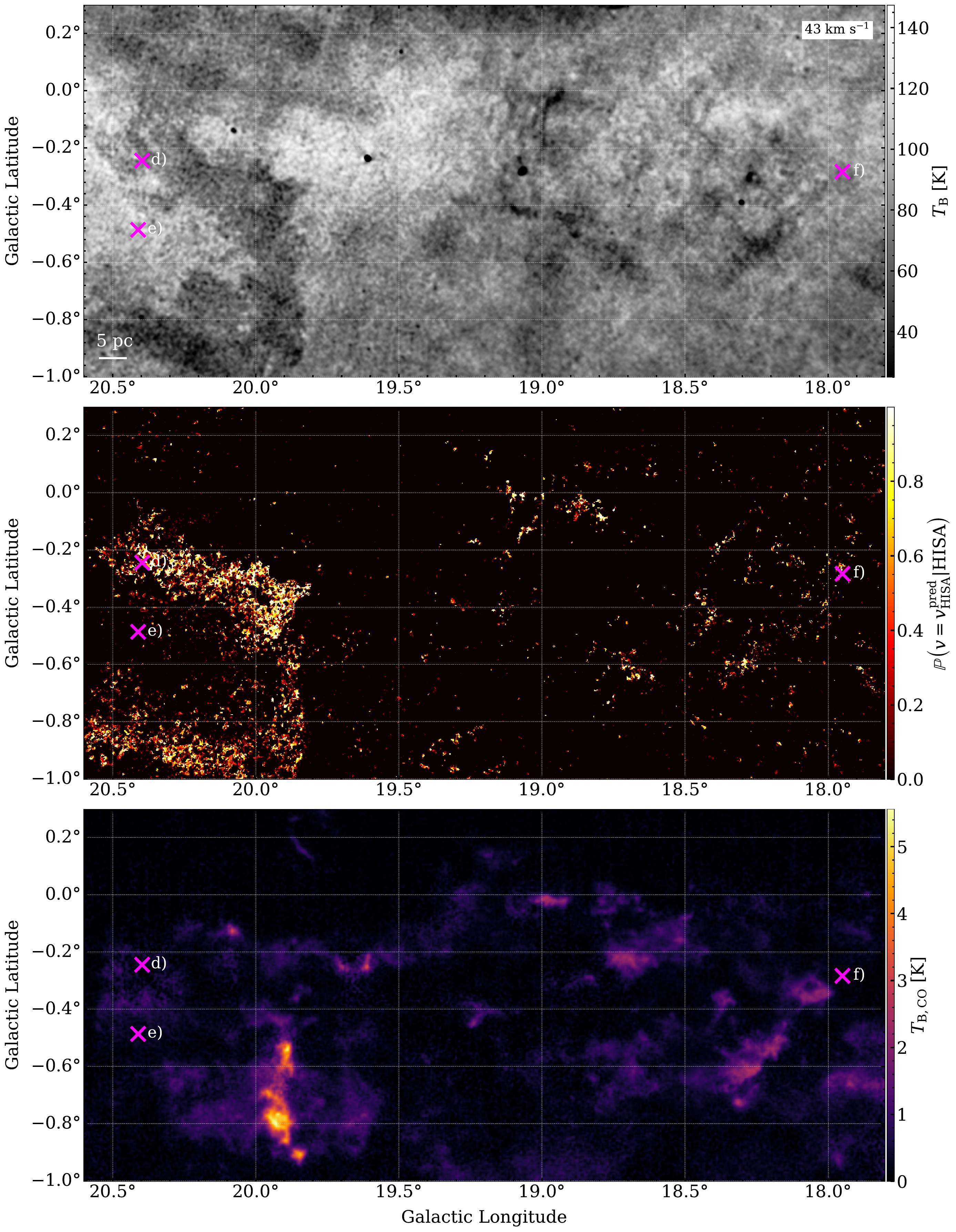}
    \caption{Neural network HISA detection in GMF20. Top: THOR+VGPS \ion{H}{i} emission slice at $43~{\rm km~s}^{-1}$. Middle: slice of the PDF activation spectra at the same velocity, showing the probability that detected HISA features in each pixel are located at $43~{\rm km~s}^{-1}$. Bottom: GRS \coion emission velocity slice at $43.1~{\rm km~s}^{-1}$. The velocity range of GMF20 from \citet{ragan2014} is $37$--$50~{\rm km~s}^{-1}$. Magenta crosses and the corresponding labels indicate the positions of three representative strong detections in the GMF field, shown in Figure~\ref{fig:gmf20 strong detections}. The vertical feature ($l=20\degr$, $b<-0\fdg3$) is present in all three maps, demonstrating the correlation between \ion{H}{i} and \coion, and the resulting HISA detections. Alt text: three graphs showing very similar images in different colours, with magenta crosses labelled D, E, and F plotted at different points on the graphs. The top panel shows a grey-scale image of some clouds, and text in the upper right gives a velocity. The middle panel shows a heat map, with most of the panel filled in black, but some bright regions that align with dark patches in the top panel. The bottom panel shows a colour image of wispy cloud structures across the entire panel, with some bright patches at the same locations as the dark patches in the top panel, and the bright regions in the middle panel.}
    \label{fig:gmf20 detection map}
\end{figure*}

Figure~\ref{fig:gmf20 detection map} shows a velocity slice of the THOR+VGPS \ion{H}{i} emission for the GMF20 region, the corresponding slice of the PDF activation spectra at the same velocity, and a GRS \coion emission slice at the same velocity. The equivalent figures for all five GMF regions are shown in Appendix~\ref{app:gmf maps}. Although we choose representative velocity slices within the filaments that best show the correlation between the \ion{H}{i}, HISA detections, and \coion in Figure~\ref{fig:gmf20 detection map} and Figures~\ref{fig:gmf26 detection map}--\ref{fig:gmf54 detection map}, within the velocity range of the filaments there are no slices of HISA detections that appear correlated in some way outside of the GMFs. Additionally, there are certainly false detections both inside and outside of the filament velocities, with observed instances following a randomly scattered morphology on sky. Despite this, there are also likely real HISA detections not associated with the filaments, which current methods relying on molecular gas as a tracer will miss.

The GMF fields expose a fundamental challenge of HISA detection in complex sightlines: a two-dimensional projection of detections in such complex fields makes it difficult to disentangle structures at different velocities. However, an advantage of our method is that the velocity localisation provided by the activation spectra simultaneously recovers where in velocity space the HISA resides, meaning the one-dimensional detector can function as a three-dimensional PPV detector without being explicitly trained to do so. By examining which sightlines are detected at a given velocity, coherent cold-gas structures can be extracted from the otherwise confusion-limited field. This is illustrated directly in the middle panels of Figure~\ref{fig:gmf20 detection map} and Figures~\ref{fig:gmf26 detection map}--\ref{fig:gmf54 detection map}, which show slices of the PDF activation spectra at the velocity of each filament: detections appear to be spatially concentrated following the morphology of the \coion emission, demonstrating that the activation spectra recover the structure of cold gas around the filaments.

Previous theoretical and observational works have shown that the presence of molecular species has been linked to regions of cold \ion{H}{i} because of the shielding from radiation and the cold, dense conditions that the CNM provides, required for molecules to form and survive \citep{knapp1974, wolfire2010, sternberg2014, park2023}. The detection rate of HISA features in the GMF fields reflects the correlation between \ion{H}{i} and \coion structures, with detections appearing more common in regions of enhanced CO emission. The HISA detection fraction across all GMF fields increases to 59.0 per cent using a threshold of 1.0 when restricted to sightlines above set integrated \coion emission thresholds, compared to the 48.6 per cent detection fraction when no emission threshold is used. Here we use the same thresholds used in \citet{syed2023} of 42, 34, 30, 20, 34, and $20~{\rm K~km~s}^{-1}$ for the GMF20, 26, 38a, 38b, 41, and 54 regions, respectively. While this enhancement is modest in absolute terms, it is consistent with spatial association between cold \ion{H}{i} gas and molecular material, which is expected of molecular gas at the centre of a shielded CNM region. Additionally, as CNM gas is expected to exist spatially outside the molecular material, and given the dependence of HISA detections on the emission configuration and S/N ratio of the spectra, a 100 per cent detection rate in bright molecular regions is not expected. The GMF38 field shows the greatest concentration of HISA detections coincident with the \coion emission, with a HISA detection fraction of 62.9 per cent in pixels with \coion emission above the 30 and $20~{\rm K~km~s}^{-1}$ for the GMF38a and GMF38b filaments (see Figures~\ref{fig:gmf38a detection map} and \ref{fig:gmf38b detection map}), while the GMF54 field shows the greatest increase in detection fraction ($+8.6$ per cent).

\subsubsection{HISA velocity localisation}\label{sec:gmf velocities}

Following the detection step, we apply the activation mapping method described in Section~\ref{sec:velocity localisation} to infer HISA feature velocities for all positively classified spectra in the five GMF fields. As an independent reference for comparison, we use \coion velocities from the Gaussian decomposition of GRS spectra by \citet{riener2020}, as described in Section~\ref{sec:thor data}. Cold atomic hydrogen co-spatial with molecular gas would be expected to share a similar line-of-sight velocity, so agreement between the predicted HISA velocities and the \coion velocities serves as a physically motivated consistency check, albeit with the caveat that a small systematic offset between atomic and molecular tracers is physically expected in at least some sightlines \citep[e.g.][]{wang2020a}.

\begin{figure}
    \centering
    \includegraphics[width=\linewidth]{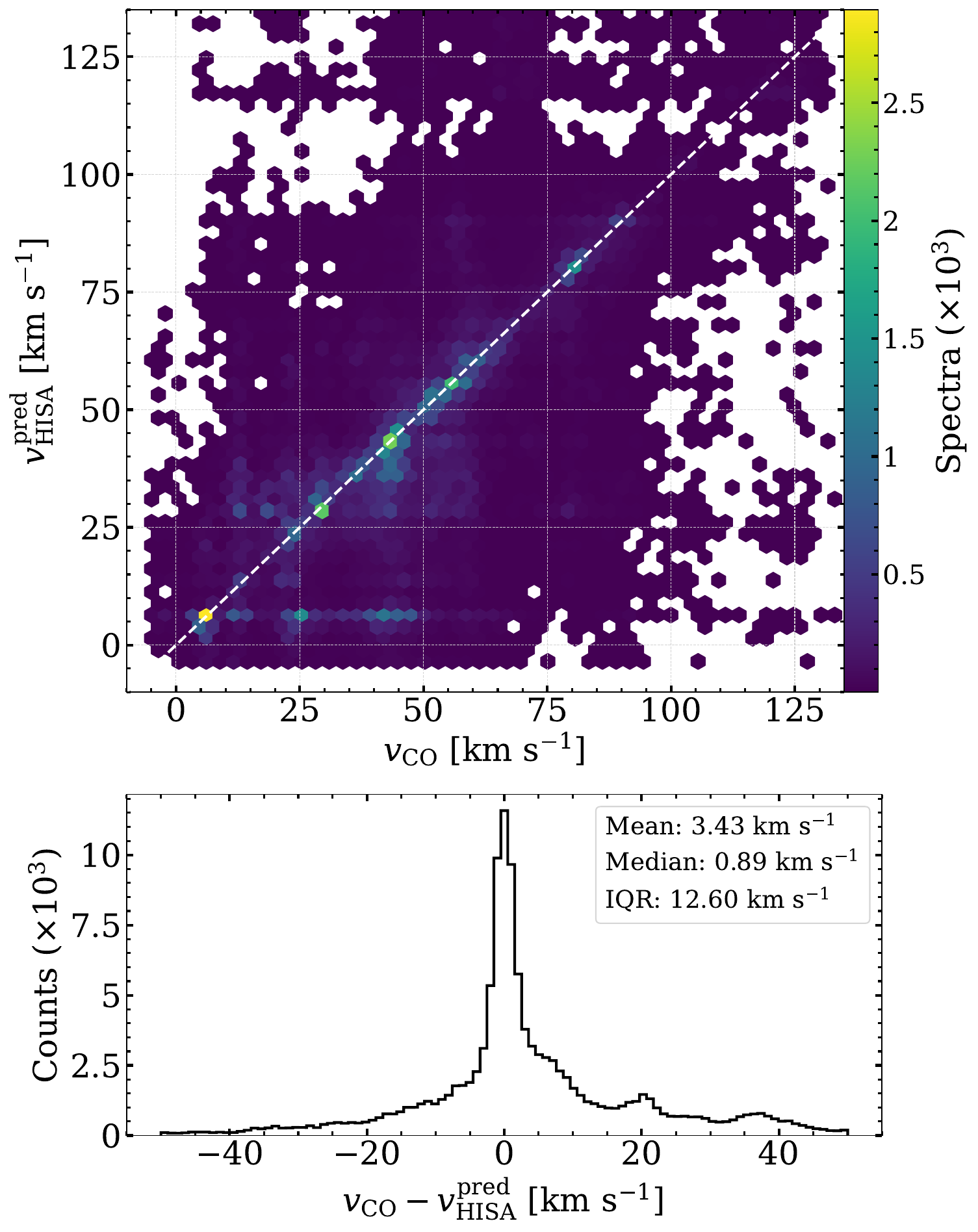}
    \caption{Top: scatter of the \coion ($v_{\rm CO}$) and predicted ($v_{\rm HISA}^{\rm pred}$) HISA feature velocities of the GMF regions. The dashed line shows the one-to-one line. Bottom: histogram of offsets between the CO and predicted HISA feature velocities of the GMF regions. Summary statistics of the offset distribution are reported in the upper-right corner. Alt text: a two-panel graph showing a two-dimensional histogram in the upper panel, and a one-dimensional histogram in the lower panel. The upper panel shows a large spread of dark pixels across the entire panel, indicating low counts. There is a strong one-to-one correlation of bright pixels at discrete points along a dashed one-to-one line, and there is a horizontal streak of bright points along the bottom edge of the panel, showing a range of \coion velocities that correspond to a single HISA velocity. The lower panel shows a narrow distribution centred at zero, that is positively-skewed and shows a pair of smaller bumps at 20 and 35 kilometres per second. Text in the upper right reports the mean, median, and interquartile range of the distribution.}
    \label{fig:gmf velocities}
\end{figure}

Figure~\ref{fig:gmf velocities} shows the scatter between the \coion velocities ($v_{\rm CO}$) and the predicted HISA feature velocities ($v_{\rm HISA}^{\rm pred}$), and the distribution of the offsets between the \coion and predicted HISA velocities. This only includes sightlines in the cubes where the velocity of the predicted HISA component from the neural network lies within the velocity range of the \coion data. While a majority of the \coion and HISA velocities show a strong correlation with each other, there are a number of spectra where the velocities do not agree. The densest correlations along the one-to-one line ($\sim45$--$55~{\rm km~s}^{-1}$) correspond with the line-of-sight velocity ranges of the filaments, as given in table 2 of \citet{ragan2014}.

The bottom panel of Figure~\ref{fig:gmf velocities} shows the velocity offsets between the HISA and CO velocities in the GMF fields. The distribution of velocity offsets shows a much broader distribution than the \rc cloud or synthetic velocity offsets, and the distribution is skewed towards positive velocity offsets as a result of a tail of predicted HISA velocities at $\sim5~{\rm km~s}^{-1}$ that do not correspond well with the \coion velocities, seen in the upper panel of Figure~\ref{fig:gmf velocities}. The offsets range from $-128$ to $130~{\rm km~s}^{-1}$, with a mean (median) of $3.43~{\rm km~s}^{-1}$ ($0.89~{\rm km~s}^{-1}$), and an interquartile range of $12.60~{\rm km~s}^{-1}$. Of the velocity offsets, 24.6 per cent lie within a single \ion{H}{i} channel width of $1.5~{\rm km~s}^{-1}$, 55.0 per cent exceed $|5|~{\rm km~s}^{-1}$ and 49.2 per cent exceed $|10|~{\rm km~s}^{-1}$. Velocity offsets between molecular tracers and cold atomic gas are physically expected, as \coion forms in the most shielded regions of the cloud. Offsets of up to $\sim5~{\rm km~s}^{-1}$ have been observed previously in similar environments \citep[e.g.][]{wang2020a, park2023}; however, 55.0 per cent of offsets exceed this. We note that any errors on the \coion velocities from \citet{riener2020} do not contribute greatly to the velocity offsets, with an average error of $0.18~{\rm km~s}^{-1}$ on the fitted $v_{\rm LSR, CO}$ values reported. The tail of catastrophic localisation failures around $v_{\rm HISA}\sim5~{\rm km~s}^{-1}$ comes primarily from the GMF20 field, and all spectra in this tail have a potential HISA feature at $\sim5~{\rm km~s}^{-1}$. The positions of these sightlines appear to be structured in the field, and may indicate candidate self-absorption features in this field that are not traced by the \coion gas.

The larger scatter in the GMF velocity comparison relative to the \rc comparison (IQR $=12.60~{\rm km~s}^{-1}$ versus $0.82~{\rm km~s}^{-1}$) reflects the greater spectral complexity of the inner Galactic Plane sightlines, as well as the velocity offset between \coion and \ion{H}{i}. In the \rc cloud, a single dominant absorption feature at $\sim0$--$10~{\rm km~s}^{-1}$ provides a well-defined activation peak, whereas GMF sightlines frequently contain multiple emission and absorption components across a wide velocity range, making the activation mapping more susceptible to latching on to the wrong spectral feature. The non-negligible fraction of sightlines with large velocity offsets between the CNN and \coion predictions is therefore consistent with the known difficulty of HISA identification in confused regions. Crucially, the median offset of $0.89~{\rm km~s}^{-1}$ is sub-channel, confirming that the central tendency of the velocity localisation is reliable even in the complex inner-plane environment.

\begin{figure}
    \centering
    \includegraphics[width=\linewidth]{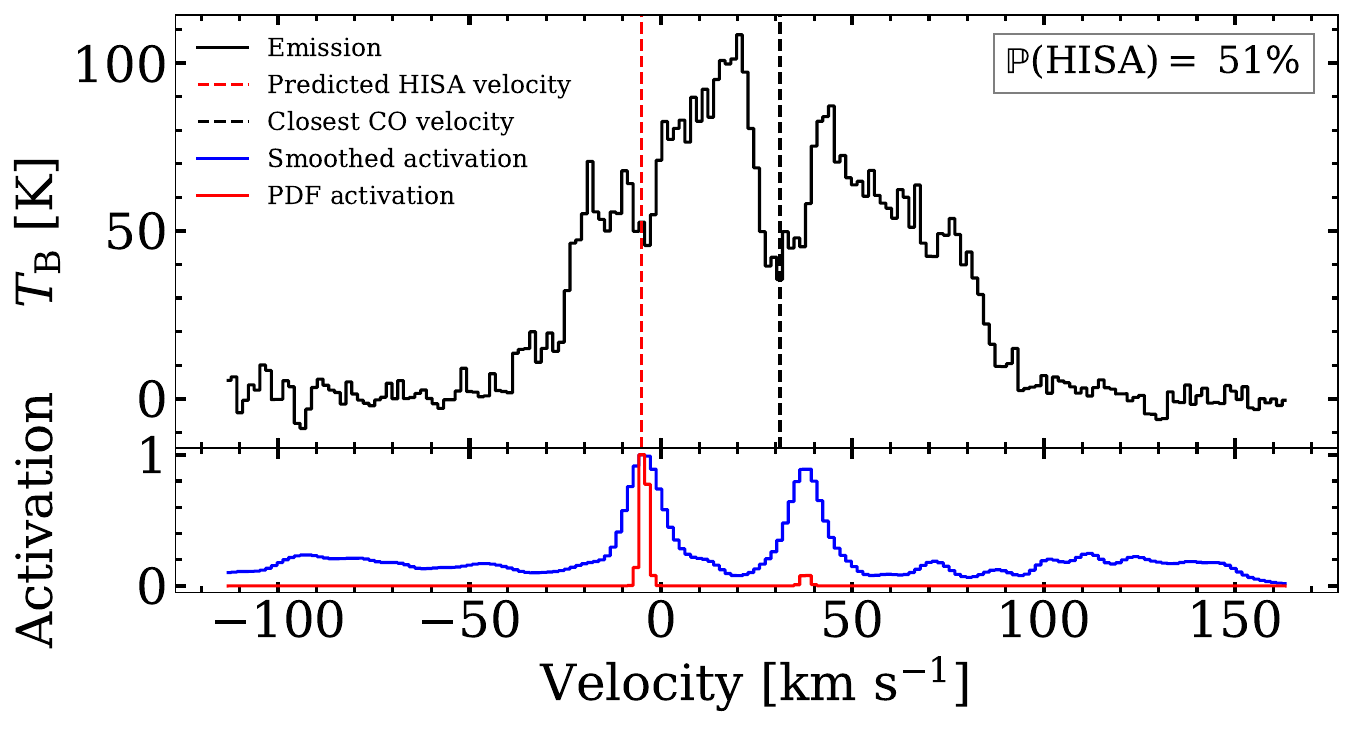}
    \caption{An emission spectrum from the GMF20 field towards $l=20\fdg6$, $b=-1\degr$, demonstrating a case where the offset between the predicted HISA velocity and the \coion velocity is large. The red dashed line shows the predicted HISA feature velocity from the neural network, while the black dashed line shows the closest \coion velocity from the Gaussian decomposition of the GRS emission data. The smoothed activation spectrum (blue line) shows a significant peak at the HISA feature detected by the neural network, as well as a secondary activation peak around the \coion velocity. The activation peaks are similar enough in magnitude that the PDF activation spectrum produces a small, secondary peak near the \coion velocity. Alt text: two-panel graph showing an emission spectrum and two activation spectra, one in red and one in blue. Text in the upper right of the graph reports the HISA probability as 51 per cent. A vertical red dashed line in the upper panel aligns with an absorption feature in the emission spectrum, and also aligns with the largest peak of the red activation spectrum. The blue activation spectrum shows two strong peaks, one centred at the same location as the large peak in the red absorption feature, and one at the location of a much broader absorption feature in the emission spectrum, where there is a very small peak in the red activation spectrum. A vertical dashed black line is located at the centre of this broad absorption feature, and is slightly offset from the second peak in the blue and red absorption spectra.}
    \label{fig:gmf20 co offset spectrum}
\end{figure}

Figure~\ref{fig:gmf20 co offset spectrum} provides a specific example from the GMF20 field at $(l=20\fdg6$, $b=-1\degr)$ that illustrates both the challenge of velocity localisation in spectrally ambiguous sightlines and the issues with using molecular tracers to define HISA searches, as well as the diagnostic value of the activation spectrum in such cases. The spectrum is assigned a HISA probability of $\mathbb{P}({\rm HISA}) = 0.51$, marginally above the default 0.5 detection threshold but below the stricter 1.0 detection threshold adopted for all GMF analyses in this work; it would therefore not appear as a detection in the maps of Section~\ref{sec:gmf detection} and is shown here as an illustration of the network's behaviour where HISA is detected at a large offset with \coion. The predicted HISA velocity is offset from the primary \coion component by approximately $40~{\rm km~s}^{-1}$, placing this sightline firmly in the extended tail of large velocity offsets in Figure~\ref{fig:gmf velocities}. Despite this, the activation spectrum is bimodal: a dominant peak drives the velocity estimate to the predicted HISA velocity, while a secondary peak is present approximately coincident with the \coion velocity, indicating that the network responds to absorption-like morphology at the molecular velocity even while ultimately assigning dominant weight to a different spectral feature. This bimodal structure suggests the sightline contains two spectral features of comparable prominence -- one at the predicted HISA velocity and one near the molecular velocity -- and that the single-velocity output of the current localisation method cannot represent both simultaneously. The primary activation peak appears to correspond to a genuine absorption feature at a velocity unrelated to the GMF, while the secondary peak corresponds to the steep velocity gradient at the edge of the broad feature centred on the \coion velocity. 

Visually, it could be argued that the detected feature more closely resembles real self-absorption in its sharp gradients and narrower widths than the feature at the \coion velocity, and this example demonstrates the potential issues with tracer-based methods. This example also motivates the development of a multi-velocity localisation output of the neural network capable of separately reporting all prominent activation peaks within a spectrum.

\begin{figure}
    \centering
    \includegraphics[width=\linewidth]{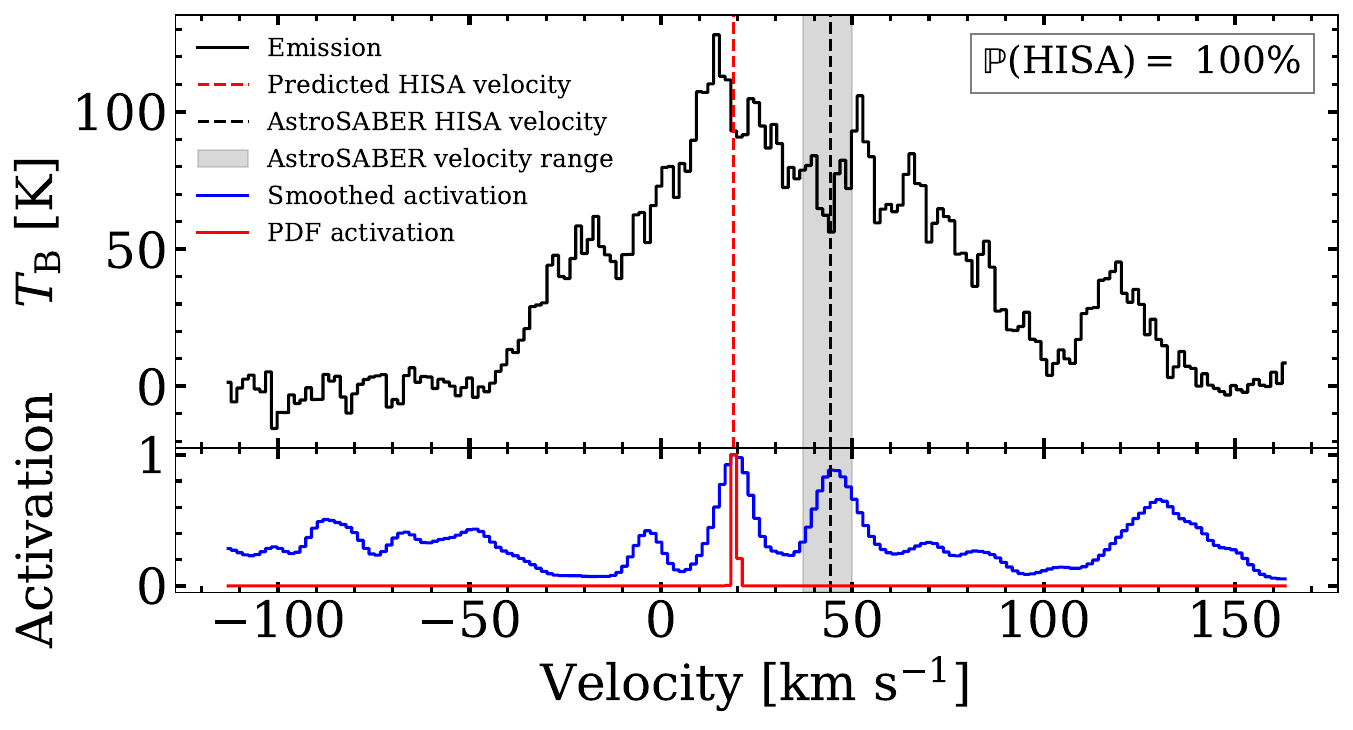}
    \caption{An emission spectrum from the GMF20 field towards $l= 20\fdg1$, $b=-0\fdg4$, demonstrating a case where the predicted HISA velocity and previous HISA localisations from \citet{syed2023} differ. The red dashed line shows the predicted HISA feature velocity from the neural network, while the black dashed line shows the HISA feature velocity from the \textsc{astroSaber} algorithm. The grey band represents the velocity range of the GMF, used by \citet{syed2023} to restrict HISA detections. The smoothed activation spectrum (blue line) shows a significant peak at the HISA feature detected by the neural network, as well as a secondary activation peak at $\sim45~{\rm km~s^{-1}}$ that corresponds exactly to the \textsc{astroSaber} detection. Despite multiple activation peaks in the smoothed activation spectrum, the softmax operation converts the activation into a probability density function with only one peak (red line). Alt text: two-panel graph showing an emission spectrum and two activation spectra, one in red and one in blue. Text in the upper right of the graph reports the HISA probability as 100 per cent. A vertical red dashed line in the upper panel aligns with a shallow absorption feature in the emission spectrum, and also aligns with the largest peak of the red and blue activation spectra. The blue activation spectrum shows two more strong peaks, one centred at the location of another absorption feature in the emission spectrum, and one in the noisy channels outside of the emission. A vertical grey band and vertical black dashed line appear at the same location as the second absorption feature, and line up with the second largest peak in the blue activation spectrum.}
    \label{fig:gmf20 astrosaber offset spectrum}
\end{figure}

Figure~\ref{fig:gmf20 astrosaber offset spectrum} highlights a sightline in the GMF20 field where the CNN identifies a HISA feature outside the velocity range used in \citet{syed2023} to constrain detections using the \textsc{astroSaber} algorithm. While the neural network detects a HISA feature (shown in red) beyond the CO velocity range of the GMF (grey band), it also assigns secondary activation to a feature at $\sim45~{\rm km~s}^{-1}$ that corresponds to the feature detected by the \textsc{astroSaber} algorithm (shown by the dashed black line). As we have chosen to limit the velocity inference to only a single velocity, it cannot represent both detections simultaneously. In its current state, the network can be applied to sightlines where there are no molecular tracers, but it is unable to distinguish between competing features in any way, whether both features are truly self-absorption or not. Future iterations incorporating multi-modal activation maps or ranked candidate lists would significantly enhance the neural network's utility in high-density regions.

\subsubsection{Performance on giant molecular filament regions}\label{sec:gmf performance}

The GMF regions present the most challenging test for the neural network. The sightlines towards the inner Galactic Plane are characterised by multiple overlapping \ion{H}{i} emission components with a wide range of velocities, and multiple genuine HISA features are often superimposed on complex, structured backgrounds. Despite these challenges, the CNN successfully recovers HISA detections that are spatially correlated with the molecular filaments. In the filaments themselves, the predicted velocities remain consistent with the \coion-traced molecular velocities. While the velocity scatter is larger in the GMF comparison (interquartile range of $12.60~{\rm km~s}^{-1}$) than in the \rc comparison (interquartile range of $0.82~{\rm km~s}^{-1}$), the distribution of velocity offsets is well-centred, and 24.6 per cent of velocity offsets are sub-channel ($<1.5~{\rm km~s}^{-1}$). However, the bulk of the offsets extend to $\sim\pm50~{\rm km~s}^{-1}$, far outside the typical offsets that would be expected from the typical offset between \ion{H}{i} and \coion, which is on the order of $\sim5~{\rm km~s}^{-1}$ for the CNM \citep{park2023}.

Despite the position of the \rc cloud near the Galactic Centre, the projected size of the cloud and the strength of its self-absorption mean that in reality, the data are far less spectrally complex than the THOR+VGPS data. The majority of the \rc cloud is located at latitudes $b\gtrsim1\degr$, away from the Galactic Centre, meaning that the background emission becomes much simpler and the velocity range of the background material becomes narrower. This velocity crowding, along with the prominence of the \rc cloud's HISA features, results in a smaller range over which the neural network needs to look for features, and they are typically easier to find.

The GMF fields sit almost entirely within $|b|\le1\degr$, and the positions of the filaments do not necessarily produce self-absorption features that are as strong as those from the \rc cloud. This increased spectral complexity of the THOR survey data influence the CNN's behaviour in two primary ways. First, classification probabilities are less distinctly separated from the 0.5 threshold. Specifically, the proportion of marginal detections (probabilities between 0.1 and 0.9) rises from 3.0 per cent in the \rc cloud to 20.3 per cent in the GMF fields. This shift demonstrates the inherent ambiguity of the inner plane, as well as the effect of increased noise in the THOR+VGPS data, and explains why GMF results are more dependent on the chosen classification threshold. Second, since the current activation mapping method returns only a single predicted velocity per spectrum, it cannot resolve multiple simultaneous HISA features. In such cases, the neural network localises only the most prominent feature, leaving secondary components unreported.

Across the five GMF fields, 70.6 per cent of CNN detections yield predicted HISA velocities lying outside the GMF velocity window for the filament as defined in table 2 of \citet{ragan2014}. While the spectral complexity of the inner plane likely introduces some false positives, the spatial coherence of a significant subset of these detections, particularly those that align with \coion emission at the same velocities (outside of that defined as the GMF), suggests they represent genuine cold atomic gas existing at velocities outside those of the molecular filament. These results indicate that the CNN can identify HISA features that existing techniques might overlook. Confirming this interpretation will require targeted follow-up, such as absorption observations towards continuum background sources (if present) at the corresponding positions, which we identify as a priority for future work.

\section{Discussion}\label{sec:discussion}

In this section we discuss the computational efficiency of the neural network, its limitations, and its prospects for application to next-generation surveys.

\subsection{Computational efficiency}\label{sec:computational speed}

A key practical advantage of the CNN approach over iterative spectral methods is its computational efficiency, which arises from the ability to evaluate the network on large batches of spectra in parallel using GPU acceleration. While the neural network was trained in 4 hours on an HPC node using a single V100 NVIDIA GPU, all further data generation and analysis was done on a 2023 MacBook Pro. For the Riegel--Crutcher datacube, the network processes 1.092 million spectra in 70 seconds, corresponding to $\sim15,600$ spectra per second. For the five GMF cubes ($\sim1.8$ million spectra combined), the processing takes 122 seconds at $\sim15,000$ spectra per second, with the marginally lower speed attributed to input/output overhead from reading five separate FITS files. The activation mapping and velocity localisation require a further 81 seconds for the \rc cloud and 174 seconds for the GMF regions, giving end-to-end (detection plus localisation) times of $\sim450$~s for $\sim2.9$ million spectra, corresponding to a rate of $\sim6,500$ spectra per second for combined detection and velocity localisation.

These speeds are several orders of magnitude faster than previous iterative methods for detecting self-absorption. The linear interpolation method of \citet{mcclure-griffiths2006} is computationally cheap overall but still requires a per-spectrum fitting step that scales linearly with data volume and does not benefit from GPU parallelisation. While more complex, the \textsc{astroSaber} algorithm \citep{syed2023} also involves iterative baseline construction but, to our knowledge, does not natively support GPU-accelerated batch processing and processes spectra at a speed of $\sim100$ spectra per second. At the data rates expected from next-generation facilities, the CNN's throughput makes real-time HISA detection operationally feasible in a way that CPU-bound iterative methods are not: the Square Kilometre Array (SKA) and its pathfinders, including ASKAP and MeerKAT, will produce \ion{H}{i} datacubes covering hundreds to thousands of square degrees, up to arcsecond resolution for the full SKA, potentially generating billions of spectra per survey. Embedding the CNN in an online data-processing pipeline would allow HISA detections and velocity maps to be produced commensurately with data acquisition, enabling rapid identification of cold-gas structures for follow-up.

While the CNN is extremely fast, it cannot be used for the analysis of self-absorption, and instead can only be used to detect strong self-absorption features. Analysis of any detected HISA must still be done with other tools such as \textsc{GaussPy+}/ROHSA or \textsc{astroSaber}; however, the activation peak from the CNN provides a useful initial velocity estimate to seed or constrain those follow-up methods, and we will explore the possibility of incorporating this information into ROHSA. This two-stage workflow combines the scalability of the CNN with the physical interpretability of parametric decomposition methods.

\subsection{Limitations}\label{sec:limitations}

The design of the present HISA detector reflects a deliberate preference for precision over completeness: the network is intended to deliver high-confidence detections of prominent HISA features, identifying sightlines that warrant follow-up spectral decomposition, rather than to provide a complete census of HISA. The current implementation of the HISA detector is subject to several constraints stemming from the design of the synthetic training dataset and the specific architecture of the neural network and, as discussed below, these constraints are largely consequences of that design choice. The primary limitation of the synthetic training dataset is the single-sightline, single-HISA-cloud model. Each training spectrum contains at most one cold foreground cloud, placed at the front of the column to ensure that the full warm emission background is available for absorption and to avoid the label ambiguity that would arise if a partially absorbed or sub-threshold cloud produced a feature indistinguishable from noise. However, lines of sight through the Galactic Plane frequently cross multiple cold structures, such as spiral arms and inter-arm features. Consequently, real-world spectra can contain multiple cold clouds at different velocities, exhibiting superimposed HISA features that are qualitatively different from the single-feature spectra on which the network was trained. The consequences of this mismatch are asymmetric for detection and velocity localisation and are worth considering separately.

For detection, the single-cloud limitation is expected to have a modest effect on recall. If any one of the multiple HISA features present in a real sightline is strong enough to dominate the spectral morphology, the network will likely assign a high HISA probability. This is because the learned convolutional filters respond to the characteristic absorption morphology regardless of whether additional weaker features are present elsewhere in the spectrum. However, detection probability will be reduced for sightlines in which no individual feature exceeds the network’s effective 8~K completeness floor, yet the combined optical depth of several weaker cold clouds collectively distorts the spectral baseline in a manner that a single-feature classifier does not recognize. In this sense the 8~K threshold, which was set to guarantee that every training label is unambiguous, simultaneously defines the lower sensitivity of the detector and biases the detected sample towards the coldest, densest self-absorption. Such cases represent a source of incompleteness intrinsic to the present single-cloud, thresholded training design that cannot be closed without extending the training set to multi-cloud spectra containing sub-threshold features. Conversely, the false positive rate is not similarly worsened; the dominant driver of false positives is inner-plane spectral complexity rather than the single-cloud assumption itself, which is consistent with the precision-over-completeness orientation of the training set.

For velocity localisation, the multi-cloud limitation is considerably more significant. When multiple HISA features are present at different velocities, the activation map will in general contain multiple peaks. The bimodal activation spectra visible in Figures~\ref{fig:gmf20 co offset spectrum} and \ref{fig:gmf20 astrosaber offset spectrum} are direct observational evidence of this effect: in both cases the network responds to two spectral features simultaneously yet can only return a single velocity, and the resulting estimate is driven to the feature that happens to produce the stronger activation rather than necessarily to the feature most closely associated with other tracers. The extended tail of large velocity offsets in Figure~\ref{fig:gmf velocities} is at least partly a consequence of this behaviour in the GMF20 field. Extending the synthetic dataset to allow multiple HISA clouds and modifying the post-hoc activation mapping of the CNN to return multiple candidate velocities per sightline, either through a direct regression head or a ranked list of candidate velocities, would better handle multi-component sightlines and improve velocity localisation. This would allow the neural network to report the full population of cold clouds along each line of sight rather than only the most prominent and would substantially increase the scientific yield of the method when applied to complex inner Galactic Plane data.

These limitations combine to define the scientific content of the detected sample. The network will preferentially identify HISA arising from a single, cold, dense cloud in front of a bright thermal background, and will systematically under-detect weaker, multi-component, or spatially extended absorption, including features in low-S/N regions of the sky and in environments in which the CNM is diffuse or clumpy rather than concentrated in a single cloud. Any subsequent analysis of the detected sample such as deriving column densities, inferring CNM fractions, or comparing with emission tracers should therefore treat the detected HISA as a lower limit on the true HISA population, biased towards the highest-column and most spectrally prominent features, rather than as a complete inventory. A future HISA catalogue, assembled either from a large observational sample or from a more elaborate simulation suite that includes multi-component and spatially resolved absorption, would relax these constraints and permit a completeness-optimised detector.

Beyond cloud count, the lack of spatial coherence in the training dataset remains a constraint. Because synthetic spectra are generated independently, the network cannot exploit the three-dimensional structure of real H I clouds. Including spatial information has been shown to improve accuracy in Gaussian decomposition methods \citep[e.g. ROHSA;][]{marchal2019a}, and spectral-line morphology alone can be used to infer the fraction of cold gas along a line of sight \citep{lei2025}. Domain shift remains a further consideration, particularly for interferometric data like THOR, where correlated noise and sidelobes from bright continuum sources may degrade performance relative to synthetic benchmarks. Extending the synthetic data generation and network architecture to the three-dimensional PPV domain, and incorporating spatially coherent emission structure, is currently being explored by the authors and is expected to improve both the sensitivity and the velocity resolution of the HISA detector.

A further limitation, arising from the architecture rather than the training data, is that the current network is restricted to a fixed spectral axis length of 185 channels. This resolution constraint arises because the convolutional kernels are tuned to specific velocity widths, while the fixed channel count is a requirement of the fully connected output layers. While replacing these with a global average pooling (GAP) operation would allow for arbitrary channel counts, our early experiments indicate this may degrade activation mapping quality, an issue the authors are currently investigating.

\subsection{Future work and applications to large-scale surveys}\label{sec:future work}

A particularly compelling near-term application is the GASKAP-HI survey \citep{dickey2013, pingel2022}, which will provide ASKAP \ion{H}{i} emission observations at $\sim30$ arcsec resolution covering the entire southern Galactic Plane and the Magellanic System. Applied to GASKAP-HI data, the CNN could produce the most spatially complete map of cold \ion{H}{i} in the southern Milky Way to date, resolving cold-gas filaments and connecting them to the molecular structures observed in follow-up CO surveys or other molecular tracers. In addition, the Magellanic Clouds present a particularly interesting extension: at low metallicity and in a lower-pressure interstellar medium, the conditions for CNM formation and self-absorption differ from those in the inner Milky Way \citep{dempsey2022, dickey2022, nguyen2024, chen2025, park2026}. Observations of the Magellanic system from the GASKAP-HI survey \citep{dickey2013, pingel2022} are available, and we plan to apply our neural network to these datasets in the future. The spatial resolution of these data at the distance of the Magellanic system is less than 10 pc, which will effectively condense structures such as the \rc cloud and the GMFs used in this work into single pixels. While this will likely prevent such detailed spatial structures from being recovered, we expect it will still allow for detections of HISA across the Magellanic system in general.

Whether the synthetic training set, calibrated to Milky Way conditions, generalises to the Magellanic Clouds without retraining is an open question that we intend to address in future work; it may require a new synthetic dataset generated following the properties of the Magellanic Clouds, particularly the velocity distribution and typical FWHM of CNM and WNM clouds. Finally, there is also the potential to fine-tune or retrain the network on a small labelled set of real observations, which was omitted in this work due to the difficulty of constructing this dataset, and that could improve performance in complex regions such as the inner Galactic Plane without requiring a full 3-D simulation framework.

\section{Conclusions}\label{sec:conclusion}

We have presented a lightweight convolutional neural network for the automated detection and velocity localisation of \ion{H}{i} self-absorption in 21-cm emission spectra. After training and validating the architecture on synthetic data, we applied the neural network to two real emission datasets. Our neural network takes a single emission spectrum as input and returns a binary classification of whether a HISA feature is present and a velocity estimate for the single strongest detected feature derived via gradient-based saliency mapping. The latter provides a physically interpretable indication of the spectral channels driving each detection, linking the network output to the actual absorption signal rather than tying it to ancillary data. As such, our approach requires no ancillary data, makes no assumptions about the form of the background emission, and operates at $\sim6,500$ spectra per second for combined detection and velocity localisation, making it viable for deployment at the data rates expected from the SKA and its pathfinders. We have demonstrated that the network successfully identifies known HISA features in challenging observed fields towards the Galactic Plane (the Riegel--Crutcher cloud and several giant molecular filaments) where complex gas morphologies and long path lengths through the Galaxy produce heavily blended emission. The principal results are as follows.
\begin{enumerate}
    \item We have designed and trained a lightweight convolutional neural network for detecting \ion{H}{i} self-absorption that achieves 96.5 per cent accuracy, 96.0 per cent precision, a false positive rate of 4.1 per cent, and a false negative rate of 3.0 per cent on the held-out synthetic test dataset.
    \item We have developed a gradient-based activation mapping method, applied post-hoc to the network's internal representations, that localises detected HISA features in velocity without any modification to the network architecture or training procedure. On synthetic data, 66.3 per cent of velocity predictions are within a single channel ($1.5~{\rm km~s}^{-1}$) of the true HISA cloud velocity, with a median offset of $0.42~{\rm km~s}^{-1}$ and an interquartile range of $1.97~{\rm km~s}^{-1}$.
    \item Applied to the Riegel--Crutcher cloud (SGPS GC data), the neural network recovers the known spatial extent and filamentary structure of the cloud, and the predicted HISA velocities agree with those from linear interpolation at the 88.9 per cent within-one-channel level (median offset $0~{\rm km~s}^{-1}$, IQR $= 0.82~{\rm km~s}^{-1}$). Spatially coherent velocity structure is recovered despite the network processing spectra entirely independently.
    \item Applied to five giant molecular filament regions in the Galactic Plane (THOR+VGPS data), the network recovers HISA detections at an elevated rate around the filaments traced by \coion, with predicted HISA velocities consistent with the molecular velocities (median offset $0.89~{\rm km~s}^{-1}$) implying they are co-spatial. At a detection threshold of 1.0, 48.6 per cent of GMF sightlines are classified as containing HISA, which is comparable to the 35.9 per cent fraction recovered by \textsc{astroSaber} \citep{syed2023} over the same fields when using molecular tracers to restrict detections. The GMF results demonstrate that cold atomic hydrogen co-spatial with molecular gas can be recovered without any CO information in the detection or localisation, in contrast to methods which use ancillary molecular data to localise features or constrain the velocity search window.
    \item The network processes $\sim6,500$ spectra per second on a single consumer laptop GPU ($\sim15,000$ spectra per second for detections only), enabling real-time HISA detection at the data rates of current and forthcoming large-area \ion{H}{i} surveys.
\end{enumerate}

The method demonstrates that a lightweight CNN trained on synthetic data can generalise to real Galactic \ion{H}{i} observations, providing both binary detection and sub-channel velocity localisation without assumptions about the background emission or reliance on ancillary tracers. Future work will focus on extending the architecture to a 3-D PPV domain to incorporate spatial and spectral coherence, replacing dense output layers with global average pooling to remove the fixed-velocity-axis constraint, and applying the method to GASKAP data of the Galactic Plane and Magellanic System.

\section*{Acknowledgements}

This research was undertaken with the assistance of resources from the National Computational Infrastructure (NCI Australia), an NCRIS enabled capability supported by the Australian Government under the ANU Merit Allocation Scheme (project xt61) and the Astronomy Supercomputer Time Allocation Committee (project fd08).  This research was partially funded by the Australian Government through an Australian Research Council Australian Laureate Fellowship (project number FL210100039) to NMc-G. FBW was supported by the European Research Council (ERC) under the European community's Seventh framework Programme, through the Advanced Grant MIST No. 742819. The authors acknowledge the Interstellar Institute's programs "II7" \& "II8" and the Paris-Saclay University's Institut Pascal for hosting discussions that nourished the development of the ideas behind this work. The authors thank the anonymous referee for their constructive suggestions and comments, which helped improve the content of this work.

\textit{Software}: Astropy \citep{Astropy2018A}, Matplotlib \citep{MatplotlibHunter2007}, NumPy \citep{vanderWalt2011}, PyTorch \citep{Pytorch2019}, SciPy \citep{Virtanen2020}.

%%%%%%%%%%%%%%%%%%%%%%%%%%%%%%%%%%%%%%%%%%%%%%%%%%

\section*{Data Availability}
The neural network design, datasets used in this study, as well as data analysis notebooks, are publicly available at doi: \href{https://doi.org/10.5281/zenodo.20827350}{10.5281/zenodo.20827350} and on \href{https://github.com/EricMullerAU/Detecting-HI-Self-Absorption-using-Neural-Networks}{GitHub}. The trained neural network weights are not included but will be shared on reasonable request to the corresponding author.



\bibliographystyle{mnras}
\bibliography{references} % if your bibtex file is called example.bib

@ARTICLE{Astropy2018A,
       author = {{Astropy Collaboration} and {Price-Whelan}, A.~M. and {Sip{\H{o}}cz}, B.~M. and {G{\"u}nther}, H.~M. and {Lim}, P.~L. and {Crawford}, S.~M. and {Conseil}, S. and {Shupe}, D.~L. and {Craig}, M.~W. and {Dencheva}, N. and {Ginsburg}, A. and {VanderPlas}, J.~T. and {Bradley}, L.~D. and {P{\'e}rez-Su{\'a}rez}, D. and {de Val-Borro}, M. and {Aldcroft}, T.~L. and {Cruz}, K.~L. and {Robitaille}, T.~P. and {Tollerud}, E.~J. and {Ardelean}, C. and {Babej}, T. and {Bach}, Y.~P. and {Bachetti}, M. and {Bakanov}, A.~V. and {Bamford}, S.~P. and {Barentsen}, G. and {Barmby}, P. and {Baumbach}, A. and {Berry}, K.~L. and {Biscani}, F. and {Boquien}, M. and {Bostroem}, K.~A. and {Bouma}, L.~G. and {Brammer}, G.~B. and {Bray}, E.~M. and {Breytenbach}, H. and {Buddelmeijer}, H. and {Burke}, D.~J. and {Calderone}, G. and {Cano Rodr{\'\i}guez}, J.~L. and {Cara}, M. and {Cardoso}, J.~V.~M. and {Cheedella}, S. and {Copin}, Y. and {Corrales}, L. and {Crichton}, D. and {D'Avella}, D. and {Deil}, C. and {Depagne}, {\'E}. and {Dietrich}, J.~P. and {Donath}, A. and {Droettboom}, M. and {Earl}, N. and {Erben}, T. and {Fabbro}, S. and {Ferreira}, L.~A. and {Finethy}, T. and {Fox}, R.~T. and {Garrison}, L.~H. and {Gibbons}, S.~L.~J. and {Goldstein}, D.~A. and {Gommers}, R. and {Greco}, J.~P. and {Greenfield}, P. and {Groener}, A.~M. and {Grollier}, F. and {Hagen}, A. and {Hirst}, P. and {Homeier}, D. and {Horton}, A.~J. and {Hosseinzadeh}, G. and {Hu}, L. and {Hunkeler}, J.~S. and {Ivezi{\'c}}, {\v{Z}}. and {Jain}, A. and {Jenness}, T. and {Kanarek}, G. and {Kendrew}, S. and {Kern}, N.~S. and {Kerzendorf}, W.~E. and {Khvalko}, A. and {King}, J. and {Kirkby}, D. and {Kulkarni}, A.~M. and {Kumar}, A. and {Lee}, A. and {Lenz}, D. and {Littlefair}, S.~P. and {Ma}, Z. and {Macleod}, D.~M. and {Mastropietro}, M. and {McCully}, C. and {Montagnac}, S. and {Morris}, B.~M. and {Mueller}, M. and {Mumford}, S.~J. and {Muna}, D. and {Murphy}, N.~A. and {Nelson}, S. and {Nguyen}, G.~H. and {Ninan}, J.~P. and {N{\"o}the}, M. and {Ogaz}, S. and {Oh}, S. and {Parejko}, J.~K. and {Parley}, N. and {Pascual}, S. and {Patil}, R. and {Patil}, A.~A. and {Plunkett}, A.~L. and {Prochaska}, J.~X. and {Rastogi}, T. and {Reddy Janga}, V. and {Sabater}, J. and {Sakurikar}, P. and {Seifert}, M. and {Sherbert}, L.~E. and {Sherwood-Taylor}, H. and {Shih}, A.~Y. and {Sick}, J. and {Silbiger}, M.~T. and {Singanamalla}, S. and {Singer}, L.~P. and {Sladen}, P.~H. and {Sooley}, K.~A. and {Sornarajah}, S. and {Streicher}, O. and {Teuben}, P. and {Thomas}, S.~W. and {Tremblay}, G.~R. and {Turner}, J.~E.~H. and {Terr{\'o}n}, V. and {van Kerkwijk}, M.~H. and {de la Vega}, A. and {Watkins}, L.~L. and {Weaver}, B.~A. and {Whitmore}, J.~B. and {Woillez}, J. and {Zabalza}, V. and {Astropy Contributors}},
        title = "{The Astropy Project: Building an Open-science Project and Status of the v2.0 Core Package}",
      journal = {\aj},
         year = 2018,
        month = sep,
       volume = {156},
       number = {3},
          eid = {123},
        pages = {123},
          doi = {10.3847/1538-3881/aabc4f},
archivePrefix = {arXiv},
       eprint = {1801.02634},
 primaryClass = {astro-ph.IM},
       adsurl = {https://ui.adsabs.harvard.edu/abs/2018AJ....156..123A}
}

@ARTICLE{Pytorch2019,
       author = {{Paszke}, Adam and {Gross}, Sam and {Massa}, Francisco and {Lerer}, Adam and {Bradbury}, James and {Chanan}, Gregory and {Killeen}, Trevor and {Lin}, Zeming and {Gimelshein}, Natalia and {Antiga}, Luca and {Desmaison}, Alban and {K{\"o}pf}, Andreas and {Yang}, Edward and {DeVito}, Zach and {Raison}, Martin and {Tejani}, Alykhan and {Chilamkurthy}, Sasank and {Steiner}, Benoit and {Fang}, Lu and {Bai}, Junjie and {Chintala}, Soumith},
        title = "{PyTorch: An Imperative Style, High-Performance Deep Learning Library}",
      journal = {arXiv e-prints},
         year = 2019,
        month = dec,
          eid = {arXiv:1912.01703},
        pages = {arXiv:1912.01703},
          doi = {10.48550/arXiv.1912.01703},
archivePrefix = {arXiv},
       eprint = {1912.01703},
 primaryClass = {cs.LG},
       adsurl = {https://ui.adsabs.harvard.edu/abs/2019arXiv191201703P}
}

@Article{MatplotlibHunter2007,
  Author    = {Hunter, J. D.},
  Title     = {Matplotlib: A 2D graphics environment},
  Journal   = {Computing in Science \& Engineering},
  Volume    = {9},
  Number    = {3},
  Pages     = {90--95},
  publisher = {IEEE COMPUTER SOC},
  doi       = {10.1109/MCSE.2007.55},
  year      = 2007
}

@ARTICLE{vanderWalt2011,
       author = {{van der Walt}, St{\'e}fan and {Colbert}, S. Chris and {Varoquaux}, Ga{\"e}l},
        title = "{The NumPy Array: A Structure for Efficient Numerical Computation}",
      journal = {Computing in Science and Engineering},
         year = 2011,
        month = mar,
       volume = {13},
       number = {2},
        pages = {22-30},
          doi = {10.1109/MCSE.2011.37},
archivePrefix = {arXiv},
       eprint = {1102.1523},
 primaryClass = {cs.MS},
       adsurl = {https://ui.adsabs.harvard.edu/abs/2011CSE....13b..22V}
}

@ARTICLE{Virtanen2020,
       author = {{Virtanen}, Pauli and {Gommers}, Ralf and {Oliphant}, Travis E. and {Haberland}, Matt and {Reddy}, Tyler and {Cournapeau}, David and {Burovski}, Evgeni and {Peterson}, Pearu and {Weckesser}, Warren and {Bright}, Jonathan and {van der Walt}, St{\'e}fan J. and {Brett}, Matthew and {Wilson}, Joshua and {Millman}, K. Jarrod and {Mayorov}, Nikolay and {Nelson}, Andrew R.~J. and {Jones}, Eric and {Kern}, Robert and {Larson}, Eric and {Carey}, C.~J. and {Polat}, {\.I}lhan and {Feng}, Yu and {Moore}, Eric W. and {VanderPlas}, Jake and {Laxalde}, Denis and {Perktold}, Josef and {Cimrman}, Robert and {Henriksen}, Ian and {Quintero}, E.~A. and {Harris}, Charles R. and {Archibald}, Anne M. and {Ribeiro}, Ant{\^o}nio H. and {Pedregosa}, Fabian and {van Mulbregt}, Paul and {SciPy 1. 0 Contributors}},
        title = "{SciPy 1.0: fundamental algorithms for scientific computing in Python}",
      journal = {Nature Methods},
         year = 2020,
        month = feb,
       volume = {17},
        pages = {261-272},
          doi = {10.1038/s41592-019-0686-2},
archivePrefix = {arXiv},
       eprint = {1907.10121},
 primaryClass = {cs.MS},
       adsurl = {https://ui.adsabs.harvard.edu/abs/2020NatMe..17..261V}
}

@ARTICLE{dickey1990,
       author = {{Dickey}, John M. and {Lockman}, Felix J.},
        title = "{H I in the galaxy.}",
      journal = {\araa},
         year = 1990,
        month = jan,
       volume = {28},
        pages = {215-261},
          doi = {10.1146/annurev.aa.28.090190.001243},
       adsurl = {https://ui.adsabs.harvard.edu/abs/1990ARA&A..28..215D}
}

@article{mcclure-griffiths2023,
    title = {Atomic {Hydrogen} in the {Milky} {Way}: {A} {Stepping} {Stone} in the {Evolution} of {Galaxies}},
    volume = {61},
    copyright = {http://creativecommons.org/licenses/by/4.0/},
    issn = {0066-4146, 1545-4282},
    shorttitle = {Atomic {Hydrogen} in the {Milky} {Way}},
    url = {https://www.annualreviews.org/content/journals/10.1146/annurev-astro-052920-104851},
    doi = {10.1146/annurev-astro-052920-104851},
    language = {en},
    number = {1},
    urldate = {2025-04-10},
    journal = {Annual Review of Astronomy and Astrophysics},
    author = {McClure-Griffiths, Naomi M. and Stanimirović, Snežana and Rybarczyk, Daniel R.},
    month = aug,
    year = {2023},
    note = {Read\_Status: New
Read\_Status\_Date: 2025-06-06T00:24:23.924Z},
    pages = {19--63},
}

@article{denes2018,
    title = {Calibrating the {HISA} temperature: {Measuring} the temperature of the {Riegel}-{Crutcher} cloud},
    volume = {479},
    issn = {0035-8711},
    shorttitle = {Calibrating the {HISA} temperature},
    url = {https://ui.adsabs.harvard.edu/abs/2018MNRAS.479.1465D},
    doi = {10.1093/mnras/sty1384},
    urldate = {2025-08-13},
    journal = {Monthly Notices of the Royal Astronomical Society},
    author = {Dénes, H. and McClure-Griffiths, N. M. and Dickey, J. M. and Dawson, J. R. and Murray, C. E.},
    month = sep,
    year = {2018},
    note = {Publisher: OUP
ADS Bibcode: 2018MNRAS.479.1465D
Read\_Status: New
Read\_Status\_Date: 2025-08-13T06:17:34.208Z},
    pages = {1465--1490},
}

@article{mcclure-griffiths2006,
    title = {Magnetically {Dominated} {Strands} of {Cold} {Hydrogen} in the {Riegel}-{Crutcher} {Cloud}},
    volume = {652},
    issn = {0004-637X},
    url = {https://ui.adsabs.harvard.edu/abs/2006ApJ...652.1339M},
    doi = {10.1086/508706},
    urldate = {2025-08-12},
    journal = {The Astrophysical Journal},
    author = {McClure-Griffiths, N. M. and Dickey, J. M. and Gaensler, B. M. and Green, A. J. and Haverkorn, Marijke},
    month = dec,
    year = {2006},
    note = {Publisher: IOP
ADS Bibcode: 2006ApJ...652.1339M
Read\_Status: New
Read\_Status\_Date: 2025-08-12T23:38:38.310Z},
    pages = {1339--1347},
}

@article{gibson2005,
    title = {An {Automated} {Method} for the {Detection} and {Extraction} of {HI} {Self}-{Absorption} in {High}-{Resolution} 21cm {Line} {Surveys}},
    volume = {626},
    issn = {0004-637X, 1538-4357},
    url = {http://arxiv.org/abs/astro-ph/0503119},
    doi = {10.1086/429871},
    number = {1},
    urldate = {2025-08-12},
    journal = {The Astrophysical Journal},
    author = {Gibson, Steven J. and Taylor, A. Russell and Higgs, Lloyd A. and Brunt, Christopher M. and Dewdney, Peter E.},
    month = jun,
    year = {2005},
    note = {arXiv:astro-ph/0503119
Read\_Status: New
Read\_Status\_Date: 2025-08-12T23:38:00.784Z},
    pages = {214--232},
}

@article{syed2023,
    title = {Cold atomic gas identified by {HI} self-absorption. {Cold} atomic clouds toward giant molecular filaments},
    volume = {679},
    issn = {0004-6361, 1432-0746},
    url = {http://arxiv.org/abs/2310.02077},
    doi = {10.1051/0004-6361/202346562},
    urldate = {2025-10-02},
    journal = {Astronomy \& Astrophysics},
    author = {Syed, J. and Beuther, H. and Goldsmith, P. F. and Henning, Th and Heyer, M. and Klessen, R. S. and Stil, J. M. and Soler, J. D. and Anderson, L. D. and Urquhart, J. S. and Rugel, M. R. and Johnston, K. G. and Brunthaler, A.},
    month = nov,
    year = {2023},
    note = {arXiv:2310.02077 [astro-ph]
Read\_Status: New
Read\_Status\_Date: 2025-10-02T00:11:04.162Z},
    pages = {A130},
}

@article{mcclure-griffiths2005,
    title = {The {Southern} {Galactic} {Plane} {Survey}: {H} {I} {Observations} and {Analysis}},
    volume = {158},
    issn = {0067-0049},
    shorttitle = {The {Southern} {Galactic} {Plane} {Survey}},
    url = {https://iopscience.iop.org/article/10.1086/430114},
    doi = {10.1086/430114},
    language = {en},
    number = {2},
    urldate = {2025-12-15},
    journal = {The Astrophysical Journal Supplement Series},
    author = {McClure-Griffiths, N. M. and Dickey, John M. and Gaensler, B. M. and Green, A. J. and Haverkorn, Marijke and Strasser, S.},
    month = jun,
    year = {2005},
    note = {Publisher: IOP Publishing
Read\_Status: New
Read\_Status\_Date: 2025-12-15T22:00:58.102Z},
    pages = {178},
}

@article{mcclure-griffiths2012,
    title = {{THE} {AUSTRALIA} {TELESCOPE} {COMPACT} {ARRAY} {H} i {SURVEY} {OF} {THE} {GALACTIC} {CENTER}},
    volume = {199},
    issn = {0067-0049},
    url = {https://doi.org/10.1088/0067-0049/199/1/12},
    doi = {10.1088/0067-0049/199/1/12},
    language = {en},
    number = {1},
    urldate = {2025-12-15},
    journal = {The Astrophysical Journal Supplement Series},
    author = {McClure-Griffiths, N. M. and Dickey, J. M. and Gaensler, B. M. and Green, A. J. and Green, J. A. and Haverkorn, M.},
    month = feb,
    year = {2012},
    note = {Publisher: The American Astronomical Society
Read\_Status: New
Read\_Status\_Date: 2025-12-15T22:01:08.976Z},
    pages = {12},
}

@article{clark2014,
    title = {{MAGNETICALLY} {ALIGNED} {H} i {FIBERS} {AND} {THE} {ROLLING} {HOUGH} {TRANSFORM}},
    volume = {789},
    issn = {0004-637X},
    url = {https://doi.org/10.1088/0004-637X/789/1/82},
    doi = {10.1088/0004-637X/789/1/82},
    language = {en},
    number = {1},
    urldate = {2025-12-15},
    journal = {The Astrophysical Journal},
    author = {Clark, S. E. and Peek, J. E. G. and Putman, M. E.},
    month = jun,
    year = {2014},
    note = {Publisher: The American Astronomical Society
Read\_Status: New
Read\_Status\_Date: 2025-12-15T22:19:24.608Z},
    pages = {82},
}

@article{crutcher1973,
    title = {Observations of {OH} {Molecules} in {Interstellar} {Dust} {Clouds}},
    volume = {185},
    issn = {0004-637X, 1538-4357},
    url = {http://adsabs.harvard.edu/doi/10.1086/152460},
    doi = {10.1086/152460},
    language = {en},
    urldate = {2025-12-15},
    journal = {The Astrophysical Journal},
    author = {Crutcher, Richard M.},
    month = nov,
    year = {1973},
    note = {Read\_Status: New
Read\_Status\_Date: 2025-12-15T22:22:26.531Z},
    pages = {857},
}

@article{dame2001,
    title = {The {Milky} {Way} in {Molecular} {Clouds}: {A} {New} {Complete} {COSurvey}},
    volume = {547},
    issn = {0004-637X},
    shorttitle = {The {Milky} {Way} in {Molecular} {Clouds}},
    url = {https://iopscience.iop.org/article/10.1086/318388},
    doi = {10.1086/318388},
    language = {en},
    number = {2},
    urldate = {2025-12-15},
    journal = {The Astrophysical Journal},
    author = {Dame, T. M. and Hartmann, Dap and Thaddeus, P.},
    month = feb,
    year = {2001},
    note = {Publisher: IOP Publishing
Read\_Status: New
Read\_Status\_Date: 2025-12-15T22:19:45.628Z},
    pages = {792},
}

@article{riegel1969,
    title = {Observations of an {Unusual} {Cold} {Cloud} in the {Galaxy}},
    volume = {157},
    issn = {0004-637X},
    url = {https://ui.adsabs.harvard.edu/abs/1969ApJ...157..563R},
    doi = {10.1086/150096},
    urldate = {2025-12-15},
    journal = {The Astrophysical Journal},
    author = {Riegel, Kurt W. and Jennings, Mark C.},
    month = aug,
    year = {1969},
    note = {Publisher: IOP
ADS Bibcode: 1969ApJ...157..563R
Read\_Status: New
Read\_Status\_Date: 2025-12-15T22:25:37.572Z},
    pages = {563},
}

@inproceedings{crutcher1984,
    title = {Distances of {Local} {Clouds} from {Optical} {Line} {Observations}},
    volume = {2345},
    url = {https://ui.adsabs.harvard.edu/abs/1984NASCP2345..117C},
    booktitle = {NASA Conference Publication},
    urldate = {2025-12-15},
    author = {Crutcher, R. M. and Lien, D. J.},
    month = nov,
    year = {1984},
    note = {ADS Bibcode: 1984NASCP2345..117C
Read\_Status: New
Read\_Status\_Date: 2025-12-15T22:26:57.793Z},
    pages = {117},
}

@article{heeschen1955,
    title = {Some {Features} of {Interstellar} {Hydrogen} in the {Section} of the {Galactic} {Center}.},
    volume = {121},
    issn = {0004-637X},
    url = {https://ui.adsabs.harvard.edu/abs/1955ApJ...121..569H},
    doi = {10.1086/146023},
    urldate = {2025-12-15},
    journal = {The Astrophysical Journal},
    author = {Heeschen, David S.},
    month = may,
    year = {1955},
    note = {Publisher: IOP
ADS Bibcode: 1955ApJ...121..569H
Read\_Status: New
Read\_Status\_Date: 2025-12-15T22:27:37.447Z},
    pages = {569},
}

@article{ragan2014,
    title = {Giant molecular filaments in the {Milky} {Way}},
    volume = {568},
    copyright = {© ESO, 2014},
    issn = {0004-6361, 1432-0746},
    url = {https://www.aanda.org/articles/aa/abs/2014/08/aa23401-14/aa23401-14.html},
    doi = {10.1051/0004-6361/201423401},
    language = {en},
    urldate = {2025-12-16},
    journal = {Astronomy \& Astrophysics},
    author = {Ragan, S. E. and Henning, Th and Tackenberg, J. and Beuther, H. and Johnston, K. G. and Kainulainen, J. and Linz, H.},
    month = aug,
    year = {2014},
    note = {Publisher: EDP Sciences
Read\_Status: New
Read\_Status\_Date: 2025-12-16T00:59:38.100Z},
    pages = {A73},
}

@article{zucker2018,
    title = {Physical {Properties} of {Large}-scale {Galactic} {Filaments}},
    volume = {864},
    issn = {0004-637X},
    url = {https://doi.org/10.3847/1538-4357/aacc66},
    doi = {10.3847/1538-4357/aacc66},
    language = {en},
    number = {2},
    urldate = {2025-12-16},
    journal = {The Astrophysical Journal},
    author = {Zucker, Catherine and Battersby, Cara and Goodman, Alyssa},
    month = sep,
    year = {2018},
    note = {Publisher: The American Astronomical Society
Read\_Status: New
Read\_Status\_Date: 2025-12-16T01:07:37.825Z},
    pages = {153},
}

@article{beuther2016,
    title = {The {HI}/{OH}/{Recombination} line survey of the inner {Milky} {Way} ({THOR}). {Survey} overview and data release 1},
    volume = {595},
    issn = {0004-6361},
    url = {https://ui.adsabs.harvard.edu/abs/2016A&A...595A..32B},
    doi = {10.1051/0004-6361/201629143},
    urldate = {2025-12-16},
    journal = {Astronomy and Astrophysics},
    author = {Beuther, H. and Bihr, S. and Rugel, M. and Johnston, K. and Wang, Y. and Walter, F. and Brunthaler, A. and Walsh, A. J. and Ott, J. and Stil, J. and Henning, Th. and Schierhuber, T. and Kainulainen, J. and Heyer, M. and Goldsmith, P. F. and Anderson, L. D. and Longmore, S. N. and Klessen, R. S. and Glover, S. C. O. and Urquhart, J. S. and Plume, R. and Ragan, S. E. and Schneider, N. and McClure-Griffiths, N. M. and Menten, K. M. and Smith, R. and Roy, N. and Shanahan, R. and Nguyen-Luong, Q. and Bigiel, F.},
    month = oct,
    year = {2016},
    note = {Publisher: EDP
ADS Bibcode: 2016A\&A...595A..32B
Read\_Status: New
Read\_Status\_Date: 2025-12-16T01:17:52.363Z},
    pages = {A32},
}

@article{stil2006,
    title = {The {VLA} {Galactic} {Plane} {Survey}},
    volume = {132},
    issn = {1538-3881},
    url = {https://iopscience.iop.org/article/10.1086/505940},
    doi = {10.1086/505940},
    language = {en},
    number = {3},
    urldate = {2025-12-16},
    journal = {The Astronomical Journal},
    author = {Stil, J. M. and Taylor, A. R. and Dickey, J. M. and Kavars, D. W. and Martin, P. G. and Rothwell, T. A. and Boothroyd, A. I. and Lockman, Felix J. and McClure-Griffiths, N. M.},
    month = jul,
    year = {2006},
    note = {Publisher: IOP Publishing
Read\_Status: New
Read\_Status\_Date: 2025-12-16T02:25:54.018Z},
    pages = {1158},
}

@article{marchal2019a,
    title = {{ROHSA}: {Regularized} {Optimization} for {Hyper}-{Spectral} {Analysis}. {Application} to phase separation of 21 cm data},
    volume = {626},
    issn = {0004-6361},
    shorttitle = {{ROHSA}},
    url = {https://ui.adsabs.harvard.edu/abs/2019A&A...626A.101M},
    doi = {10.1051/0004-6361/201935335},
    urldate = {2024-10-15},
    journal = {Astronomy and Astrophysics},
    author = {Marchal, Antoine and Miville-Deschênes, Marc-Antoine and Orieux, François and Gac, Nicolas and Soussen, Charles and Lesot, Marie-Jeanne and d'Allonnes, Adrien Revault and Salomé, Quentin},
    month = jun,
    year = {2019},
    note = {ADS Bibcode: 2019A\&A...626A.101M},
    pages = {A101},
}

@misc{kingma2017a,
    title = {Adam: {A} {Method} for {Stochastic} {Optimization}},
    shorttitle = {Adam},
    url = {http://arxiv.org/abs/1412.6980},
    doi = {10.48550/arXiv.1412.6980},
    urldate = {2026-02-20},
    publisher = {arXiv},
    author = {Kingma, Diederik P. and Ba, Jimmy},
    month = jan,
    year = {2017},
    note = {arXiv:1412.6980 [cs]
Read\_Status: New
Read\_Status\_Date: 2026-02-20T05:11:45.967Z},
}

@article{seifried2022a,
    title = {On the accuracy of {H} {I} observations in molecular clouds - {More} cold {H} {I} than thought?},
    volume = {512},
    issn = {0035-8711},
    url = {https://ui.adsabs.harvard.edu/abs/2022MNRAS.512.4765S},
    doi = {10.1093/mnras/stac607},
    urldate = {2026-02-24},
    journal = {Monthly Notices of the Royal Astronomical Society},
    publisher = {OUP},
    author = {Seifried, D. and Beuther, H. and Walch, S. and Syed, J. and Soler, J. D. and Girichidis, P. and Wünsch, R.},
    month = jun,
    year = {2022},
    note = {ADS Bibcode: 2022MNRAS.512.4765S
Read\_Status: New
Read\_Status\_Date: 2026-02-24T00:46:24.204Z},
    pages = {4765--4784},
}

@article{heiner2008,
    title = {The volume densities of giant molecular clouds in {M} 83},
    volume = {489},
    copyright = {© ESO, 2008},
    issn = {0004-6361, 1432-0746},
    url = {https://www.aanda.org/articles/aa/abs/2008/38/aa10430-08/aa10430-08.html},
    doi = {10.1051/0004-6361:200810430},
    language = {en},
    number = {2},
    urldate = {2026-02-24},
    journal = {Astronomy \& Astrophysics},
    publisher = {EDP Sciences},
    author = {Heiner, J. S. and Allen, R. J. and Wong, O. I. and Kruit, P. C. van der},
    month = oct,
    year = {2008},
    note = {Read\_Status: New
Read\_Status\_Date: 2026-02-24T00:59:06.376Z},
    pages = {533--541},
}

@article{wenger2024,
    title = {bayes\_spec: {A} {Bayesian} {Spectral} {Line} {Modeling} {Framework} for {Astrophysics}},
    volume = {9},
    issn = {2475-9066},
    shorttitle = {bayes\_spec},
    url = {https://joss.theoj.org/papers/10.21105/joss.07201},
    doi = {10.21105/joss.07201},
    language = {en},
    number = {103},
    urldate = {2026-02-24},
    journal = {Journal of Open Source Software},
    author = {Wenger, Trey V.},
    month = nov,
    year = {2024},
    note = {Read\_Status: New
Read\_Status\_Date: 2026-02-24T01:02:03.073Z},
    pages = {7201},
}

@article{riener2020,
    title = {Autonomous {Gaussian} decomposition of the {Galactic} {Ring} {Survey}. {I}. {Global} statistics and properties of the {13CO} emission data},
    volume = {633},
    issn = {0004-6361},
    url = {https://ui.adsabs.harvard.edu/abs/2020A&A...633A..14R},
    doi = {10.1051/0004-6361/201936814},
    urldate = {2025-11-18},
    journal = {Astronomy and Astrophysics},
    publisher = {EDP},
    author = {Riener, M. and Kainulainen, J. and Beuther, H. and Henshaw, J. D. and Orkisz, J. H. and Wang, Y.},
    month = jan,
    year = {2020},
    note = {ADS Bibcode: 2020A\&A...633A..14R
Read\_Status: New
Read\_Status\_Date: 2025-11-18T00:26:14.111Z},
    pages = {A14},
}

@article{nguyen2025,
    title = {{TPCNet}: representation learning for {H} i mapping},
    volume = {536},
    issn = {0035-8711},
    shorttitle = {{TPCNet}},
    url = {https://doi.org/10.1093/mnras/stae2631},
    doi = {10.1093/mnras/stae2631},
    number = {1},
    urldate = {2025-05-22},
    journal = {Monthly Notices of the Royal Astronomical Society},
    author = {Nguyen, Hiep and Tang, Haiyang and Alger, Matthew and Marchal, Antoine and Muller, Eric G M and Ong, Cheng Soon and McClure-Griffiths, N M},
    month = jan,
    year = {2025},
    note = {Read\_Status: New
Read\_Status\_Date: 2025-06-06T00:24:23.923Z},
    pages = {962--987},
}

@article{murray2020,
    title = {Extracting the cold neutral medium from {HI} emission with deep learning: {Implications} for {Galactic} foregrounds at high latitude},
    volume = {899},
    issn = {0004-637X, 1538-4357},
    shorttitle = {Extracting the cold neutral medium from {HI} emission with deep learning},
    url = {http://arxiv.org/abs/2006.16490},
    doi = {10.3847/1538-4357/aba19b},
    number = {1},
    urldate = {2024-05-30},
    journal = {The Astrophysical Journal},
    author = {Murray, Claire E. and Peek, J. E. G. and Kim, Chang-Goo},
    month = aug,
    year = {2020},
    note = {arXiv:2006.16490 [astro-ph]},
    pages = {15},
}

@misc{selvaraju2016,
    title = {Grad-{CAM}: {Visual} {Explanations} from {Deep} {Networks} via {Gradient}-based {Localization}},
    shorttitle = {Grad-{CAM}},
    url = {https://ui.adsabs.harvard.edu/abs/2016arXiv161002391S},
    doi = {10.48550/arXiv.1610.02391},
    urldate = {2026-02-24},
    publisher = {arXiv},
    author = {Selvaraju, Ramprasaath R. and Cogswell, Michael and Das, Abhishek and Vedantam, Ramakrishna and Parikh, Devi and Batra, Dhruv},
    month = oct,
    year = {2016},
    note = {ADS Bibcode: 2016arXiv161002391S
Read\_Status: New
Read\_Status\_Date: 2026-02-24T07:27:56.733Z},
}

@misc{simonyan2014,
    title = {Deep {Inside} {Convolutional} {Networks}: {Visualising} {Image} {Classification} {Models} and {Saliency} {Maps}},
    shorttitle = {Deep {Inside} {Convolutional} {Networks}},
    url = {http://arxiv.org/abs/1312.6034},
    doi = {10.48550/arXiv.1312.6034},
    language = {en},
    urldate = {2025-11-12},
    publisher = {arXiv},
    author = {Simonyan, Karen and Vedaldi, Andrea and Zisserman, Andrew},
    month = apr,
    year = {2014},
    note = {arXiv:1312.6034 [cs]
Read\_Status: New
Read\_Status\_Date: 2025-11-12T21:55:13.603Z},
}

@article{jackson2006,
    title = {The {Boston} {University}-{Five} {College} {Radio} {Astronomy} {Observatory} {Galactic} {Ring} {Survey}},
    volume = {163},
    issn = {0067-0049},
    url = {https://ui.adsabs.harvard.edu/abs/2006ApJS..163..145J},
    doi = {10.1086/500091},
    urldate = {2026-02-16},
    journal = {The Astrophysical Journal Supplement Series},
    publisher = {IOP},
    author = {Jackson, J. M. and Rathborne, J. M. and Shah, R. Y. and Simon, R. and Bania, T. M. and Clemens, D. P. and Chambers, E. T. and Johnson, A. M. and Dormody, M. and Lavoie, R. and Heyer, M. H.},
    month = mar,
    year = {2006},
    note = {ADS Bibcode: 2006ApJS..163..145J
Read\_Status: New
Read\_Status\_Date: 2026-02-16T07:27:38.228Z},
    pages = {145--159},
}

@article{field1969,
    title = {Cosmic-{Ray} {Heating} of the {Interstellar} {Gas}},
    volume = {155},
    issn = {0004-637X},
    url = {https://ui.adsabs.harvard.edu/abs/1969ApJ...155L.149F},
    doi = {10.1086/180324},
    urldate = {2024-10-14},
    journal = {The Astrophysical Journal},
    publisher = {IOP},
    author = {Field, G. B. and Goldsmith, D. W. and Habing, H. J.},
    month = mar,
    year = {1969},
    note = {ADS Bibcode: 1969ApJ...155L.149F},
    pages = {L149},
}

@article{mckee1977,
    title = {A theory of the interstellar medium: three components regulated by supernova explosions in an inhomogeneous substrate.},
    volume = {218},
    issn = {0004-637X},
    shorttitle = {A theory of the interstellar medium},
    url = {https://ui.adsabs.harvard.edu/abs/1977ApJ...218..148M},
    doi = {10.1086/155667},
    urldate = {2024-10-14},
    journal = {The Astrophysical Journal},
    publisher = {IOP},
    author = {McKee, C. F. and Ostriker, J. P.},
    month = nov,
    year = {1977},
    note = {ADS Bibcode: 1977ApJ...218..148M},
    pages = {148--169},
}

@article{wolfire2003,
    title = {Neutral {Atomic} {Phases} of the {Interstellar} {Medium} in the {Galaxy}},
    volume = {587},
    issn = {0004-637X},
    url = {https://ui.adsabs.harvard.edu/abs/2003ApJ...587..278W},
    doi = {10.1086/368016},
    urldate = {2024-06-13},
    journal = {The Astrophysical Journal},
    publisher = {IOP},
    author = {Wolfire, Mark G. and McKee, Christopher F. and Hollenbach, David and Tielens, A. G. G. M.},
    month = apr,
    year = {2003},
    note = {ADS Bibcode: 2003ApJ...587..278W},
    pages = {278--311},
}

@article{li2003,
    title = {H {I} {Narrow} {Self}-{Absorption} in {Dark} {Clouds}},
    volume = {585},
    issn = {0004-637X},
    url = {https://iopscience.iop.org/article/10.1086/346227},
    doi = {10.1086/346227},
    language = {en},
    number = {2},
    urldate = {2026-03-24},
    journal = {The Astrophysical Journal},
    publisher = {IOP Publishing},
    author = {Li, D. and Goldsmith, P. F.},
    month = mar,
    year = {2003},
    note = {Read\_Status: New
Read\_Status\_Date: 2026-03-24T01:06:34.227Z},
    pages = {823},
}

@article{krco2008,
	title = {An {Improved} {Technique} for {Measurement} of {Cold} {H} {I} in {Molecular} {Cloud} {Cores}},
	volume = {689},
	issn = {0004-637X},
	url = {https://ui.adsabs.harvard.edu/abs/2008ApJ...689..276K},
	doi = {10.1086/592553},
	urldate = {2025-08-13},
	journal = {The Astrophysical Journal},
	publisher = {IOP},
	author = {Krčo, Marko and Goldsmith, Paul F. and Brown, Robert L. and Li, D.},
	month = dec,
	year = {2008},
	note = {ADS Bibcode: 2008ApJ...689..276K},
	pages = {276--289},
}

@article{wang2020a,
	title = {Cloud formation in the atomic and molecular phase: {HI} self absorption ({HISA}) towards a {Giant} {Molecular} {Filament}},
	volume = {634},
	issn = {0004-6361, 1432-0746},
	shorttitle = {Cloud formation in the atomic and molecular phase},
	url = {http://arxiv.org/abs/2001.00953},
	doi = {10.1051/0004-6361/201935866},
	urldate = {2025-11-05},
	journal = {Astronomy \& Astrophysics},
	author = {Wang, Y. and Bihr, S. and Beuther, H. and Rugel, M. R. and Soler, J. D. and Ott, J. and Kainulainen, J. and Schneider, N. and Klessen, R. S. and Glover, S. C. O. and McClure-Griffiths, N. M. and Goldsmith, P. F. and Johnston, K. G. and Menten, K. M. and Ragan, S. and Anderson, L. D. and Urquhart, J. S. and Linz, H. and Roy, N. and Smith, R. J. and Bigiel, F. and Henning, T. and Longmore, S. N.},
	month = feb,
	year = {2020},
	note = {arXiv:2001.00953 [astro-ph]},
	pages = {A139},
}

@article{dickey2013,
	title = {{GASKAP}-{The} {Galactic} {ASKAP} {Survey}},
	volume = {30},
	issn = {1323-3580},
	url = {https://ui.adsabs.harvard.edu/abs/2013PASA...30....3D},
	doi = {10.1017/pasa.2012.003},
	urldate = {2024-10-15},
	journal = {Publications of the Astronomical Society of Australia},
	author = {Dickey, John M. and McClure-Griffiths, Naomi and Gibson, Steven J. and Gómez, José F. and Imai, Hiroshi and Jones, Paul and Stanimirović, Snežana and Van Loon, Jacco Th. and Walsh, Andrew and Alberdi, A. and Anglada, G. and Uscanga, L. and Arce, H. and Bailey, M. and Begum, A. and Wakker, B. and Bekhti, N. Ben and Kalberla, P. and Winkel, B. and Bekki, K. and For, B. -Q. and Staveley-Smith, L. and Westmeier, T. and Burton, M. and Cunningham, M. and Dawson, J. and Ellingsen, S. and Diamond, P. and Green, J. A. and Hill, A. S. and Koribalski, B. and McConnell, D. and Rathborne, J. and Voronkov, M. and Douglas, K. A. and English, J. and Ford, H. Alyson and Lockman, F. J. and Foster, T. and Gomez, Y. and Green, A. and Bland-Hawthorn, J. and Gulyaev, S. and Hoare, M. and Joncas, G. and Kang, J. -H. and Kerton, C. R. and Koo, B. -C. and Leahy, D. and Lo, N. and Migenes, V. and Nakashima, J. and Zhang, Y. and Nidever, D. and Peek, J. E. G. and Tafoya, D. and Tian, W. and Wu, D.},
	month = jan,
	year = {2013},
	note = {ADS Bibcode: 2013PASA...30....3D},
	pages = {e003},
}

@article{gibson2000,
	title = {A {New} {View} of {Cold} {H} {I} {Clouds} in the {Milky} {Way}},
	volume = {540},
	issn = {0004-637X},
	url = {https://ui.adsabs.harvard.edu/abs/2000ApJ...540..851G},
	doi = {10.1086/309364},
	urldate = {2026-03-25},
	journal = {The Astrophysical Journal},
	publisher = {IOP},
	author = {Gibson, Steven J. and Taylor, A. Russell and Higgs, Lloyd A. and Dewdney, Peter E.},
	month = sep,
	year = {2000},
	note = {ADS Bibcode: 2000ApJ...540..851G},
	pages = {851--862},
}

@article{dickey2022,
    title = {{GASKAP} {Pilot} {Survey} {Science}. {II}. {ASKAP} {Zoom} {Observations} of {Galactic} 21 cm {Absorption}},
    volume = {926},
    issn = {0004-637X},
    url = {https://ui.adsabs.harvard.edu/abs/2022ApJ...926..186D},
    doi = {10.3847/1538-4357/ac3a89},
    urldate = {2026-03-27},
    journal = {The Astrophysical Journal},
    publisher = {IOP},
    author = {Dickey, John M. and Dempsey, J. M. and Pingel, N. M. and McClure-Griffiths, N. M. and Jameson, K. and Dawson, J. R. and Dénes, H. and Clark, S. E. and Joncas, G. and Leahy, D. and Lee, Min-Young and Miville-Deschênes, M.-A. and Stanimirović, S. and Tremblay, C. D. and van Loon, J. Th.},
    month = feb,
    year = {2022},
    note = {ADS Bibcode: 2022ApJ...926..186D
Read\_Status: New
Read\_Status\_Date: 2026-03-27T05:30:12.658Z},
    pages = {186},
}

@article{nguyen2024,
    title = {Local {Hi} {Absorption} towards the {Magellanic} {Cloud} foreground using {ASKAP}},
    doi = {10.1093/mnras/stae2274},
    language = {en},
    journal = {MNRAS},
    author = {Nguyen, Hiep and McClure-Griffiths, N M and Dempsey, James and Dickey, John M and Lee, Min-Young and Lynn, Callum and Murray, Claire and Stanimirović, Snežana and Busch, Michael P and Clark, Susan E and Dawson, J R and Dénes, Helga and Gibson, Steven and Jameson, Katherine and Joncas, Gilles and Kemp, Ian and Leahy, Denis and Ma, Yik Ki and Marchal, Antoine and Miville-Deschênes, Marc-Antoine and Pingel, Nickolas M and Seta, Amit and Soler, Juan D and van Loon, Jacco Th},
    year = {2024},
}

@article{riener2019,
    title = {{GAUSSPY}+: {A} fully automated {Gaussian} decomposition package for emission line spectra},
    volume = {628},
    issn = {0004-6361},
    shorttitle = {{GAUSSPY}+},
    url = {https://ui.adsabs.harvard.edu/abs/2019A&A...628A..78R},
    doi = {10.1051/0004-6361/201935519},
    urldate = {2024-10-15},
    journal = {Astronomy and Astrophysics},
    author = {Riener, M. and Kainulainen, J. and Henshaw, J. D. and Orkisz, J. H. and Murray, C. E. and Beuther, H.},
    month = aug,
    year = {2019},
    note = {ADS Bibcode: 2019A\&A...628A..78R},
    pages = {A78},
}

@article{knapp1974,
    title = {Observations of {HI} in dense interstellar dust clouds: {II}. {The} cloud {Khavtassi} 3},
    volume = {79},
    issn = {0004-6256},
    shorttitle = {Observations of {HI} in dense interstellar dust clouds},
    url = {https://ui.adsabs.harvard.edu/abs/1974AJ.....79..541K},
    doi = {10.1086/111575},
    urldate = {2026-04-06},
    journal = {The Astronomical Journal},
    publisher = {IOP},
    author = {Knapp, G. R.},
    month = may,
    year = {1974},
    note = {ADS Bibcode: 1974AJ.....79..541K
Read\_Status: New
Read\_Status\_Date: 2026-04-06T23:24:21.286Z},
    pages = {541},
}

@misc{kessler2025,
    title = {Identification of molecular line emission using {Convolutional} {Neural} {Networks}},
    url = {http://arxiv.org/abs/2510.09119},
    doi = {10.48550/arXiv.2510.09119},
    urldate = {2025-10-13},
    publisher = {arXiv},
    author = {Kessler, Nina and Csengeri, Timea and Cornu, David and Bontemps, Sylvain and Bouscasse, Laure},
    month = oct,
    year = {2025},
    note = {arXiv:2510.09119 [astro-ph]
Read\_Status: New
Read\_Status\_Date: 2025-10-13T23:26:42.369Z},
}

@misc{lei2025,
    title = {Neutral gas phase distribution from {HI} morphology: phase separation with scattering spectra and variational autoencoders},
    shorttitle = {Neutral gas phase distribution from {HI} morphology},
    url = {http://arxiv.org/abs/2505.20407},
    doi = {10.48550/arXiv.2505.20407},
    urldate = {2025-05-29},
    publisher = {arXiv},
    author = {Lei, Minjie and Clark, S. E. and Morel, Rudy and Allys, E. and Butsky, Iryna S. and Redshaw, Caleb and Fielding, Drummond B.},
    month = may,
    year = {2025},
    note = {arXiv:2505.20407 [astro-ph]
Read\_Status: New
Read\_Status\_Date: 2025-06-06T00:23:32.982Z},
}

@article{lecun1998a,
    title = {Gradient-{Based} {Learning} {Applied} to {Document} {Recognition}},
    volume = {86},
    doi = {10.1109/5.726791},
    journal = {Proceedings of the IEEE},
    author = {Lecun, Yann and Haffner, Patrick and Rachmad, Yoesoep and Bottou, Leon},
    month = dec,
    year = {1998},
    note = {Read\_Status: New
Read\_Status\_Date: 2026-04-16T02:39:35.995Z},
    pages = {2278--2324},
}

@misc{he2015,
    title = {Deep {Residual} {Learning} for {Image} {Recognition}},
    url = {http://arxiv.org/abs/1512.03385},
    doi = {10.48550/arXiv.1512.03385},
    urldate = {2026-04-16},
    publisher = {arXiv},
    author = {He, Kaiming and Zhang, Xiangyu and Ren, Shaoqing and Sun, Jian},
    month = dec,
    year = {2015},
    note = {arXiv:1512.03385 [cs]
Read\_Status: New
Read\_Status\_Date: 2026-04-16T02:41:58.788Z},
}

@article{chen2025,
    title = {A {Neutral} {Hydrogen} {Absorption} {Study} of {Cold} {Gas} in the {Outskirts} of the {Magellanic} {Clouds} {Using} the {GASKAP}-{H} i {Survey}},
    volume = {169},
    issn = {1538-3881},
    url = {https://doi.org/10.3847/1538-3881/adc67c},
    doi = {10.3847/1538-3881/adc67c},
    language = {en},
    number = {5},
    urldate = {2026-02-16},
    journal = {The Astronomical Journal},
    publisher = {The American Astronomical Society},
    author = {Chen, Hongxing and Stanimirović, Snežana and Pingel, Nickolas M. and Dempsey, James and Buckland-Willis, Frances and Clark, Susan E. and Dénes, Helga and Dickey, John M. and Gibson, Steven and Jameson, Katherine and Kemp, Ian and Leahy, Denis and Lee, Min-Young and Lynn, Callum and Ma, Yik Ki and McClure-Griffiths, N. M. and Murray, Claire E. and Nguyen, Hiep and Uscanga, Lucero and van Loon, Jacco Th. and Vázquez-Semadeni, Enrique},
    month = apr,
    year = {2025},
    note = {Read\_Status: New
Read\_Status\_Date: 2026-02-16T07:26:14.725Z},
    pages = {284},
}

@article{pingel2022,
    title = {{GASKAP}-{HI} pilot survey science {I}: {ASKAP} zoom observations of {HI} emission in the {Small} {Magellanic} {Cloud}},
    volume = {39},
    issn = {1323-3580},
    shorttitle = {{GASKAP}-{HI} pilot survey science {I}},
    url = {https://ui.adsabs.harvard.edu/abs/2022PASA...39....5P},
    doi = {10.1017/pasa.2021.59},
    urldate = {2026-04-21},
    journal = {Publications of the Astronomical Society of Australia},
    author = {Pingel, N. M. and Dempsey, J. and McClure-Griffiths, N. M. and Dickey, J. M. and Jameson, K. E. and Arce, H. and Anglada, G. and Bland-Hawthorn, J. and Breen, S. L. and Buckland-Willis, F. and Clark, S. E. and Dawson, J. R. and Dénes, H. and Di Teodoro, E. M. and For, B.-Q. and Foster, Tyler J. and Gómez, J. F. and Imai, H. and Joncas, G. and Kim, C.-G. and Lee, M.-Y. and Lynn, C. and Leahy, D. and Ma, Y. K. and Marchal, A. and McConnell, D. and Miville-Deschènes, M.-A. and Moss, V. A. and Murray, C. E. and Nidever, D. and Peek, J. and Stanimirović, S. and Staveley-Smith, L. and Tepper-Garcia, T. and Tremblay, C. D. and Uscanga, L. and van Loon, J. Th. and Vázquez-Semadeni, E. and Allison, J. R. and Anderson, C. S. and Ball, Lewis and Bell, M. and Bock, D. C.-J. and Bunton, J. and Cooray, F. R. and Cornwell, T. and Koribalski, B. S. and Gupta, N. and Hayman, D. B. and Harvey-Smith, L. and Lee-Waddell, K. and Ng, A. and Phillips, C. J. and Voronkov, M. and Westmeier, T. and Whiting, M. T.},
    month = feb,
    year = {2022},
    note = {ADS Bibcode: 2022PASA...39....5P
Read\_Status: New
Read\_Status\_Date: 2026-04-21T01:10:24.033Z},
    pages = {e005},
}

@article{wolfire1995,
    title = {The {Neutral} {Atomic} {Phases} of the {Interstellar} {Medium}},
    volume = {443},
    issn = {0004-637X},
    url = {https://ui.adsabs.harvard.edu/abs/1995ApJ...443..152W},
    doi = {10.1086/175510},
    urldate = {2024-06-13},
    journal = {The Astrophysical Journal},
    publisher = {IOP},
    author = {Wolfire, M. G. and Hollenbach, D. and McKee, C. F. and Tielens, A. G. G. M. and Bakes, E. L. O.},
    month = apr,
    year = {1995},
    note = {ADS Bibcode: 1995ApJ...443..152W},
    pages = {152},
}

@article{sternberg2014,
    title = {H {I}-to-{H2} {Transitions} and {H} {I} {Column} {Densities} in {Galaxy} {Star}-forming {Regions}},
    volume = {790},
    issn = {0004-637X},
    url = {https://ui.adsabs.harvard.edu/abs/2014ApJ...790...10S},
    doi = {10.1088/0004-637X/790/1/10},
    urldate = {2024-11-10},
    journal = {The Astrophysical Journal},
    publisher = {IOP},
    author = {Sternberg, Amiel and Le Petit, Franck and Roueff, Evelyne and Le Bourlot, Jacques},
    month = jul,
    year = {2014},
    note = {ADS Bibcode: 2014ApJ...790...10S},
    pages = {10},
}

@article{wolfire2010,
    title = {The {Dark} {Molecular} {Gas}},
    volume = {716},
    issn = {0004-637X},
    url = {https://ui.adsabs.harvard.edu/abs/2010ApJ...716.1191W},
    doi = {10.1088/0004-637X/716/2/1191},
    urldate = {2026-04-23},
    journal = {The Astrophysical Journal},
    publisher = {IOP},
    author = {Wolfire, Mark G. and Hollenbach, David and McKee, Christopher F.},
    month = jun,
    year = {2010},
    note = {ADS Bibcode: 2010ApJ...716.1191W
Read\_Status: New
Read\_Status\_Date: 2026-04-23T01:43:44.396Z},
    pages = {1191--1207},
}

@article{moss2012,
    title = {{GSH} 006-15+7: {A} local {Galactic} supershell featuring transition from {HI} emission to absorption},
    volume = {421},
    issn = {00358711},
    shorttitle = {{GSH} 006-15+7},
    url = {http://arxiv.org/abs/1201.2700},
    doi = {10.1111/j.1365-2966.2012.20538.x},
    number = {4},
    urldate = {2026-04-23},
    journal = {Monthly Notices of the Royal Astronomical Society},
    author = {Moss, Vanessa A. and McClure-Griffiths, Naomi M. and Braun, Robert and Hill, Alex S. and Madsen, Greg J.},
    month = apr,
    year = {2012},
    note = {arXiv:1201.2700 [astro-ph]
Read\_Status: New
Read\_Status\_Date: 2026-04-23T01:48:46.978Z},
    pages = {3159--3169},
}

@article{park2023,
    title = {Probing the {Conditions} for the {H} i-to-{H2} {Transition} in the {Interstellar} {Medium}},
    volume = {955},
    issn = {0004-637X},
    url = {https://doi.org/10.3847/1538-4357/ace164},
    doi = {10.3847/1538-4357/ace164},
    language = {en},
    number = {2},
    urldate = {2026-02-16},
    journal = {The Astrophysical Journal},
    publisher = {The American Astronomical Society},
    author = {Park, Gyueun and Lee, Min-Young and Bialy, Shmuel and Burkhart, Blakesley and Dawson, J. R. and Heiles, Carl and Li, Di and Murray, Claire and Nguyen, Hiep and Hafner, Anita and Rybarczyk, Daniel R. and Stanimirović, Snežana},
    month = sep,
    year = {2023},
    note = {Read\_Status: New
Read\_Status\_Date: 2026-02-16T07:25:38.526Z},
    pages = {145},
}

@article{beuther2011,
    title = {The {Coalsack} near and far},
    volume = {533},
    copyright = {© ESO, 2011},
    issn = {0004-6361, 1432-0746},
    url = {https://www.aanda.org/articles/aa/abs/2011/09/aa16746-11/aa16746-11.html},
    doi = {10.1051/0004-6361/201116746},
    language = {en},
    urldate = {2026-04-24},
    journal = {Astronomy \& Astrophysics},
    publisher = {EDP Sciences},
    author = {Beuther, H. and Kainulainen, J. and Henning, Th and Plume, R. and Heitsch, F.},
    month = sep,
    year = {2011},
    note = {Read\_Status: New
Read\_Status\_Date: 2026-04-24T03:26:54.523Z},
    pages = {A17},
}

@article{ge2023,
    title = {Large-scale velocity-coherent filaments in the {SEDIGISM} survey: {Association} with spiral arms and the fraction of dense gas},
    volume = {675},
    copyright = {© The Authors 2023},
    issn = {0004-6361, 1432-0746},
    shorttitle = {Large-scale velocity-coherent filaments in the {SEDIGISM} survey},
    url = {https://www.aanda.org/articles/aa/abs/2023/07/aa45784-22/aa45784-22.html},
    doi = {10.1051/0004-6361/202245784},
    language = {en},
    urldate = {2026-04-24},
    journal = {Astronomy \& Astrophysics},
    publisher = {EDP Sciences},
    author = {Ge, Y. and Wang, K. and Duarte-Cabral, A. and Pettitt, A. R. and Dobbs, C. L. and Sánchez-Monge, A. and Neralwar, K. R. and Urquhart, J. S. and Colombo, D. and Durán-Camacho, E. and Beuther, H. and Bronfman, L. and Rigby, A. J. and Eden, D. and Neupane, S. and Barnes, P. and Henning, T. and Yang, A. Y.},
    month = jul,
    year = {2023},
    note = {Read\_Status: New
Read\_Status\_Date: 2026-04-24T03:50:21.293Z},
    pages = {A119},
}

@article{abreu-vicente2016,
    title = {Giant molecular filaments in the {Milky} {Way} - {II}. {The} fourth {Galactic} quadrant},
    volume = {590},
    copyright = {© ESO, 2016},
    issn = {0004-6361, 1432-0746},
    url = {https://www.aanda.org/articles/aa/abs/2016/06/aa27674-15/aa27674-15.html},
    doi = {10.1051/0004-6361/201527674},
    language = {en},
    urldate = {2026-04-24},
    journal = {Astronomy \& Astrophysics},
    publisher = {EDP Sciences},
    author = {Abreu-Vicente, J. and Ragan, S. and Kainulainen, J. and Henning, Th and Beuther, H. and Johnston, K.},
    month = jun,
    year = {2016},
    note = {Read\_Status: New
Read\_Status\_Date: 2026-04-24T03:51:40.640Z},
    pages = {A131},
}

@article{rozanski2022,
    title = {{SUPPNet}: {Neural} network for stellar spectrum normalisation},
    volume = {659},
    copyright = {© ESO 2022},
    issn = {0004-6361, 1432-0746},
    shorttitle = {{SUPPNet}},
    url = {https://www.aanda.org/articles/aa/abs/2022/03/aa41480-21/aa41480-21.html},
    doi = {10.1051/0004-6361/202141480},
    language = {en},
    urldate = {2026-04-27},
    journal = {Astronomy \& Astrophysics},
    publisher = {EDP Sciences},
    author = {Różański, T. and Niemczura, E. and Lemiesz, J. and Posiłek, N. and Różański, P.},
    month = mar,
    year = {2022},
    note = {Read\_Status: New
Read\_Status\_Date: 2026-04-27T03:19:30.695Z},
    pages = {A199},
}

@book{sparke2007,
    title = {Galaxies in the {Universe}},
    url = {https://ui.adsabs.harvard.edu/abs/2007gaun.book.....S},
    urldate = {2026-04-30},
    author = {Sparke, Linda S. and Gallagher, III, John S.},
    month = feb,
    year = {2007},
    note = {Publication Title: Galaxies in the Universe
ADS Bibcode: 2007gaun.book.....S
Read\_Status: New
Read\_Status\_Date: 2026-04-30T01:59:39.644Z},
}

@article{park2026,
    title = {A {High}-resolution {Study} of the {Cold} {Neutral} {Medium} in and around 30 {Doradus}},
    volume = {997},
    issn = {0004-637X},
    url = {https://ui.adsabs.harvard.edu/abs/2026ApJ...997...50P},
    doi = {10.3847/1538-4357/ae28d7},
    urldate = {2026-04-30},
    journal = {The Astrophysical Journal},
    publisher = {IOP},
    author = {Park, Gyueun and Lee, Min-Young and Dickey, John M. and Pingel, Nick M. and Dempsey, James and Dénes, Helga and Gibson, Steven and Jameson, Katie and Kemp, Ian and Kim, Chang-Goo and Leahy, Denis and Lee, Bumhyun and Lynn, Callum and Ma, Yik Ki and Marchal, Antoine and McClure-Griffiths, Naomi M. and Muller, Eric and Nguyen, Hiep and Stanimirović, Snežana and van Loon, Jacco Th. and {The Gaskap-H I Collaboration}},
    month = jan,
    year = {2026},
    note = {ADS Bibcode: 2026ApJ...997...50P
Read\_Status: New
Read\_Status\_Date: 2026-04-30T04:23:51.361Z},
    pages = {50},
}

@article{dempsey2022,
    title = {{GASKAP}-{HI} {Pilot} {Survey} {Science} {III}: {An} unbiased view of cold gas in the {Small} {Magellanic} {Cloud}},
    volume = {39},
    issn = {1323-3580},
    shorttitle = {{GASKAP}-{HI} {Pilot} {Survey} {Science} {III}},
    url = {https://ui.adsabs.harvard.edu/abs/2022PASA...39...34D},
    doi = {10.1017/pasa.2022.18},
    urldate = {2024-10-19},
    journal = {Publications of the Astronomical Society of Australia},
    author = {Dempsey, James and McClure-Griffiths, N. M. and Murray, Claire and Dickey, John M. and Pingel, Nickolas M. and Jameson, Katherine and Dénes, Helga and van Loon, Jacco Th. and Leahy, D. and Lee, Min-Young and Stanimirović, S. and Breen, Shari and Buckland-Willis, Frances and Gibson, Steven J. and Imai, Hiroshi and Lynn, Callum and Tremblay, C. D.},
    month = aug,
    year = {2022},
    note = {ADS Bibcode: 2022PASA...39...34D},
    pages = {e034},
}

@article{song2023,
    title = {Learning {From} {Noisy} {Labels} {With} {Deep} {Neural} {Networks}: {A} {Survey}},
    volume = {34},
    issn = {2162-2388},
    shorttitle = {Learning {From} {Noisy} {Labels} {With} {Deep} {Neural} {Networks}},
    doi = {10.1109/TNNLS.2022.3152527},
    language = {eng},
    number = {11},
    journal = {IEEE transactions on neural networks and learning systems},
    author = {Song, Hwanjun and Kim, Minseok and Park, Dongmin and Shin, Yooju and Lee, Jae-Gil},
    month = nov,
    year = {2023},
    note = {Read\_Status: New
Read\_Status\_Date: 2026-05-24T11:32:08.076Z},
    pages = {8135--8153},
}

@article{frenay2014,
    title = {Classification in the {Presence} of {Label} {Noise}: {A} {Survey}},
    volume = {25},
    issn = {2162-2388},
    shorttitle = {Classification in the {Presence} of {Label} {Noise}},
    url = {https://ieeexplore.ieee.org/document/6685834},
    doi = {10.1109/TNNLS.2013.2292894},
    number = {5},
    urldate = {2026-05-24},
    journal = {IEEE Transactions on Neural Networks and Learning Systems},
    author = {Frenay, Benoit and Verleysen, Michel},
    month = may,
    year = {2014},
    note = {Read\_Status: New
Read\_Status\_Date: 2026-05-24T11:31:42.108Z},
    pages = {845--869},
}

@article{audit2005a,
    title = {Thermal condensation in a turbulent atomic hydrogen flow},
    volume = {433},
    issn = {0004-6361},
    url = {https://ui.adsabs.harvard.edu/abs/2005A&A...433....1A},
    doi = {10.1051/0004-6361:20041474},
    urldate = {2026-06-16},
    journal = {Astronomy and Astrophysics},
    publisher = {EDP},
    author = {Audit, E. and Hennebelle, P.},
    month = apr,
    year = {2005},
    note = {ADS Bibcode: 2005A\&A...433....1A
Read\_Status: New
Read\_Status\_Date: 2026-06-16T00:17:26.515Z},
    pages = {1--13},
}

@inproceedings{inutsuka2016,
    title = {The {Formation} and {Destruction} of {Molecular} {Clouds} and {Galactic} {Star} {Formation}},
    volume = {315},
    url = {https://ui.adsabs.harvard.edu/abs/2016IAUS..315...61I},
    doi = {10.1017/S1743921316007262},
    urldate = {2026-06-16},
    author = {Inutsuka, Shu-Ichiro and Inoue, Tsuyoshi and Iwasaki, Kazunari and Hosokawa, Takashi and Kobayashi, Masato I. N.},
    month = jan,
    year = {2016},
    note = {ADS Bibcode: 2016IAUS..315...61I
Read\_Status: New
Read\_Status\_Date: 2026-06-16T00:17:40.872Z},
    pages = {61--68},
}

@article{montgomery1995,
    title = {Studies of {H} {I} self-absorption in the {Riegel} \& {Crutcher} cold cloud},
    volume = {273},
    issn = {0035-8711},
    url = {https://doi.org/10.1093/mnras/273.2.449},
    doi = {10.1093/mnras/273.2.449},
    number = {2},
    urldate = {2026-06-19},
    journal = {Monthly Notices of the Royal Astronomical Society},
    author = {Montgomery, A. S. and Bates, B. and Davies, R. D.},
    month = mar,
    year = {1995},
    note = {Read\_Status: New
Read\_Status\_Date: 2026-06-19T05:36:47.487Z},
    pages = {449--460},
}

@article{lindner2015,
    title = {Autonomous {Gaussian} {Decomposition}},
    volume = {149},
    issn = {0004-6256},
    url = {https://ui.adsabs.harvard.edu/abs/2015AJ....149..138L},
    doi = {10.1088/0004-6256/149/4/138},
    urldate = {2024-10-15},
    journal = {The Astronomical Journal},
    publisher = {IOP},
    author = {Lindner, Robert R. and Vera-Ciro, Carlos and Murray, Claire E. and Stanimirović, Snežana and Babler, Brian and Heiles, Carl and Hennebelle, Patrick and Goss, W. M. and Dickey, John},
    month = apr,
    year = {2015},
    note = {ADS Bibcode: 2015AJ....149..138L},
    pages = {138},
}

@article{zhou2006,
    title = {Training cost-sensitive neural networks with methods addressing the class imbalance problem},
    volume = {18},
    issn = {1558-2191},
    url = {https://ieeexplore.ieee.org/document/1549828},
    doi = {10.1109/TKDE.2006.17},
    number = {1},
    urldate = {2026-06-24},
    journal = {IEEE Transactions on Knowledge and Data Engineering},
    author = {Zhou, Zhi-Hua and Liu, Xu-Ying},
    month = jan,
    year = {2006},
    note = {Read\_Status: New
Read\_Status\_Date: 2026-06-24T03:07:58.496Z},
    pages = {63--77},
}

@misc{smilkov2017,
    title = {{SmoothGrad}: removing noise by adding noise},
    shorttitle = {{SmoothGrad}},
    url = {http://arxiv.org/abs/1706.03825},
    doi = {10.48550/arXiv.1706.03825},
    urldate = {2026-06-24},
    publisher = {arXiv},
    author = {Smilkov, Daniel and Thorat, Nikhil and Kim, Been and Viégas, Fernanda and Wattenberg, Martin},
    month = jun,
    year = {2017},
    note = {arXiv:1706.03825 [cs.LG]
Read\_Status: New
Read\_Status\_Date: 2026-06-24T03:59:00.218Z},
}

@article{kim2017,
    title = {Three-phase {Interstellar} {Medium} in {Galaxies} {Resolving} {Evolution} with {Star} {Formation} and {Supernova} {Feedback} ({TIGRESS}): {Algorithms}, {Fiducial} {Model}, and {Convergence}},
    volume = {846},
    issn = {0004-637X},
    shorttitle = {Three-phase {Interstellar} {Medium} in {Galaxies} {Resolving} {Evolution} with {Star} {Formation} and {Supernova} {Feedback} ({TIGRESS})},
    url = {https://ui.adsabs.harvard.edu/abs/2017ApJ...846..133K},
    doi = {10.3847/1538-4357/aa8599},
    urldate = {2024-11-20},
    journal = {The Astrophysical Journal},
    publisher = {IOP},
    author = {Kim, Chang-Goo and Ostriker, Eve C.},
    month = sep,
    year = {2017},
    note = {ADS Bibcode: 2017ApJ...846..133K
Read\_Status: New
Read\_Status\_Date: 2024-11-20T08:08:03.022Z},
    pages = {133},
}

@article{kado-fong2020,
    title = {Diffuse {Ionized} {Gas} in {Simulations} of {Multiphase}, {Star}-forming {Galactic} {Disks}},
    volume = {897},
    issn = {0004-637X},
    url = {https://ui.adsabs.harvard.edu/abs/2020ApJ...897..143K},
    doi = {10.3847/1538-4357/ab9abd},
    urldate = {2026-08-26},
    journal = {The Astrophysical Journal},
    publisher = {IOP},
    author = {Kado-Fong, Erin and Kim, Jeong-Gyu and Ostriker, Eve C. and Kim, Chang-Goo},
    month = jul,
    year = {2020},
    note = {ADS Bibcode: 2020ApJ...897..143K
Read\_Status: New
Read\_Status\_Date: 2026-08-26T06:13:03.134Z},
    pages = {143},
}

@misc{guo2017,
    title = {On {Calibration} of {Modern} {Neural} {Networks}},
    url = {http://arxiv.org/abs/1706.04599},
    doi = {10.48550/arXiv.1706.04599},
    urldate = {2026-08-27},
    publisher = {arXiv},
    author = {Guo, Chuan and Pleiss, Geoff and Sun, Yu and Weinberger, Kilian Q.},
    month = aug,
    year = {2017},
    note = {arXiv:1706.04599 [cs.LG]
Read\_Status: New
Read\_Status\_Date: 2026-08-27T07:45:47.958Z},
}

@misc{nixon2020,
    title = {Measuring {Calibration} in {Deep} {Learning}},
    url = {http://arxiv.org/abs/1904.01685},
    doi = {10.48550/arXiv.1904.01685},
    urldate = {2026-08-27},
    publisher = {arXiv},
    author = {Nixon, Jeremy and Dusenberry, Mike and Jerfel, Ghassen and Nguyen, Timothy and Liu, Jeremiah and Zhang, Linchuan and Tran, Dustin},
    month = aug,
    year = {2020},
    note = {arXiv:1904.01685 [cs.LG]
Read\_Status: New
Read\_Status\_Date: 2026-08-27T07:44:47.081Z},
}

%%%%%%%%%%%%%%%%%%%%%%%%%%%%%%%%%%%%%%%%%%%%%%%%%%

%%%%%%%%%%%%%%%%% APPENDICES %%%%%%%%%%%%%%%%%%%%%

\appendix

\section{Model Design}\label{app:model design}

\begin{figure}
    \centering
    \includegraphics[width=0.9\linewidth]{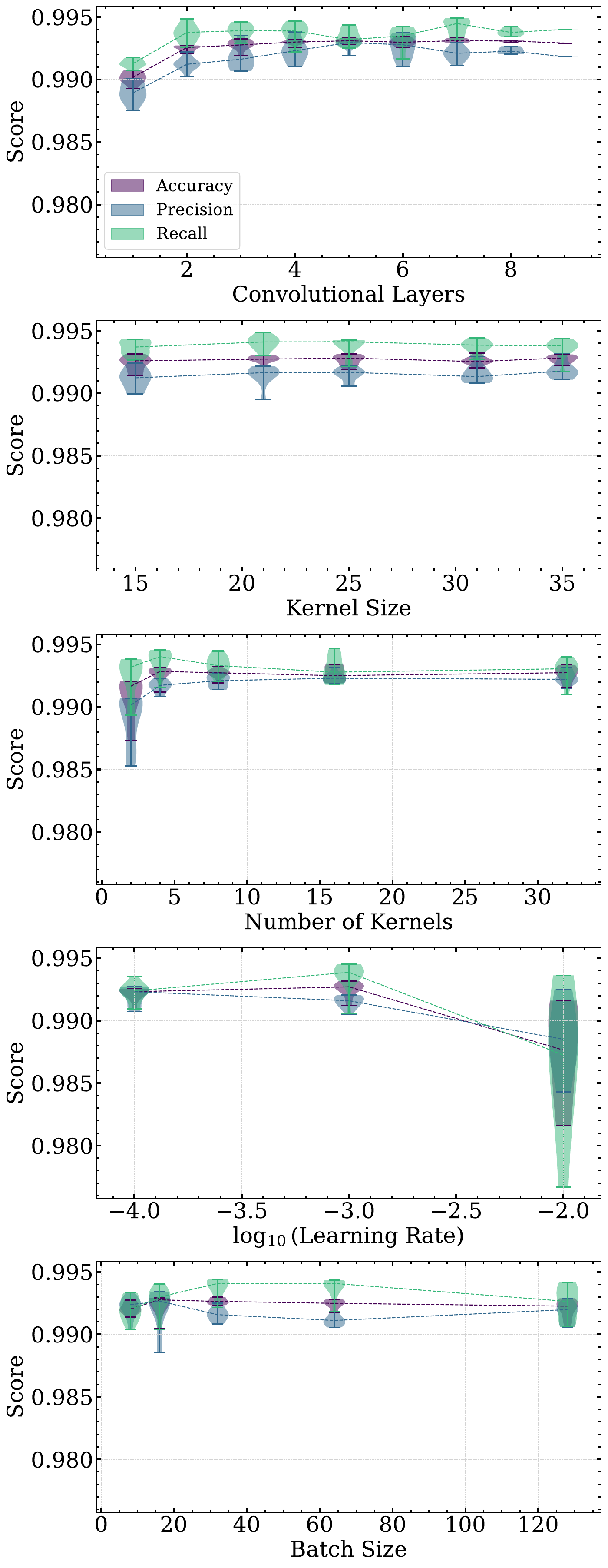}
    \caption{Violin plots for the testing metrics of the grid search networks. From top to bottom: number of convolutional layers, size of convolutional kernels in units of channels, number of convolutional kernels, learning rate, and batch size. Metrics of accuracy, precision, and recall are reported in black, red, and purple respectively. Each version of the network is trained ten times, with the distribution shown by the violins and the median value shown by the dashed lines. Alt text: graphs comparing score against various hyperparameters, for three sets of coloured lines representing different metrics.}
    \label{fig:grid search}
\end{figure}

\begin{figure}
    \centering
    \includegraphics[width=\linewidth]{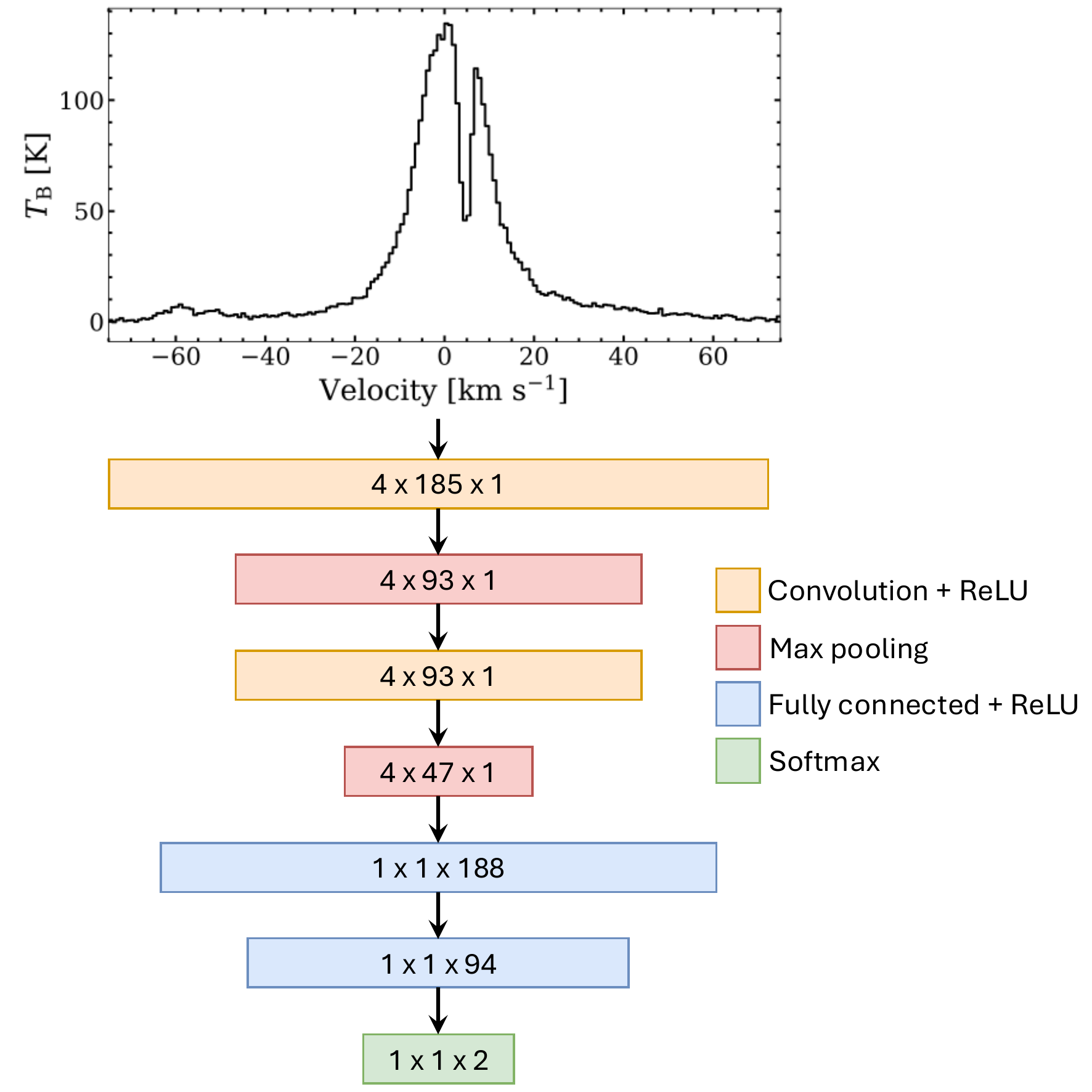}
    \caption{The architecture of the convolutional neural network used for HISA detection in this work. The final layer returns the probabilities of the two classes, $\mathbb{P}(\neg{\rm HISA})$ and $\mathbb{P}({\rm HISA})$. Alt text: block flowchart showing the architecture of a neural network, from a spectrum input to a two-element vector output. Blocks are colour-coded according to their type, representing convolutional, pooling, fully connected, and softmax layers.}
    \label{fig:cnn architecture}
\end{figure}

As detailed in Section~\ref{sec:machine learning}, we use a sequential grid search to determine the ideal combination of hyperparameters for the convolutional neural network. The optimised hyperparameters were the number of convolutional layers, the kernel size, the number of kernels, the learning rate, and the batch size. Each combination of hyperparameters was used to train ten neural networks on the training dataset (described in Section~\ref{sec:synthetic data}). Figure~\ref{fig:grid search} shows the results of the grid search used to determine the ideal hyperparameters. Figure~\ref{fig:cnn architecture} shows the best-performing architecture of the neural network.

\section{Riegel--Crutcher Linear Interpolation HISA Detections}\label{app:rc linear interpolation}

\begin{figure*}
    \centering
    \includegraphics[width=\linewidth]{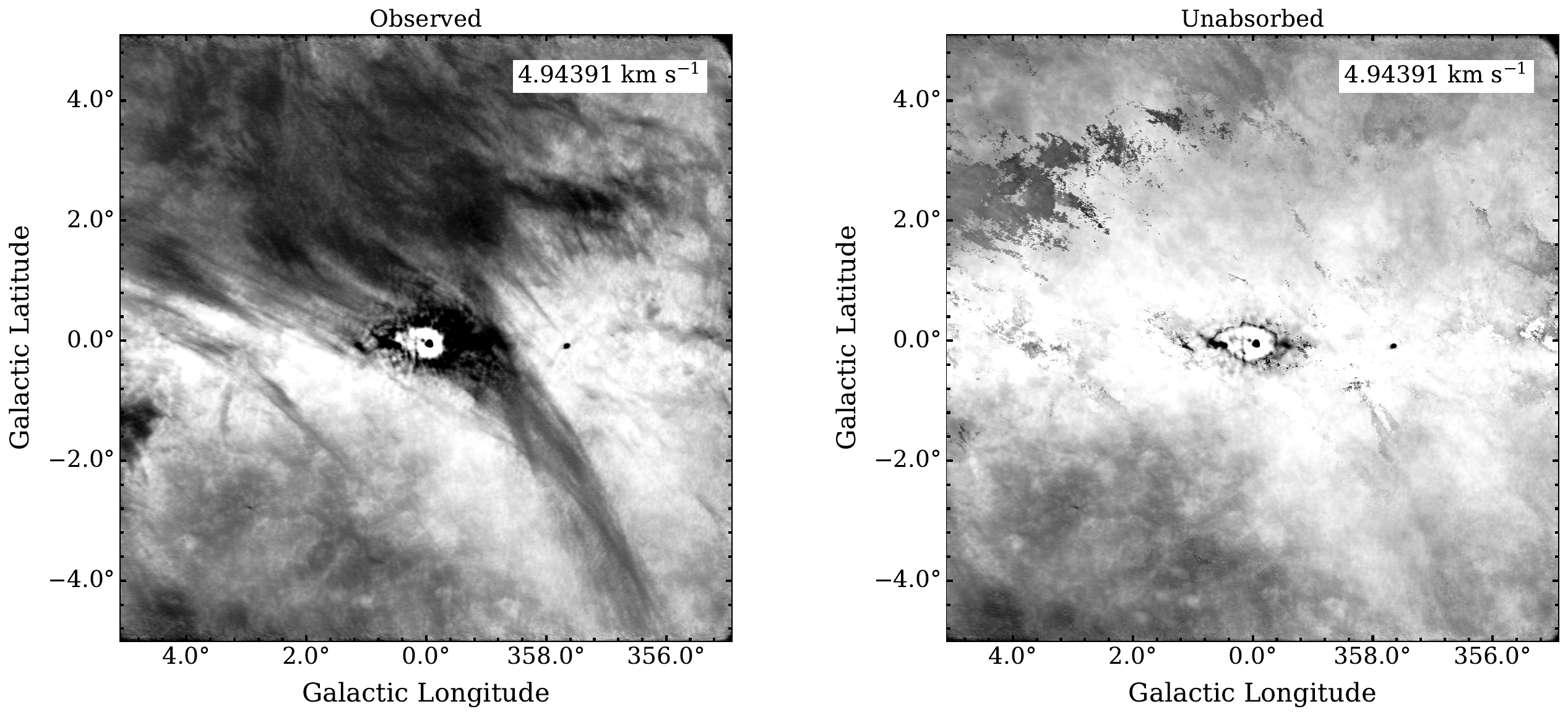}
    \caption{Left: observed velocity slice of the SGPS GC \ion{H}{i} emission data at $v=4.94~{\rm km~s}^{-1}$. Right: interpolated estimation of the unabsorbed emission. The grey scale is linear between 30 (white) and 150~K (black). This figure is analogous to fig. 1 in \citet{mcclure-griffiths2006}. Alt text: two-panel graph that shows a grey-scale image of a cloud on the left, and the same field with the cloud removed on the right. Text in the upper right of each panel reports the velocity. The right panel shows a coherent image of the background field, except for a small dark region towards the top left of the panel.}
    \label{fig:rc linear interpolation slice}
\end{figure*}

Here we present the results of the linear interpolation method used as a comparative HISA detection technique in the Riegel--Crutcher cloud. The method used is discussed in Section~\ref{sec:rc interpolation comparison}, and loosely follows the methodology originally used to detect cold \ion{H}{i} structures in the cloud by \citet{mcclure-griffiths2006}. Figure~\ref{fig:rc linear interpolation slice} shows the comparison between an observed and `unabsorbed' velocity slice of the cloud at $4.9~{\rm km~s}^{-1}$. This is analogous to fig. 1 in \citet{mcclure-griffiths2006}, and demonstrates that our reimplementation of the interpolation method generally produces similar results. The differences along the Galactic Plane and in particular the Galactic Centre between the two implementations is likely primarily caused by the difference in the observed data, with the SGPS cube used in our implementation featuring the prominent emission from the Galactic Centre, while this has been removed in the data used by \citet{mcclure-griffiths2006}.

\begin{figure*}
    \centering
    \includegraphics[width=0.85\linewidth]{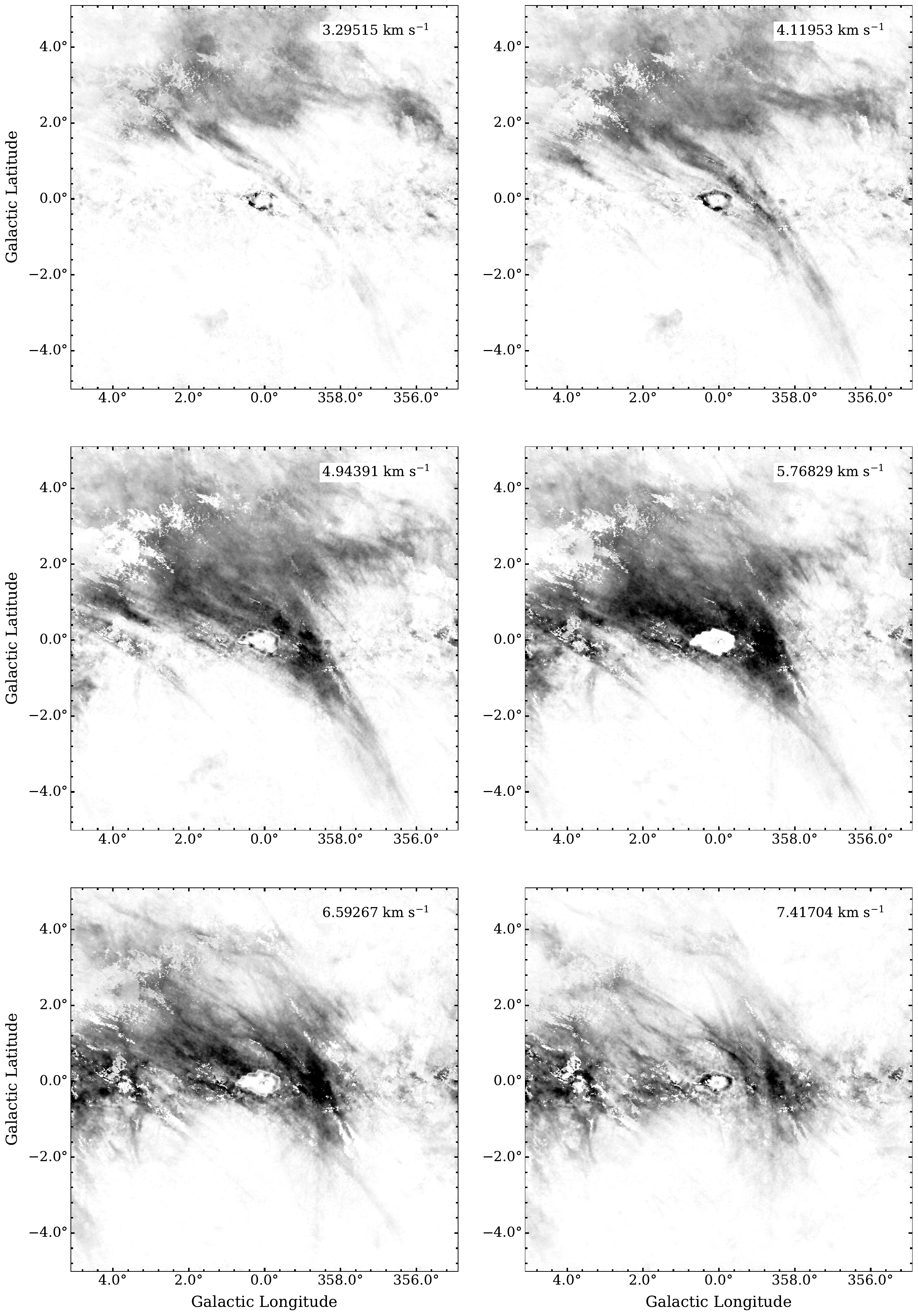}
    \caption{Velocity slices of $\Delta T_{\rm B}$, calculated as the difference between the observed and unabsorbed Riegel--Crutcher emission. Velocity slices from 3.3 to $7.4~{\rm km~s}^{-1}$ are shown. The grey scale is linear between 0 (white) and $-120~{\rm K}$ (black). This figure is analogous to fig. 3 in \citet{mcclure-griffiths2006}. Alt text: graphs showing six different grey-scale images of the same field. From left to right, top to bottom the graphs show a cloud structure begin to form, fill the majority of the field, then dissipate into filamentary structures. Text in the upper right of each graph reports the velocity of the image.}
    \label{fig:rc deltaT slices}
\end{figure*}

Figure~\ref{fig:rc deltaT slices} shows velocity slices of $\Delta T_{\rm B}$, calculated as the difference between the observed and unabsorbed Riegel--Crutcher emission, using our linear interpolation method adapted from \citet{mcclure-griffiths2006}. This figure is analogous to fig. 3 in \citet{mcclure-griffiths2006}, except again our implementation contains emission from the Galactic Centre that is not present in the original implementation.

\begin{figure}
    \centering
    \includegraphics[width=\linewidth]{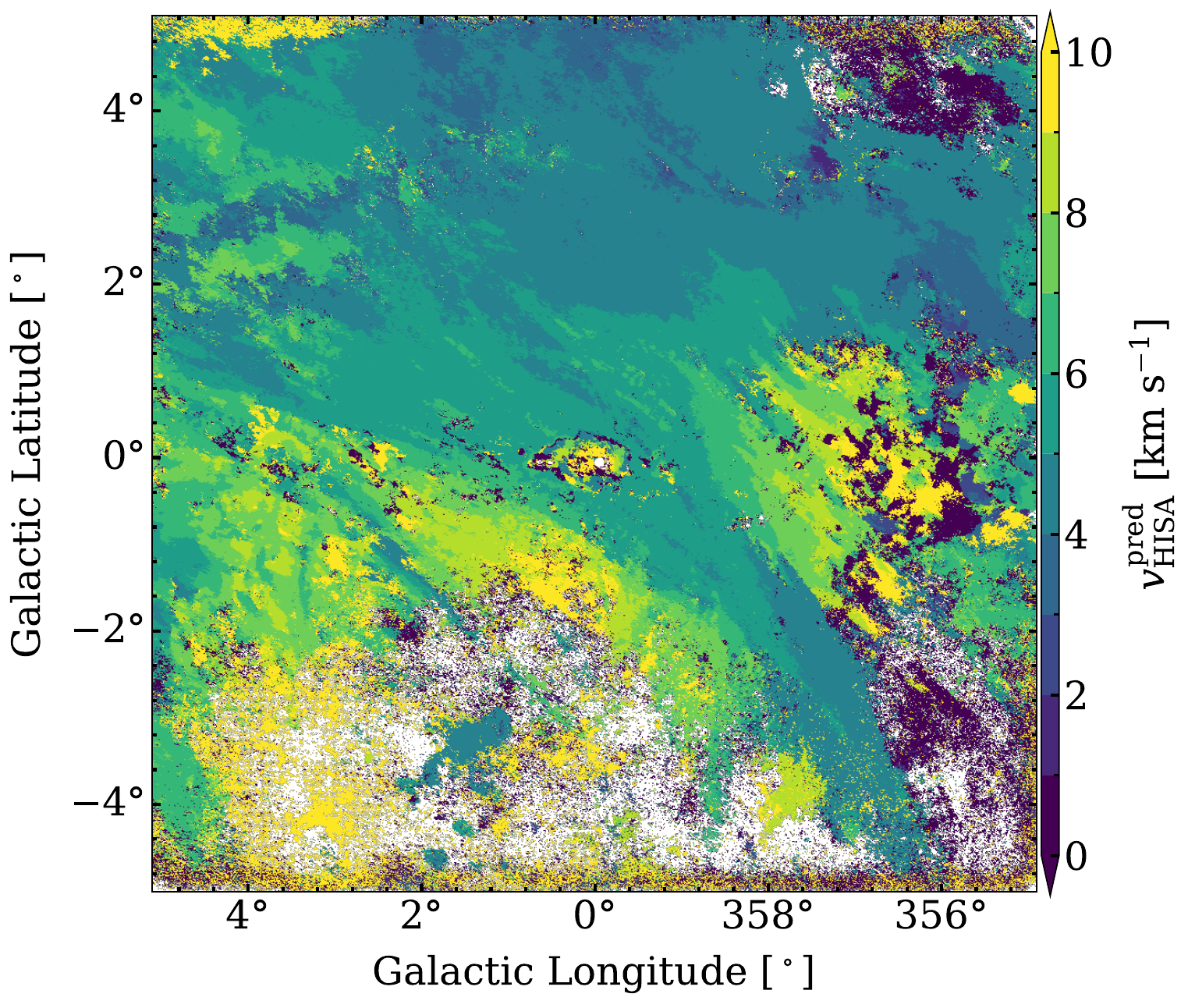}
    \caption{Predicted HISA velocity map for the \rc cloud produced by the linear interpolation method described in Section~\ref{sec:rc data}. Only pixels where the interpolation identifies an absorption feature are shown, coloured by the velocity of the deepest interpolated channel. The method recovers a similar velocity gradient to the neural network (Figure~\ref{fig:rc velocity map}), but produces a higher detection fraction along the Galactic Plane and towards the Galactic Centre, where complex emission is misidentified as self-absorption. Alt text: graph showing a cloud, with a colour gradient that changes across the cloud from top to bottom, representing the velocity of the cloud. While the main body of the cloud appears coherent, there are many coloured regions outside the cloud that appear incoherent and change colours rapidly.}
    \label{fig:rc linear interpolation velocity map}
\end{figure}

Figure~\ref{fig:rc linear interpolation velocity map} shows the predicted HISA velocity map, generated by taking the velocity of the minimum value of $\Delta T_{\rm B}$ for each spectrum in the cube. While the predicted velocities largely recovers the same physical structure and velocity gradient of the cloud as the neural network, it contains more lines of sight where a potential absorption feature is detected. Many of these features lie at velocities beyond the structure of the \rc cloud, and are especially common along the Galactic Plane and towards the Galactic Centre, where the neural network produces non-detections.

\begin{figure}
    \centering
    \includegraphics[width=0.95\linewidth]{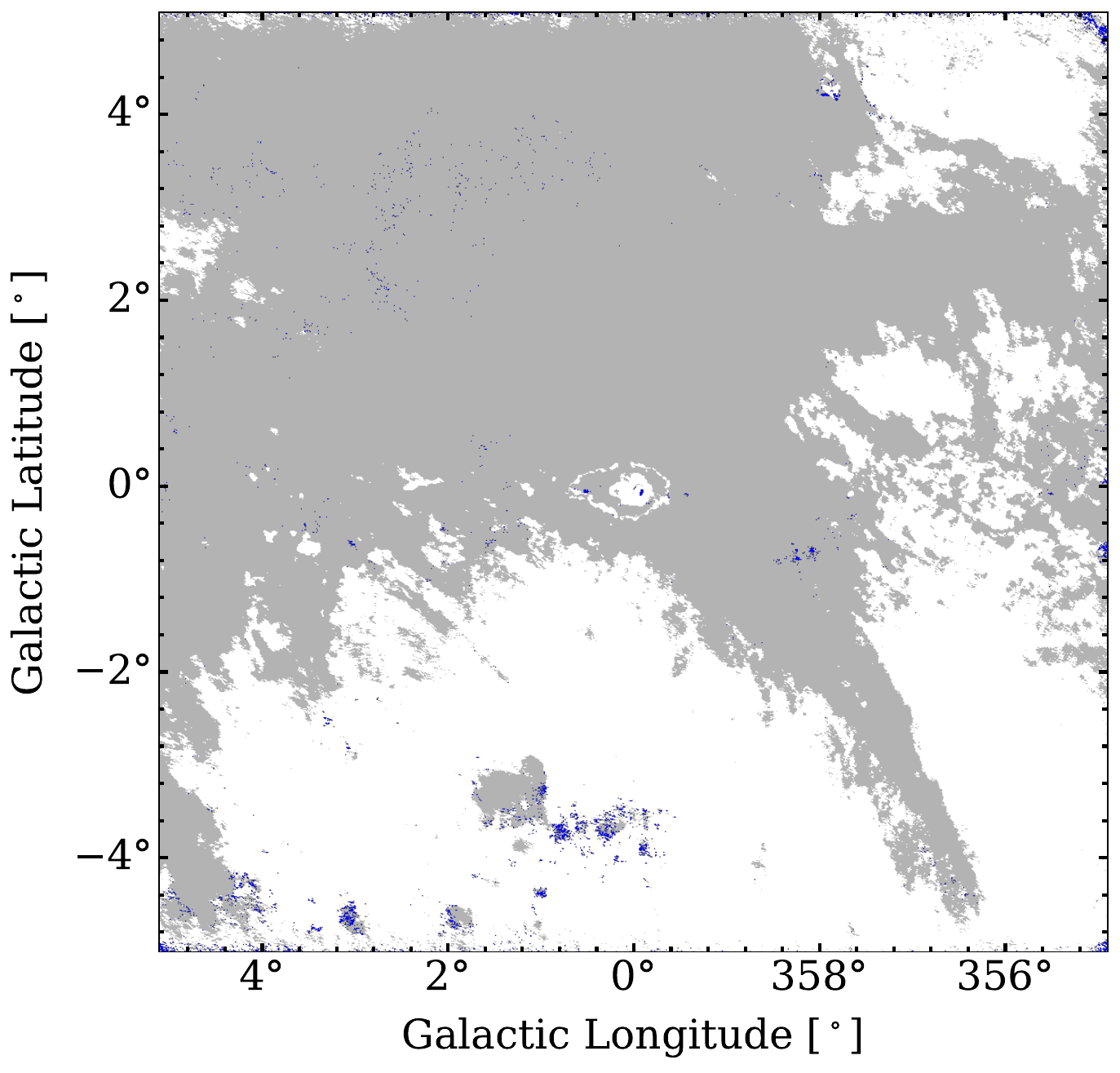}
    \caption{Map of the CNN-only HISA detections of the \rc cloud. The locations of CNN-only detections near the clumps of HISA detections in the bottom of the field indicate a physical origin of the detections. The CNN detection map is shown in grey. Alt text: graph showing a grey image of a cloud, with blue pixels scattered around the image, concentrated around small grey clouds at the bottom of the image.}
    \label{fig:cnn only detection map}
\end{figure}

\begin{figure}
    \centering
    \includegraphics[width=0.95\linewidth]{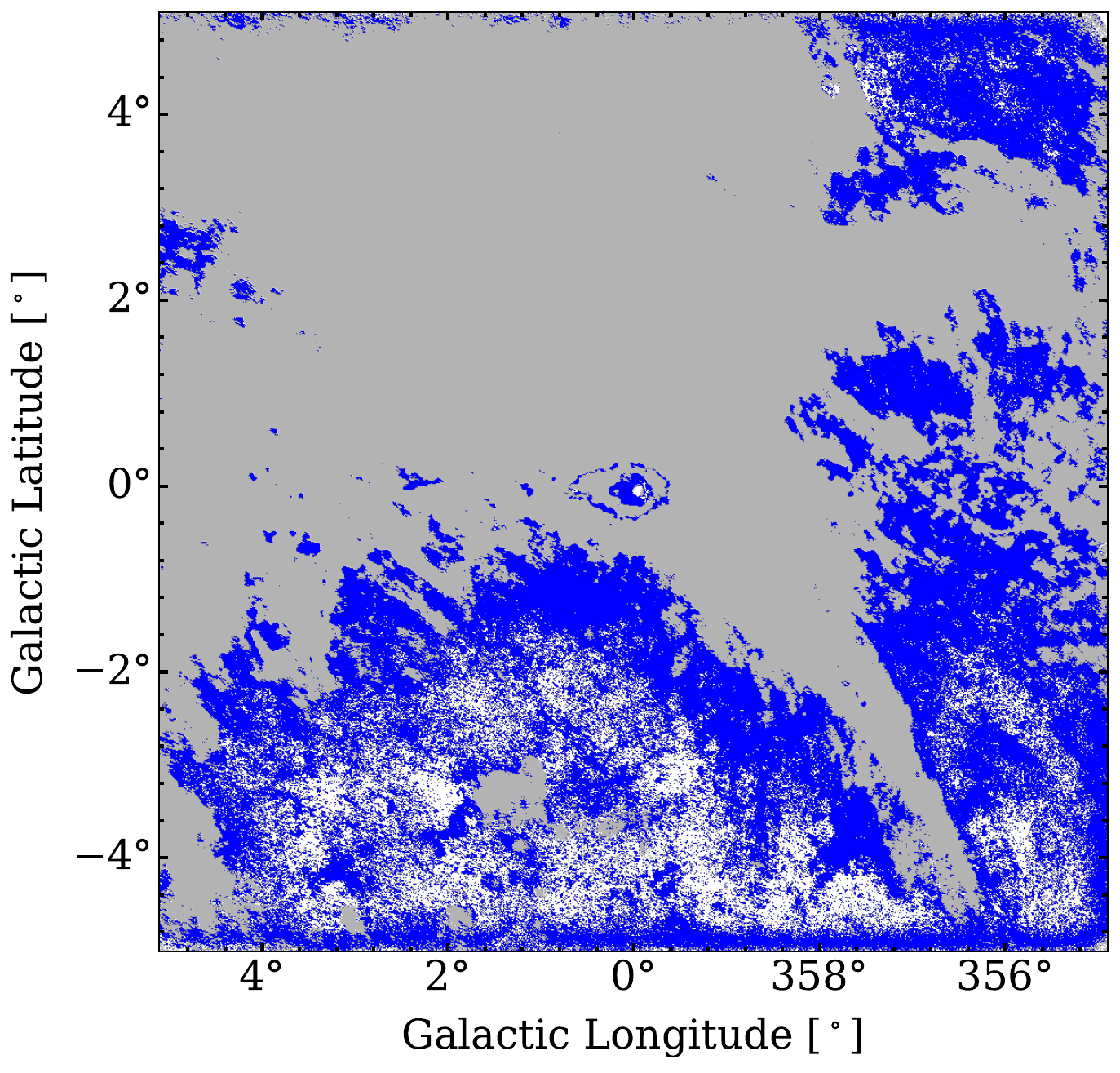}
    \caption{Map of the interpolation-only HISA detections of the \rc cloud. The interpolation-only detections fill much of the space not covered by the CNN detections, shown in grey, but are most prominent around the edges of the body of the cloud. Alt text: graph showing a grey image of a cloud, with blue pixels scattered around the image, filling most of the space around the cloud. The blue pixels are concentrated around the edge of the cloud.}
    \label{fig:linear interpolation only map}
\end{figure}

Figures~\ref{fig:cnn only detection map} and \ref{fig:linear interpolation only map} show the locations of CNN-only and interpolation-only detections in the \rc cloud.

\section{GMF Region Maps}\label{app:gmf maps}

Here we present HISA detections predicted by the neural network for the five giant molecular filaments not shown in Section~\ref{sec:gmf results} (GMF26, 38a, 38b, 41, and 54). Each figure shows a velocity slice  of the THOR+VGPS \ion{H}{i} emission data at a representative velocity of each filament, a slice of the PDF activation spectra produced by the neural network at the same velocity, and a slice of \coion emission from the GRS data at the same velocity.

\begin{figure*}
    \centering
    \includegraphics[width=0.95\linewidth]{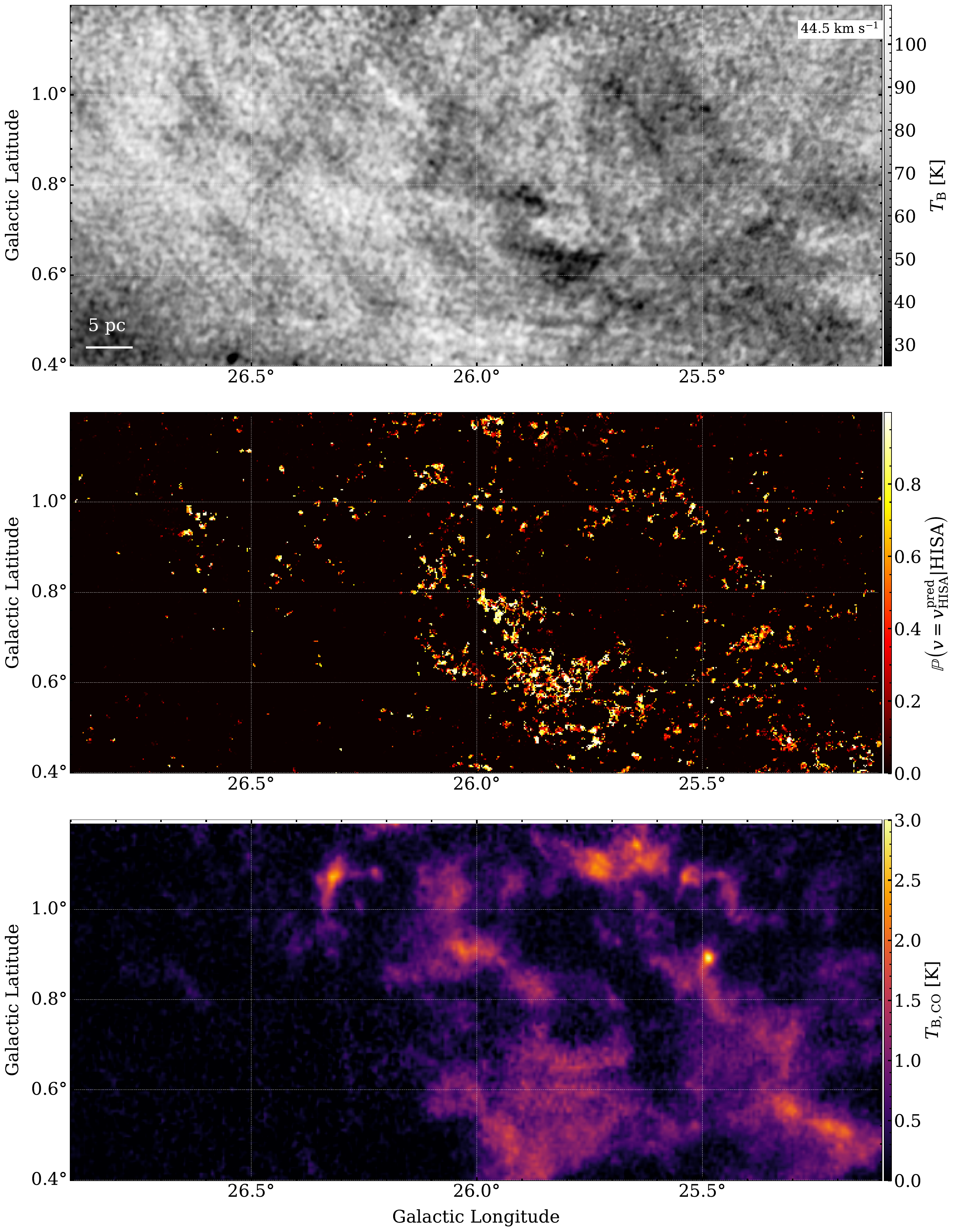}
    \caption{Neural network HISA detection in GMF26. Top: THOR+VGPS \ion{H}{i} emission slice at $44.5~{\rm km~s}^{-1}$. Middle: slice of the PDF activation spectra at the same velocity, showing the probability that detected HISA features in each pixel are located at $44.5~{\rm km~s}^{-1}$. Bottom: GRS \coion emission velocity slice at $44.6~{\rm km~s}^{-1}$. The velocity range of GMF26 is 41--$51~{\rm km~s}^{-1}$ \citep{ragan2014}. Alt text: three graphs showing very similar images in different colours. The top panel shows a grey-scale image of some clouds, and text in the upper right gives a velocity. The middle panel shows a heat map, with most of the panel filled in black, but some bright regions that align with dark patches in the top panel. The bottom panel shows a colour image of wispy cloud structures across the entire panel, with some bright patches at the same locations as the dark patches in the top panel, and the bright regions in the middle panel.}
    \label{fig:gmf26 detection map}
\end{figure*}

\begin{figure*}
    \centering
    \includegraphics[width=0.95\linewidth]{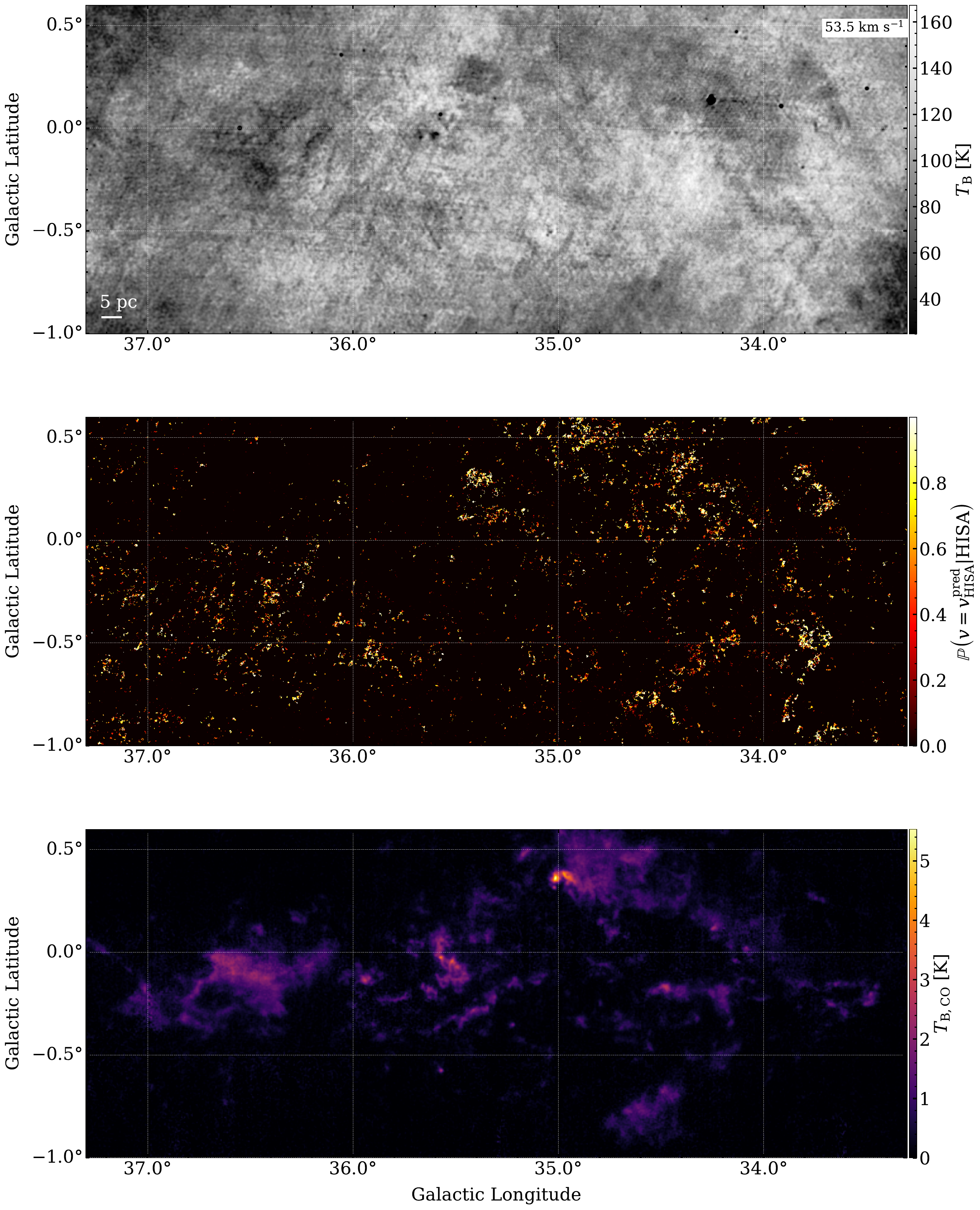}
    \caption{Neural network HISA detection in GMF38a. Top: THOR+VGPS \ion{H}{i} emission slice at $53.5~{\rm km~s}^{-1}$. Middle: slice of the PDF activation spectra at the same velocity, showing the probability that detected HISA features in each pixel are located at $53.5~{\rm km~s}^{-1}$. Bottom: GRS \coion emission velocity slice at $53.5~{\rm km~s}^{-1}$. The velocity range of GMF38a is 50--$60~{\rm km~s}^{-1}$ \citep{ragan2014}. Alt text: three graphs showing very similar images in different colours. The top panel shows a grey-scale image of some sparse clouds, and text in the upper right gives a velocity. The middle panel shows a heat map, with most of the panel filled in black, but some sparse bright regions that align with dark patches in the top panel. The bottom panel shows a colour image of wispy cloud structures horizontally across the middle of the panel, with some bright patches at the same locations as the dark patches in the top panel, and the bright regions in the middle panel.}
    \label{fig:gmf38a detection map}
\end{figure*}

\begin{figure*}
    \centering
    \includegraphics[width=0.95\linewidth]{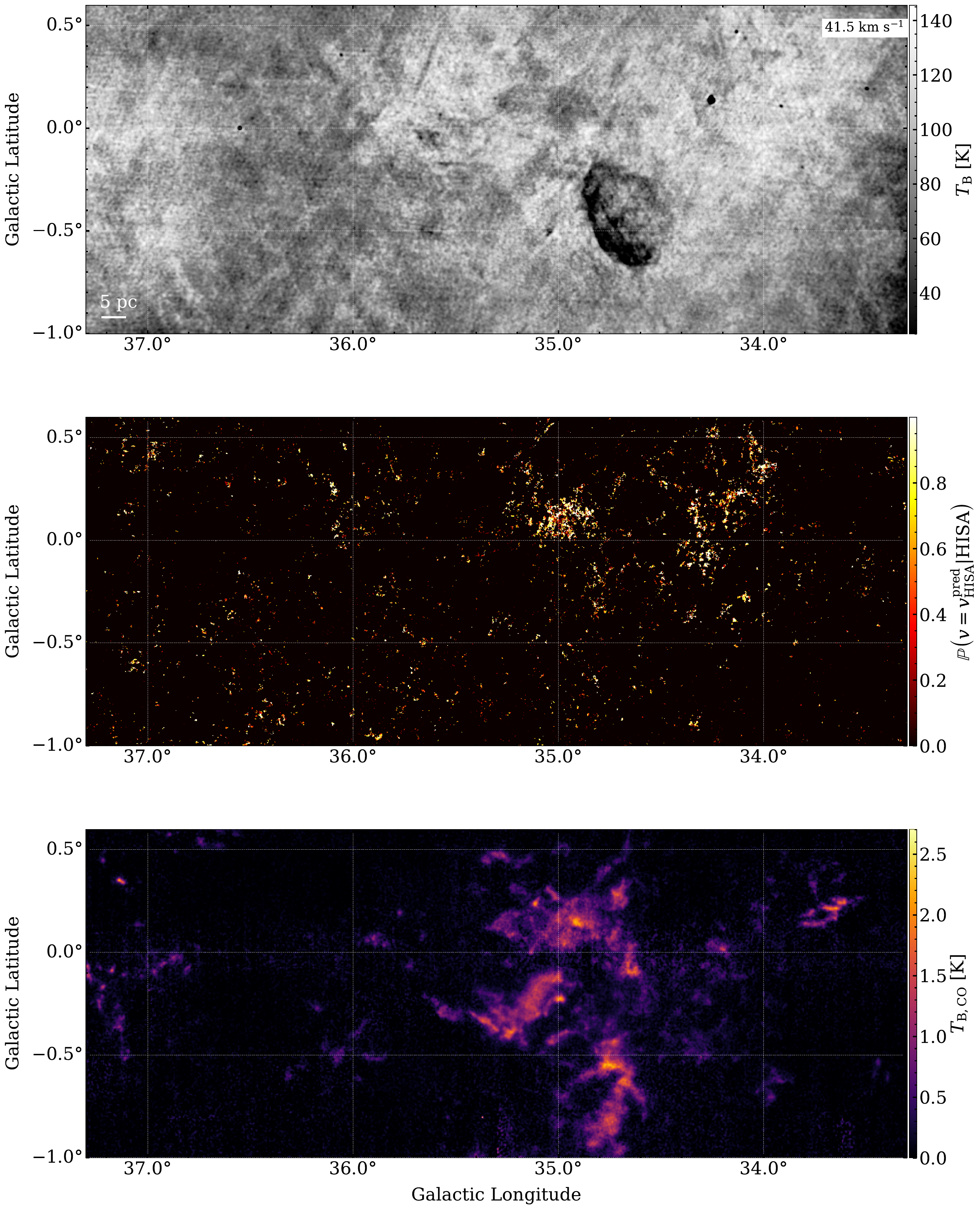}
    \caption{Neural network HISA detection in GMF38b. Top: THOR+VGPS \ion{H}{i} emission slice at $41.5~{\rm km~s}^{-1}$. Middle: slice of the PDF activation spectra at the same velocity, showing the probability that detected HISA features in each pixel are located at $41.5~{\rm km~s}^{-1}$. Bottom: GRS \coion emission velocity slice at $41.6~{\rm km~s}^{-1}$. The velocity range of GMF38b is 43--$46~{\rm km~s}^{-1}$ \citep{ragan2014}. Alt text: three graphs showing very similar images in different colours. The top panel shows a grey-scale image of some clouds with a very prominent dark circular structure near the middle of the image, and text in the upper right gives a velocity. The middle panel shows a heat map, with most of the panel filled in black, but some bright regions that align with dark patches in the top panel, but that do not align with the circular structure. The bottom panel shows a colour image of wispy cloud structures vertically across the middle of the panel, with some bright patches at the same locations as the dark patches in the top panel, and the bright regions in the middle panel.}
    \label{fig:gmf38b detection map}
\end{figure*}

\begin{figure*}
    \centering
    \includegraphics[width=0.995\linewidth]{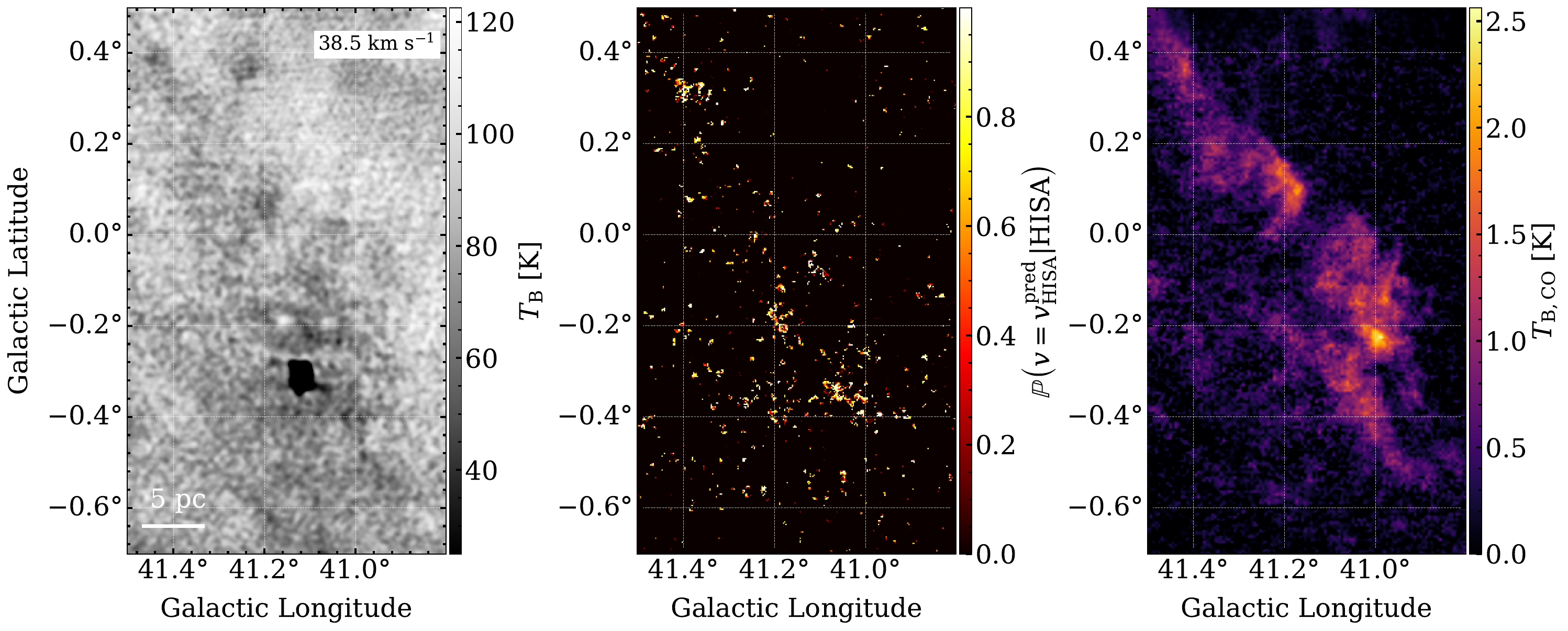}
    \caption{Neural network HISA detection in GMF41. Left: THOR+VGPS \ion{H}{i} emission slice at $38.5~{\rm km~s}^{-1}$. Middle: slice of the PDF activation spectra at the same velocity, showing the probability that detected HISA features in each pixel are located at $38.5~{\rm km~s}^{-1}$. Right: GRS \coion emission velocity slice at $38.4~{\rm km~s}^{-1}$. The velocity range of GMF41 is 34--$42~{\rm km~s}^{-1}$ \citep{ragan2014}. Alt text: three graphs showing very similar images in different colours. The left panel shows a grey-scale image of some dark structures that stretch from the top left to the bottom right of the image, with a dark spot along the structure towards the bottom of the image. Text in the upper right reports a velocity of the image. The middle panel shows a heat map, with most of the panel filled in black, but some sparse bright regions that align with the diagonal structure in the left panel. The right panel shows a colour image of thick cloud structures that trace the same dark patches in the left panel and the bright regions in the middle panel.}
    \label{fig:gmf41 detection map}
\end{figure*}

\begin{figure*}
    \centering
    \includegraphics[width=0.9\linewidth]{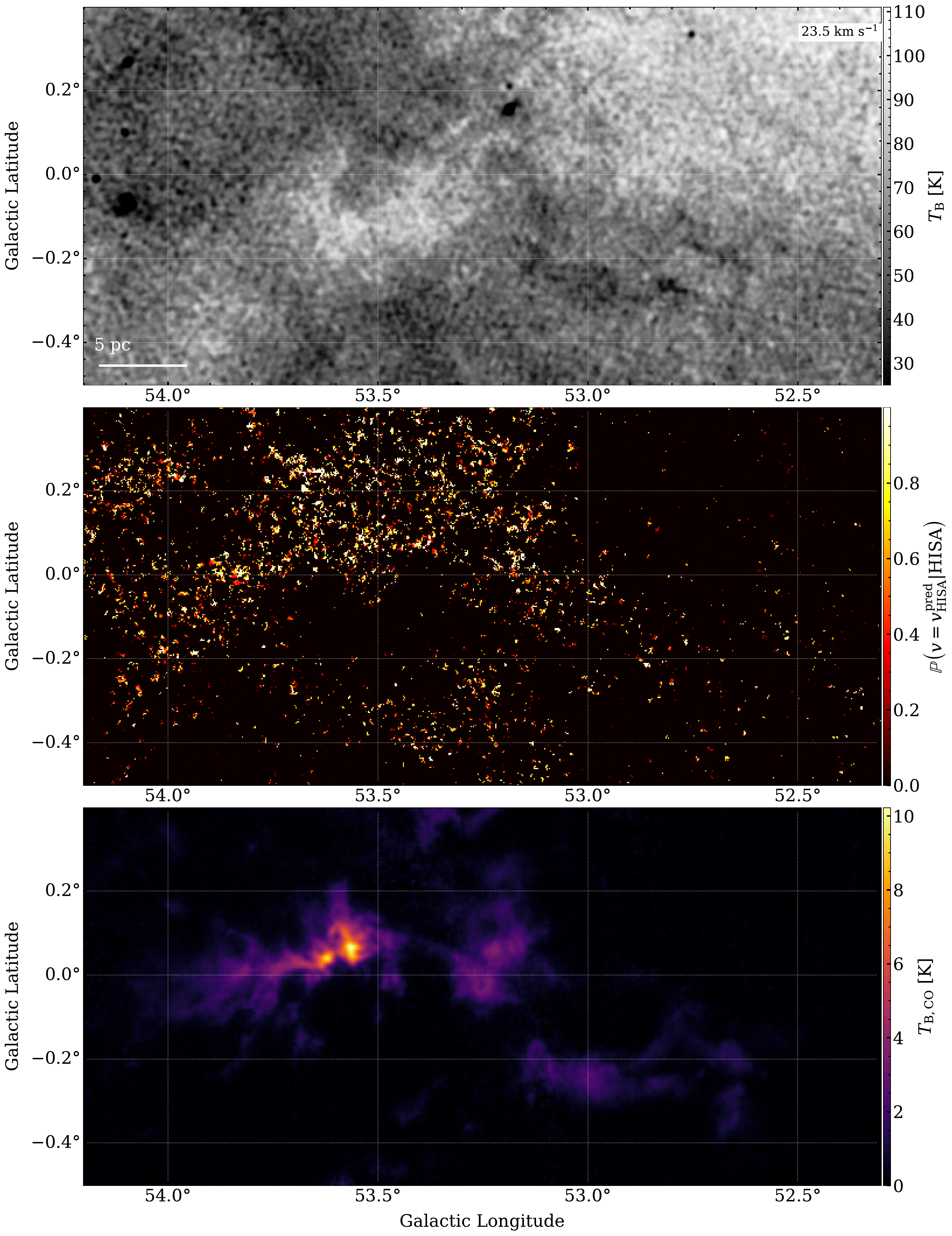}
    \caption{Neural network HISA detection in GMF54. Top: THOR+VGPS \ion{H}{i} emission slice at $23.5~{\rm km~s}^{-1}$. Middle: slice of the PDF activation spectra at the same velocity, showing the probability that detected HISA features in each pixel are located at $23.5~{\rm km~s}^{-1}$. Bottom: GRS \coion emission velocity slice at $23.5~{\rm km~s}^{-1}$. The velocity range of GMF54 is 20--$26~{\rm km~s}^{-1}$\citep{ragan2014}. Alt text: three graphs showing very similar images in different colours. The top panel shows a grey-scale image of some thin filaments, and text in the upper right reports the velocity of the image. The middle panel shows a heat map, with most of the panel filled in black, but some bright regions that roughly cover the same area as the dark filaments in the top panel. The bottom panel shows a colour image of cloud structures horizontally across the left of the panel, with some bright patches at the same locations as the dark patches in the top panel and the bright regions in the middle panel.}
    \label{fig:gmf54 detection map}
\end{figure*}

%%%%%%%%%%%%%%%%%%%%%%%%%%%%%%%%%%%%%%%%%%%%%%%%%%

% Don't change these lines
\bsp	% typesetting comment
\label{lastpage}
\end{document}